%% file: main.tex
\documentclass{SciPost}

\hypersetup{
    colorlinks,
    linkcolor={red!50!black},
    citecolor={blue!50!black},
    urlcolor={blue!80!black}
}

\usepackage[bitstream-charter]{mathdesign}
\makeatletter
\DeclareRobustCommand*{\doi}{\hyper@normalise\geodesiq@doi}
\def\geodesiq@doi#1{\doitext\hyper@linkurl{\Hurl{#1}}{\doiurl#1}}
\makeatother

\DeclareSymbolFont{usualmathcal}{OMS}{cmsy}{m}{n}
\DeclareSymbolFontAlphabet{\mathcal}{usualmathcal}

\fancypagestyle{SPstyle}{
\fancyhf{}
\lhead{\colorbox{scipostblue}{\bf \color{white} ~SciPost Physics Codebases }}
\rhead{{\bf \color{scipostdeepblue} ~Submission }}

\fancyfoot[C]{\textbf{\thepage}}
}

\usepackage{xcolor}
\usepackage[most]{tcolorbox}
\usepackage[frozencache]{minted}
\tcbuselibrary{minted}
\usepackage{needspace}

\usepackage{tikz}
\usetikzlibrary{calc,positioning,backgrounds,shapes.geometric,arrows.meta,shadows,fit}
\usepackage{standalone}

\definecolor{LightGray}{gray}{0.97}
\definecolor{terminalBg}{RGB}{240, 240, 250}
\definecolor{codePurple}{RGB}{142, 68, 173}
\definecolor{codeblue}{HTML}{E8F4FC}
\definecolor{codeborder}{HTML}{5dade2}
\definecolor{methodgray}{HTML}{f5f5f5}

\usetikzlibrary{shapes.geometric, arrows.meta, positioning, calc, shadows, fit, backgrounds}

\definecolor{navyblue}{RGB}{28, 54, 94}
\definecolor{controlblue}{RGB}{41, 128, 185}
\definecolor{pulsepurple}{RGB}{142, 68, 173}
\definecolor{dynamicsgreen}{RGB}{39, 174, 96}
\definecolor{lightbg}{RGB}{248, 249, 250}
\definecolor{cardbg}{RGB}{255, 255, 255}
\definecolor{bordergrey}{RGB}{200, 205, 210}
\definecolor{darktext}{RGB}{44, 62, 80}

\newcommand{\code}[1]{{\color{codePurple}\footnotesize\mintinline{python}{#1}}}
\setmintedinline{breaklines=true}
\setminted[python]{fontsize=\footnotesize,breaklines=true,breakindent=1em}
\usepackage{physics}

\newtcblisting{apisignature}{
    enhanced,
    breakable,
    listing only,
    listing engine=minted,
    minted language=python,
    minted options={fontsize=\footnotesize,breaklines=true,breakindent=1em},
    colback=codeblue,
    colframe=codeborder,
    boxrule=0pt,
    toprule=1.5pt,
    arc=0pt,
    outer arc=0pt,
    left=3mm,
    right=3mm,
    top=2mm,
    bottom=2mm
}

\newtcolorbox{methodbox}[1]{
    enhanced,
    breakable,
    colback=methodgray,
    colframe=black!25,
    boxrule=0.5pt,
    arc=1mm,
    left=4mm,
    right=4mm,
    top=3mm,
    bottom=3mm,
    title={\textbf{#1}},
    fonttitle=\normalsize,
    coltitle=black,
    colbacktitle=black!5,
    attach boxed title to top left={xshift=3mm,yshift=-2mm},
    boxed title style={
        boxrule=0pt,
        colframe=black!5,
        colback=black!5
    },
    before skip=8pt,
    after skip=10pt,
    left skip=10mm,
    before upper=\raggedright
}

\newcommand{\apiheading}[1]{%
    \par\Needspace{4\baselineskip}\vspace{0.8em}
    {\large\bfseries #1}
    \par\vspace{0.4em}
}

\newcommand{\apiitem}[3]{%
    \par\addvspace{0.35em}\noindent\hspace*{2em}%
    \begin{minipage}{\dimexpr\linewidth-2em\relax}
        \raggedright\code{#1}: #2\par\smallskip
        \noindent\hspace*{2em}%
        \begin{minipage}{\dimexpr\linewidth-2em\relax}
            \raggedright #3
        \end{minipage}
    \end{minipage}\par\vspace{0.5em}
}

\begin{document}

\pagestyle{SPstyle}

\begin{center}{\Large \textbf{\color{scipostdeepblue}{
geodesiq: a Python package for geometric \\ diabatic--adiabatic optimal state transfer\\
}}}\end{center}

\begin{center}\textbf{
David Fern\'andez-Fern\'andez\textsuperscript{1$\star$$\ddagger$},
Chris Ventura-Meinersen\textsuperscript{2$\dagger$$\ddagger$},
Maximilian Rimbach-Russ\textsuperscript{2}
and
Gloria Platero\textsuperscript{3,4}
}\end{center}

\begin{center}
{\bf 1} Institute of Physics, University of Augsburg, 86159 Augsburg, Germany
\\
{\bf 2} QuTech and Kavli Institute of Nanoscience, Delft University of Technology, PO Box 5046, 2600 GA Delft, The Netherlands
\\
{\bf 3} Instituto de Ciencia de Materiales de Madrid ICMM-CSIC, 28049 Madrid, Spain
\\
{\bf 4} Quantum Advanced Research Center QUARC-CSIC, 28006 Madrid, Spain
\\[\baselineskip]
$\star$ \href{mailto:david.fernandez.fernandez@uni-a.de}{\small david.fernandez.fernandez@uni-a.de}\,,\quad
$\dagger$ \href{mailto:c.venturameinersen@tudelft.nl}{\small c.venturameinersen@tudelft.nl}
\\
$\ddagger$ {\small These authors contributed equally to this work.}
\end{center}

\section*{\color{scipostdeepblue}{Abstract}}
\textbf{\boldmath{%
We introduce the \href{https://github.com/geodesiq-control/geodesiq-control}{\texttt{geodesiq}} Python package\footnote{The software version described here is \texttt{geodesiq} v0.5.0, DOI \href{https://doi.org/10.5281/zenodo.24321210}{\nolinkurl{10.5281/zenodo.24321210}}.}, which implements a geometric framework for shaping control pulses that achieve high-fidelity state transfer in multilevel quantum systems.
The underlying method combines adiabatic and diabatic control into a single metric, suppressing unwanted excitations while allowing selected diabatic passages.
Finding the pulse thereby reduces to solving a single first-order ordinary differential equation.
We explain how to use \texttt{geodesiq} and illustrate its use explicitly with Landau--Zener transitions, spin-qubit initialization, adiabatic many-body ground-state preparation, and a diabatic--adiabatic transfer in a four-level system.
}}

\vspace{\baselineskip}

\noindent\textcolor{white!90!black}{%
\fbox{\parbox{0.975\linewidth}{%
\textcolor{white!40!black}{\begin{tabular}{lr}%
  \begin{minipage}{0.6\textwidth}%
    {\small Copyright attribution to authors.\newline
    This work is a submission to SciPost Physics Codebases.\newline
    License information to appear upon publication.\newline
    Publication information to appear upon publication.}
  \end{minipage} & \begin{minipage}{0.4\textwidth}
    {\small Received Date \newline Accepted Date \newline Published Date}%
  \end{minipage}
\end{tabular}}
}}
}

\vspace{10pt}
\noindent\rule{\textwidth}{1pt}
\tableofcontents
\noindent\rule{\textwidth}{1pt}
\vspace{10pt}

\section{Introduction}
\label{sec:intro}

Harnessing a quantum advantage requires precise preparation of complex quantum states~\cite{glaserTrainingSchrodingersCat2015}.
A common challenge is to steer a state with high fidelity through the spectrum of the system Hamiltonian, in particular when the spectrum is dense.
Dense energy spectra are present in many quantum platforms like superconducting circuits~\cite{izmalkovObservationMacroscopicLandauZener2004, berkeTransmonPlatformQuantum2022}, semiconductor quantum dots~\cite{fernandez-fernandezFlyingSpinQubits2024, fernandez-fernandezLongrangeTransportAsymmetric2025, fernandez-fernandezSpinorbitenabledRealizationArbitrary2025, farinaSiteresolvedMagnonTriplon2025, jirovecManybodyInterferometrySemiconductor2026}, neutral atoms in optical lattices~\cite{karchDynamicalPreparationU12026} and Rydberg-atom arrays~\cite{bernienProbingManybodyDynamics2017}, among many others.
Quantum optimal control provides protocols that maximize a figure of merit, such as a fidelity, under experimental constraints.
Many approaches exist in the literature, including mathematical control theory~\cite{khanejaTimeOptimalControl2001, carliniTimeOptimalQuantumEvolution2006,pontryaginMathematicalTheoryOptimal2017,carliniTimeoptimalUnitaryOperations2007,boscainIntroductionPontryaginMaximum2021,koikeQuantumBrachistochrone2022, rezakhaniQuantumAdiabaticBrachistochrone2009, stepanenkoTimeOptimalTransferQuantum2025}, numerical optimization~\cite{khanejaOptimalControlCoupled2005,goodwinAcceleratedNewtonRaphsonGRAPE2023, doriaOptimalControlTechnique2011, canevaChoppedRandombasisQuantum2011, fauquenotOpenClosedLoop2025, katiraee-farUnifiedEvolutionaryOptimization2025}, spectral methods~\cite{martinisFastAdiabaticQubit2014, rimbach-russSimpleFrameworkSystematic2023, wuSimultaneousHighFidelitySingleQubit2025, polatPulseShapingUltrafast2026}, and techniques that accelerate adiabatic processes, known as shortcuts to adiabaticity~\cite{bergmannCoherentPopulationTransfer1998,motzoiSimplePulsesElimination2009,berryTransitionlessQuantumDriving2009,degrandiAdiabaticPerturbationTheory2010,ivakhnenkoNonadiabaticLandauZener2023,vuthaSimpleApproachLandauZener2010,motzoiOptimalControlMethods2011,chenLewisRiesenfeldInvariantsTransitionless2011,ribeiroSystematicMagnusBasedApproach2017,selsMinimizingIrreversibleLosses2017,theisCounteractingSystemsDiabaticities2018,banFastLongrangeCharge2018,takahashiHamiltonianEngineeringAdiabatic2019,banSpinEntangledState2019,guery-odelinShortcutsAdiabaticityConcepts2019,setiawanAnalyticDesignAccelerated2021,takahashiDynamicalInvariantFormalism2022,glasbrennerLandauZenerFormula2023,dengisAcceleratedCreationNOON2025,romeroOptimizingEdgestateTransfer2024, liuAcceleratedAdiabaticPassage2024,xuImprovingCoherentPopulation2019,fehseGeneralizedFastQuasiadiabatic2023,limaPartialLandauZenerTransitions2025,richermeExperimentalPerformanceQuantum2013, rolandQuantumSearchLocal2002,martinez-garaotFastQuasiadiabaticDynamics2015, baksicSpeedingAdiabaticQuantum2016, stepanenkoQuantumAdvantageTransfer2026, tangExploringGroundStates2024, tangWeightedNestedCommutators2026}.
Several open-source packages implement numerical optimal control, for example \texttt{Krotov}~\cite{goerzKrotovPythonImplementation2019}, \texttt{QuOCS}~\cite{rossignoloQuOCSQuantumOptimal2023}, \texttt{C3}~\cite{wittlerIntegratedToolSet2021}, \texttt{Qiskit Dynamics}~\cite{puzzuoliQiskitDynamicsPython2023}, and the optimal control packages of \texttt{QuTiP}~\cite{lambertQuTiP5Quantum2026}.
These packages iteratively optimize a cost function such as the fidelity, whereas \texttt{geodesiq} constructs the pulse directly from the spectrum of the Hamiltonian, without iterating on the fidelity.
A key requirement for experimental platforms is the generality and immediate applicability of any method.
Ideally, a protocol should not rely on special symmetries, be directly usable through modulation or pulse shaping of experimentally accessible parameters, and be numerically efficient to implement.
In this work, we focus on optimizing state-transfer protocols, where the initial and target states are fixed, and the overlap between the time-evolved initial state and the target state is maximized.

We introduce the \texttt{geodesiq} Python package, which recasts optimal state transfer as a geometric problem, based on the quantum metric tensor and its extensions~\cite{chenSpeedingQuantumAdiabatic2022, ventura-meinersenQuantumGeometricProtocols2025, meinersenUnifyingAdiabaticStatetransfer2025, ventura-meinersenMultilevelSpectralNavigation2026a, dengisTimeindependentCounterdiabaticDriving2026}.
The quantum metric tensor~\cite{kolodrubetzClassifyingMeasuringGeometry2013, houLocalGeometryQuantum2024, liskaHiddenSymmetriesBianchi2021} underlies many phenomena in physics, including quantum sensing~\cite{sidhuGeometricPerspectiveQuantum2020, valvoProbingElectrostaticDisorder2026}, quantum phase transitions~\cite{lambertClassicalQuantumInformation2023, camposvenutiQuantumCriticalScaling2007}, the onset of many-body ergodicity~\cite{ventura-meinersenGeometricSignaturesOnset2026}, and the training of neural quantum states~\cite{dashEfficiencyNeuralQuantum2025, dawidMachineLearningQuantum2025}.
The \texttt{geodesiq} Python package can be installed with \texttt{pip}:
\begin{minted}
    [bgcolor=terminalBg, fontsize=\footnotesize]{bash}
pip install geodesiq
\end{minted}
The \texttt{geodesiq} library\footnote{The library is released under the GNU Lesser General Public License, version 2.1 only (SPDX: LGPL-2.1-only); see Appendix~\ref{app: LGPL}.} implements the algorithms that compute optimal state-transfer protocols for a single control parameter\footnote{The extension to several control parameters is discussed as future work in Sec.~\ref{sec: conclusion}.} of the system Hamiltonian.
Beyond the pulse solver, the package provides tools to analyze and post-process the pulse and to simulate the resulting time evolution, including open-system dynamics.
The manuscript is structured as follows:
\begin{itemize}
    \item Sec.~\ref{sec: theory} reviews the theory behind the algorithm;
    \item Sec.~\ref{sec: geodesiq} presents the API using the two-level Landau--Zener model;
    \item Sec.~\ref{sec: examples} applies it to spin-qubit initialization, many-body ground-state preparation, and a diabatic--adiabatic transfer in a four-level system;
    \item Sec.~\ref{sec: conclusion} concludes; and
    \item the appendices contain the license note, formal derivations, numerical benchmarks, and the API documentation.
\end{itemize}

We note that the actual use of the \texttt{geodesiq} Python library does not require an understanding of the theoretical underpinnings of the pulse generation.
Hence, to obtain practical intuition first, we refer the reader to Sec.~\ref{sec: geodesiq} and then to the examples in Sec.~\ref{sec: examples}.

\section{Theory of optimal state transfer}
\label{sec: theory}
In this section, we outline protocols that speed up adiabatic evolution, allowing an eigenstate to be transported while reducing undesired excitations to other eigenstates.
We review how several existing adiabatic protocols are unified in a single geometric framework~\cite{meinersenUnifyingAdiabaticStatetransfer2025} and how this framework extends beyond minimizing undesired transitions to target specific desired transitions while remaining adiabatic with respect to others.
We set $\hbar=1$ throughout, except in the description of the \code{Dynamics} API, which accepts a user-defined $\hbar$.

\subsection{Introduction to quantum state transfer}

As discussed in the introduction, quantum optimal control encompasses a wide range of optimization tasks.
In general, it engineers a time-evolution operator that implements a specific operation (in quantum computing, a gate) under environmental or experimental constraints such as noise.
Gate design constrains the full unitary, because the gate must act correctly on arbitrary input states.

\begin{figure}[!ht]
    \centering
    \includegraphics[width=\linewidth]{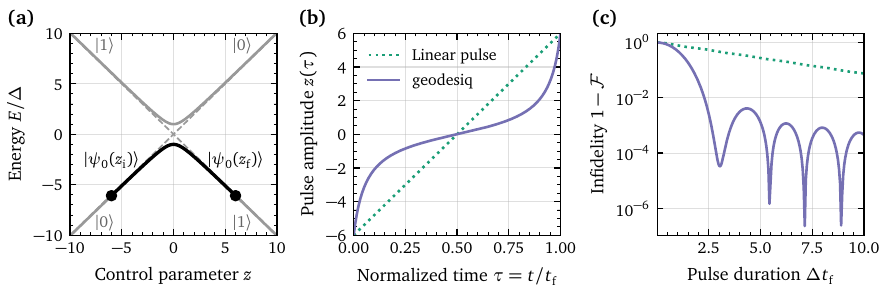}
    \caption{Quantum state transfer with the \texttt{geodesiq} Python library.
    Panel \textbf{(a)} shows the eigenenergies (gray solid lines) of the two-level Hamiltonian $\hat{H}=z\,\sigma^z+\Delta\,\sigma^x$ as a function of the control parameter $z$, in units of $\Delta=1$.
    The gray dashed lines are the diabatic energies $\pm z$, and the gray labels mark the diabatic state that each branch approaches far from the anticrossing.
    The state transfer follows the ground state (bold black line) from $\ket{\psi_0(z_\mathrm{i})}$ at $z_\mathrm{i}=-6\Delta$ to $\ket{\psi_0(z_\mathrm{f})}$ at $z_\mathrm{f}=6\Delta$ (black markers).
    The linear ramp [$(\alpha,\beta)=(0,0)$, green dotted line] and the \texttt{geodesiq}-optimized pulse shape for $(\alpha,\beta)=(2,2)$ (purple) of this sweep are shown in panel \textbf{(b)} as functions of the normalized time $\tau=t/t_\mathrm{f}$.
    Panel \textbf{(c)} shows the resulting final-state infidelity $1-\mathcal{F}$ on a logarithmic scale as a function of the total pulse time $\Delta t_\mathrm{f}$, sampled in $[0.01, 10]$.}
    \label{fig: introduction}
\end{figure}

Quantum state transfer deals with engineering a time evolution that maps a known initial state to a desired final state.
Therefore, the resulting time-evolution operator is not required to perform a known gate, although for some systems it can also be used to design gates.
A ubiquitous example of state transfer is the Landau--Zener (LZ) model~\cite{ivakhnenkoNonadiabaticLandauZener2023}, described by the two-level time-dependent Hamiltonian $\hat{H}(t)=z(t)\,\sigma^z+\Delta\,\sigma^x$, where $z(t)=vt$ is a linear ramp with speed $v$, $\Delta$ is a constant coupling between the two levels, $\sigma^z:=\ketbra{0}-\ketbra{1}$ and $\sigma^x:=\ketbra{0}{1}+\ketbra{1}{0}$.
For $\Delta=0$, the two basis states $\ket{0},\ket{1}$ are decoupled and form the diabatic eigenstates of the system Hamiltonian.
For $\Delta\neq 0$, the Hamiltonian eigenstates are time-dependent and are described by the instantaneous (adiabatic) eigenstates $\ket{\psi_0(t)}, \ket{\psi_1(t)}$ with instantaneous eigenenergies $E_{0,1}(t)=\mp \sqrt{z(t)^2+\Delta^2}$, respectively.
Figure~\ref{fig: introduction}(a) shows that the adiabatic eigenstates approach the diabatic states for $|z(t)|\gg |\Delta|$.
The index $n$ of $\ket{\psi_n}$ counts the instantaneous levels from the bottom of the spectrum and does not label a basis state: the ground state $\ket{\psi_0}$ approaches $\ket{0}$ for $z\ll-|\Delta|$ and $\ket{1}$ for $z\gg|\Delta|$.
Within this model, the state-transfer task is usually stated as follows: \textit{Starting from the diabatic state $\ket{0}$, can we shape $z(t)$ such that the final state, after some total pulse time $t_\mathrm{f}$, is given by $\ket{1}$?}
One route is adiabatic state transfer, where one follows the ground state without inducing any excitations.
For a linear ramp $z(t)=vt$, the transition probability has the closed form $P_{0\to1}=1-\exp(-\pi \Delta^2/v)$ (the LZ formula; see Ref.~\cite{ivakhnenkoNonadiabaticLandauZener2023} for derivations)\footnote{For the derivation of the formula one takes $t_\mathrm{i}\to-\infty$ and $t_\mathrm{f}\to+\infty$.}.
Adiabatic transfer ($P_{0\to1}\approx1$) therefore requires $\Delta^2/v\gg1$, i.e., a sweep rate that is small compared with the square of the minimum gap.
For a linear sweep from $z_\mathrm{i}$ to $z_\mathrm{f}$ in a time $t_\mathrm{f}$, the LZ formula gives an infidelity $1-P_{0\to1}\approx\exp[-\pi \Delta^2 t_\mathrm{f}/(z_\mathrm{f}-z_\mathrm{i})]$, so the linear ramp only approaches a high-fidelity state transfer for long pulse times $t_\mathrm{f}\propto 1/v$.
For the sweep from $z_\mathrm{i}=-6\Delta$ to $z_\mathrm{f}=6\Delta$ used in Fig.~\ref{fig: introduction}, this infidelity, $\approx e^{-\pi \Delta t_\mathrm{f}/12}$, remains above $7\times10^{-2}$ over the whole window of panel (c).

The linear ramp has an immediate drawback: in the presence of decoherence, long pulse times reduce the fidelity.
A faster ramp does not help either, because it induces non-adiabatic transitions that also reduce the fidelity.
The \texttt{geodesiq} Python package allows for an optimization of the pulse shape that goes beyond a simple linear ramp.
Figure~\ref{fig: introduction}(b) compares a \texttt{geodesiq} pulse (purple) with the linear ramp (dotted green).
Reducing the sweep rate $dz/dt$ near the anticrossing yields a state-transfer infidelity below $10^{-3}$ for all $\Delta t_\mathrm{f}\gtrsim 6.5$ [Fig.~\ref{fig: introduction}(c)], whereas the LZ estimate above requires $\Delta t_\mathrm{f}\approx 26$ for the linear ramp.

The following subsections describe the theoretical foundations of the \texttt{geodesiq} pulse-shaping algorithm: adiabatic perturbation theory and extensions of the quantum metric tensor.

\subsection{Adiabatic perturbation theory and FAQUAD}
The adiabatic theorem guarantees that a system follows an instantaneous eigenstate with arbitrary accuracy if the gap stays open and the evolution is slow enough~\cite{albashAdiabaticQuantumComputation2018}.
Because gaps typically shrink with system size, the required evolution time usually grows unfavorably.
Hence, methods to speed up the adiabatic evolution have been proposed, collectively known as shortcuts to adiabaticity~\cite{guery-odelinShortcutsAdiabaticityConcepts2019}.
The ideal adiabatic evolution can be enforced with an additional counterdiabatic driving term~\cite{demirplakAdiabaticPopulationTransfer2003, berryTransitionlessQuantumDriving2009}, which cancels all non-adiabatic transitions.
In generic systems, however, this term involves operators (often non-local or many-body) that are not available as experimental controls.

Another approach is to develop approximate methods that can be applied by shaping and modulating existing control parameters of the Hamiltonian.
This is also the starting point of the \texttt{geodesiq} package.
We start with a Hamiltonian $\hat{H}(x^\mu(t))$, where $x^\mu$ corresponds to a set of experimentally accessible control parameters.
To obtain a condition for approximate adiabaticity, we expand the state in the instantaneous eigenbasis,
\begin{align*}
    \ket{\Psi(t)}=\sum_n c_n(t)\,e^{-i\int_0^t E_n(t')\,dt'}\ket{\psi_n(t)}.
\end{align*}
The coefficients obey
\begin{align}
    \dot{c}_m=-c_m\braket{\psi_m}{\dot{\psi}_m}-\sum_{n\neq m}c_n\frac{\mel{\psi_m}{\dot{\hat{H}}}{\psi_n}}{E_n-E_m}e^{-i\int_0^t dt'\,[E_n(t')-E_m(t')]},
\end{align}
where the second term on the right-hand side couples $\ket{\psi_m}$ to the other eigenstates $\ket{\psi_n}$ with energy $E_n$.
The dot denotes the total time derivative: $\dot{c}_m\equiv dc_m/dt$.
For a system that occupies the eigenstate $\ket{\psi_k}$, adiabaticity requires
\begin{align}
    \frac{|\mel{\psi_k}{\dot{\hat{H}}}{\psi_n}|}{|E_n-E_k|}=\frac{|\mel{\psi_k}{\partial_x{\hat{H}}}{\psi_n}|}{|E_n-E_k|}\left|\dv{x}{t}\right|\ll |E_n-E_k|,    \hspace{1cm}  \forall n\neq k.
\end{align}
From here on, $k$ always labels the occupied eigenstate.
This motivates the following condition for approximate adiabaticity of a single control parameter $x(t)$, known as fast quasiadiabatic dynamics (FAQUAD)~\cite{martinez-garaotFastQuasiadiabaticDynamics2015, fernandez-fernandezQuantumControlHole2022}:
\begin{align}
\label{eqn: faquad}
    \sum_{n\neq k}\frac{|\mel{\psi_k}{\dot{\hat{H}}}{\psi_n}|}{(E_n-E_k)^2}=\sum_{n\neq k}\frac{|\mel{\psi_k}{\partial_x{\hat{H}}}{\psi_n}|}{(E_n-E_k)^2}\left|\dv{x}{t}\right|=\delta_\mathrm{adiabatic},
\end{align}
where the constant $\delta_\mathrm{adiabatic}$ is fixed by the boundary values $x_\mathrm{i}\equiv x(0)$, $x_\mathrm{f}\equiv x(t_\mathrm{f})$ and the total pulse time $t_\mathrm{f}$.
For a monotonic sweep, it is
\begin{align}
    \label{eqn: faquad delta}
    \delta_\mathrm{adiabatic}=\frac{1}{t_\mathrm{f}}\left|\int_{x_\mathrm{i}}^{x_\mathrm{f}}dx \sum_{n\neq k}\frac{|\mel{\psi_k}{\partial_x{\hat{H}}}{\psi_n}|}{(E_n-E_k)^2}\right|.
\end{align}
In this language, we expect an adiabatic state transfer if $\delta_\mathrm{adiabatic}\ll 1$.
Solving the differential equation, Eq.~\eqref{eqn: faquad}, for a given $t_\mathrm{f}$ yields a pulse that distributes the adiabaticity parameter uniformly along the path, i.e., the sweep slows down where the gaps are small and speeds up elsewhere.
Requiring $\delta_\mathrm{adiabatic}\ll 1$ in Eq.~\eqref{eqn: faquad delta} then gives the shortest acceptable pulse time.
This approach was introduced by Martinez-Garaot et al.~\cite{martinez-garaotFastQuasiadiabaticDynamics2015} as FAQUAD and also appears as fast-QUAD, e.g., in Ref.~\cite{fehseGeneralizedFastQuasiadiabatic2023}.

In a multilevel system, Eq.~\eqref{eqn: faquad} combines the adiabaticity parameters of the individual transitions,
\begin{align}
    \label{eqn: faquad eta}
    \eta_{kn}=\frac{|\mel{\psi_k}{\partial_x{\hat{H}}}{\psi_n}|}{(E_n-E_k)^2}\left|\dv{x}{t}\right|,
\end{align}
through their sum, i.e., the $\ell^1$ norm of the vector $(\eta_{kn})_{n\neq k}$.
This is the multilevel FAQUAD protocol of Ref.~\cite{palmeroGenerationCatStates2019}, which includes the $K$ transitions that couple the occupied state appreciably (FAQUAD-$K$).
The sum is not the only way to obtain a single condition: keeping the Euclidean ($\ell^2$) norm $(\sum_{n\neq k}\eta_{kn}^2)^{1/2}$ or the largest parameter $\max_{n\neq k}\eta_{kn}$ ($\ell^\infty$ norm) constant is equally consistent with the local adiabaticity condition.
The latter is the generalized FAQUAD of Ref.~\cite{garcia-ripollQuantumControlFrequencyTunable2020}, which takes the maximum over all transitions out of a subset of eigenstates to preserve that subset.
The three conditions coincide when a single transition contributes, as in a two-level system, but yield different pulses when several transitions contribute.
We return to this choice in Sec.~\ref{sec: transition norm}.
\subsection{Optimal control with the quantum metric tensor}
\label{sec: geometric control}
A second approach, mentioned in the introduction, is geometric.
Here, the task of optimizing the fidelity of state transfer $\ket{\psi_k(x)}\to \ket{\psi_k(x')}$ is described by a parameter-space trajectory $x\to x'$ through the geometry induced by the quantum metric tensor $g_{\mu\nu}$, where the indices refer to all possible parameters $x^\mu=(x^1,x^2, \dots, x^{N_\mathrm{c}})$ of the system Hamiltonian.
The quantum metric tensor $g_{\mu \nu}$ describes the infinitesimal distance between two pure states via the local infidelity (up to second order in parameter changes)~\cite{liskaHiddenSymmetriesBianchi2021}
\begin{align}
    \label{eqn: fidelity susceptibility}
    1-|\braket{\psi_k(x)}{\psi_k(x+dx)}|^2\approx g_{\mu \nu}(x)dx^\mu dx^\nu.
\end{align}
The quantum metric tensor of the occupied eigenstate $\ket{\psi_k}$ can be written in terms of the eigenstates and eigenenergies of the Hamiltonian~\cite{provostRiemannianStructureManifolds1980, kolodrubetzGeometryNonadiabaticResponse2017}
\begin{align}
    g_{\mu \nu}=\Re \sum_{n\neq k} \frac{\mel{\psi_k}{\partial_\mu \hat{H}}{\psi_n}\mel{\psi_n}{\partial_\nu \hat{H}}{\psi_k}}{(E_n-E_k)^2},
\end{align}
where $\partial_\mu = \partial/\partial x^\mu$ is the derivative with respect to the parameters of the Hamiltonian.
A state transfer corresponds to a path connecting the set of initial parameter values $x^\mu_\mathrm{i}\equiv x^\mu(0)$ to some final set $x^\mu_\mathrm{f}\equiv x^\mu(t_\mathrm{f})$.
Fast, high-fidelity state transfer then amounts to finding the path $x^\mu_\mathrm{geo}(t)$ between these points that minimizes the length functional
\begin{align}
    \label{eqn: length functional}
    \mathcal{L}[x,\dot{x}]=\int_{0}^{t_\mathrm{f}} dt \,\sqrt{g_{\mu \nu}(x)\dv{x^\mu}{t}\dv{x^\nu}{t}}.
\end{align}

Since $\mathcal{L}$ is invariant under reparametrizations of $t$, for a single control parameter all monotonic sweeps between $x(0)$ and $x(t_\mathrm{f})$ have the same length.
The time dependence is instead fixed by minimizing the action $\mathcal{S}[x]=\int_0^{t_\mathrm{f}}dt\,g_{\mu\nu}\dot{x}^\mu\dot{x}^\nu\geq \mathcal{L}^2/t_\mathrm{f}$, whose minimizers are the constant-speed (affinely parametrized) geodesics,
\begin{align}
    \label{eqn: adiabatic quantum geo protocol}
    g_{\mu \nu}(x)\dv{x^\mu}{t}\dv{x^\nu}{t}=\delta_\mathrm{geometric}^2.
\end{align}
Here, $\delta_\mathrm{geometric}$ is a constant fixed by the boundary conditions.
Unlike $\delta_\mathrm{adiabatic}$, it is a rate (an energy for $\hbar=1$), and an adiabatic transfer requires it to be small compared with the gaps $|E_n-E_k|$ along the path, $\delta_\mathrm{geometric}\ll\min_{x,\,n\neq k}|E_n(x)-E_k(x)|$.
Along the occupied instantaneous eigenstate $\ket{\psi_k(x)}$, the variance of $\hat{H}$ itself vanishes.
The quantity set by the quantum metric is instead the variance of the generator of the transport, the counterdiabatic Hamiltonian $\hat{H}_\mathrm{CD}=\dot{x}^\mu\hat{\mathcal{A}}_\mu$ with the adiabatic gauge potential $\hat{\mathcal{A}}_\mu=i\partial_\mu$~\cite{bukovGeometricSpeedLimit2019}:
\begin{align}
    \sigma^2_\mathrm{CD}=\expval{\hat{H}_\mathrm{CD}^2}-\expval{\hat{H}_\mathrm{CD}}^2=g_{\mu \nu}(x)\dv{x^\mu}{t}\dv{x^\nu}{t}.
\end{align}
The local infidelity accumulated along the sweep, Eq.~\eqref{eqn: fidelity susceptibility}, is therefore proportional to the action $\mathcal{S}=\int_0^{t_\mathrm{f}}dt\,\sigma^2_\mathrm{CD}$.
For a fixed $t_\mathrm{f}$, a constant-speed geodesic does not reduce the local infidelity pointwise.
Instead, it distributes it evenly along the sweep, $\sigma^2_\mathrm{CD}=\delta^2_\mathrm{geometric}$, and thereby minimizes both its time integral, $\mathcal{S}=\mathcal{L}^2/t_\mathrm{f}$, and its maximum, since $\max_t\sigma_\mathrm{CD}\geq\mathcal{L}/t_\mathrm{f}$ for any monotonic sweep.
The value $\delta_\mathrm{geometric}=\mathcal{L}/t_\mathrm{f}$ itself is not optimized: it is fixed by the endpoints of the sweep and by $t_\mathrm{f}$.
For a single parameter $x^\mu=x(t)$, we can express $\delta_\mathrm{geometric}$ in terms of the pulse boundary conditions
\begin{align}
    \delta_\mathrm{geometric} = \frac{1}{t_\mathrm{f}}\left|\int_{x_\mathrm{i}}^{x_\mathrm{f}}dx \, \sqrt{g_{xx}}\right|=\frac{\mathcal{L}[x]}{t_\mathrm{f}}.
\end{align}
Hence, energy-variance-minimizing protocols may be understood as trajectories that locally minimize the length of the path $\mathcal{L}[x]$ that they trace out, i.e., geodesics traversed at constant metric speed in time $t_\mathrm{f}$.
From the perspective of adiabatic perturbation theory,
\begin{align*}
    g_{xx}\dot{x}^2=\sum_{n\neq k}|\braket{\psi_n}{\partial_t\psi_k}|^2
\end{align*}
is the squared non-adiabatic coupling in the moving frame, which a geodesic pulse keeps constant along the sweep.
This allows us to draw a connection between the FAQUAD protocol and geometry.
To find an optimal time evolution of $x(t)$, we need to solve
\begin{align}
    \label{eqn: fQ equation}
    g_{xx}\dot{x}^2=\sum_{n\neq k}\frac{|\mel{\psi_k}{\partial_x \hat{H}}{\psi_n}|^2}{(E_n-E_k)^2}\left(\dv{x}{t}\right)^2=\delta_\mathrm{geometric}^2.
\end{align}
For a single transition, Eq.~\eqref{eqn: fQ equation} differs from the square of the FAQUAD condition, Eq.~\eqref{eqn: faquad}~\cite{martinez-garaotFastQuasiadiabaticDynamics2015, fehseGeneralizedFastQuasiadiabatic2023, xuImprovingCoherentPopulation2019}, only in the power of the energy splitting in the denominator: 2 here and 4 for FAQUAD~\cite{chenSpeedingQuantumAdiabatic2022}.
For several transitions, Eq.~\eqref{eqn: fQ equation} sums the squares of the individual terms ($\ell^2$ norm), whereas Eq.~\eqref{eqn: faquad} squares their sum ($\ell^1$ norm), so that the two conditions also differ in how the transitions are combined.
\subsection{Generalized metrics and diabatic--adiabatic protocols}
\subsubsection{Unifying adiabatic protocols}
This similarity suggests a common framework for both.
The quantum metric depends on two ingredients: the coupling matrix elements $\mel{\psi_k}{\partial_\mu \hat{H}}{\psi_n}$ and the energy splittings $(E_n-E_k)$.
These factors determine the resulting control pulse and the associated error.
Introducing two exponents $(\alpha,\beta)$, we define an extended quantum geometric tensor~\cite{meinersenUnifyingAdiabaticStatetransfer2025}
\begin{subequations}
    \label{eqn: generalized qgt}
    \begin{align}
    \mathcal{Q}^{(\alpha, \beta)}_{\mu \nu}&:= \sum_{n\neq k} \frac{P^{(kn)}_{\mu\nu}\,\big|P^{(kn)}_{\mu\nu}\big|^{\beta/2-1}}{|E_n-E_k|^{\alpha}},\\
    P^{(kn)}_{\mu\nu}&=\mel{\psi_k}{\partial_\mu \hat{H}}{\psi_n}\mel{\psi_n}{\partial_\nu \hat{H}}{\psi_k},\\
    G^{(\alpha, \beta)}_{\mu \nu}&:= \Re \mathcal{Q}^{(\alpha, \beta)}_{\mu \nu}= \sum_{n\neq k} G^{kn,(\alpha, \beta)}_{\mu \nu},\\
    G^{kn,(\alpha,\beta)}_{\mu\nu}&=\frac{\Re\big(P^{(kn)}_{\mu\nu}\big|P^{(kn)}_{\mu\nu}\big|^{\beta/2-1}\big)}{|E_n-E_k|^{\alpha}}.
    \end{align}
\end{subequations}
We call $\mathcal{Q}^{(\alpha,\beta)}_{\mu\nu}$ the \textit{$(\alpha,\beta)$-hypergeometric tensor}, where $\alpha,\beta \in \mathbb{R}$, and $G^{(\alpha,\beta)}_{\mu\nu}$ the associated \textit{hypermetric tensor}, with components $G^{kn,(\alpha,\beta)}_{\mu\nu}$.
For $(\alpha,\beta)=(2,2)$, $G^{(\alpha,\beta)}_{\mu\nu}$ reduces to $g_{\mu\nu}$.
For general exponents, the diagonal elements read
\begin{equation}
    G^{(\alpha,\beta)}_{\mu\mu}=\sum_{n\neq k}\frac{|\mel{\psi_k}{\partial_\mu \hat{H}}{\psi_n}|^{\beta}}{|E_n-E_k|^{\alpha}}.
\end{equation}
The products $P^{(kn)}_{\mu\nu}$ are invariant under the gauge transformations $\ket{\psi_n}\mapsto e^{i\chi_n}\ket{\psi_n}$, and raising only their modulus to a non-integer power avoids the branch ambiguity that would appear if the complex matrix elements were raised to the power $\beta/2$ individually.
Moreover, since $P^{(kn)}_{\nu\mu}=(P^{(kn)}_{\mu\nu})^*$, the hypergeometric tensor is Hermitian, and the hypermetric tensor is real and symmetric.
The diagonal elements need a convention for transitions that $\partial_\mu\hat{H}$ does not couple, such as those to levels of another symmetry sector: their numerator is then $0^0$ for $\beta=0$ and diverges for $\beta<0$.
For $\beta=0$, we set $0^0=1$, so that every transition contributes $|E_n-E_k|^{-\alpha}$, irrespective of its matrix element.
The weight then depends only on the spectrum, as in the local-adiabaticity condition, and does not jump where a matrix element vanishes; in particular, $(\alpha,\beta)=(0,0)$ gives a constant metric, $G^{(0,0)}_{\mu\mu}=d-1$ for $d$ levels, and hence the linear ramp of Table~\ref{tab: adiabatic protocols and hypergeo}.
For $\beta<0$, these terms are set to zero: a transition that $\partial_\mu\hat{H}$ does not couple cannot be driven, and should not slow down the pulse.
For $\beta>0$, they vanish anyway.

Carrying the geometric picture of $g_{\mu\nu}$ over to the hypermetric tensor $G_{\mu\nu}^{(\alpha,\beta)}$ heuristically, we propose the analogous pulse equation~\cite{meinersenUnifyingAdiabaticStatetransfer2025}
\begin{align}
    \label{eq: hypergeo protocol}
    G_{\mu \nu}^{(\alpha,\beta)}\dv{x^\mu}{\tau}\dv{x^\nu}{\tau}=\big(\mathcal{L}^{(\alpha,\beta)}\big)^2,
\end{align}
where $\mathcal{L}^{(\alpha,\beta)}$ is the length of the path measured with $G^{(\alpha,\beta)}_{\mu\nu}$.
Unlike Eqs.~\eqref{eqn: faquad}--\eqref{eqn: fQ equation}, Eq.~\eqref{eq: hypergeo protocol} is written in the normalized time $\tau=t/t_\mathrm{f}\in[0,1]$, so that $dx^\mu/d\tau=t_\mathrm{f}\,dx^\mu/dt$.
As before, $\mathcal{L}^{(\alpha,\beta)}$ is fixed by the boundary conditions, but it no longer depends on $t_\mathrm{f}$: for a monotonic single-parameter sweep,
\begin{align}
    \label{eqn: path length}
    \mathcal{L}^{(\alpha,\beta)}=\left|\int_{x_\mathrm{i}}^{x_\mathrm{f}}dx\,\sqrt{G^{(\alpha,\beta)}_{xx}}\right|,
\end{align}
and for $(\alpha,\beta)=(2,2)$ it is the length $\mathcal{L}^{(2,2)}=\mathcal{L}[x]=t_\mathrm{f}\,\delta_\mathrm{geometric}$ of Eq.~\eqref{eqn: length functional}.
It only normalizes the pulse; the speed in physical time is $\mathcal{L}^{(\alpha,\beta)}/t_\mathrm{f}$, which equals $\delta_\mathrm{geometric}$ for $(2,2)$ and the $\ell^2$ form of $\delta_\mathrm{adiabatic}$ (Sec.~\ref{sec: transition norm}) for $(4,2)$, and is dimensionless only in the latter case.
The pulse $x^\mu(\tau)$ is therefore computed once; for any pulse time $t_\mathrm{f}$, the physical pulse at time $t$ is its value at $\tau=t/t_\mathrm{f}$.
Different choices of $(\alpha,\beta)$ recover the protocols listed in Table~\ref{tab: adiabatic protocols and hypergeo}; see Ref.~\cite{meinersenUnifyingAdiabaticStatetransfer2025} for details.
\begin{table}[t!]
    \centering
    \begin{tabular}{|c|c|c|}\hline
         $(\alpha, \beta)$  & Protocol & Reference \\ \hline \hline 
         $(0, 0)$ & Linear & \cite{degrandiAdiabaticPerturbationTheory2010,ivakhnenkoNonadiabaticLandauZener2023, vuthaSimpleApproachLandauZener2010, glasbrennerLandauZenerFormula2023} \\ \hline
         $(2, 2)$ & Geometric & \cite{ventura-meinersenQuantumGeometricProtocols2025} \\ \hline
         $(4, 2)$ & FAQUAD & \cite{martinez-garaotFastQuasiadiabaticDynamics2015, fernandez-fernandezQuantumControlHole2022, fehseGeneralizedFastQuasiadiabatic2023, chenSpeedingQuantumAdiabatic2022} \\ \hline
         $(2\,\alpha_\mathrm{LA}, 0)$ & Local adiabaticity & \cite{fernandez-fernandezQuantumControlHole2022, guery-odelinShortcutsAdiabaticityConcepts2019, richermeExperimentalPerformanceQuantum2013, rolandQuantumSearchLocal2002} \\ \hline
    \end{tabular}
    \caption{Adiabatic protocols and the corresponding exponents $(\alpha,\beta)$ of the hypermetric tensor.
    $\alpha_\mathrm{LA}$ is the exponent of the energy gap in the local-adiabaticity condition.
    For several transitions, the exponents $(4,2)$ with the default $p=2$ in Eq.~\eqref{eqn: transition norm} give the $\ell^2$ form of FAQUAD, and $p=1$ gives the sum of Eq.~\eqref{eqn: faquad}.
    Table adapted from~\cite{meinersenUnifyingAdiabaticStatetransfer2025}.}
    \label{tab: adiabatic protocols and hypergeo}
\end{table}

\subsubsection{Combining transitions in multilevel systems}
\label{sec: transition norm}
The hypermetric tensor, Eq.~\eqref{eqn: generalized qgt}, sums the contributions of the transitions out of the occupied state.
For a single control, the pulse speed in Eq.~\eqref{eq: hypergeo protocol} is therefore set by $\sqrt{G^{(\alpha,\beta)}_{xx}}$, the $\ell^2$ norm of the amplitudes $(G^{kn,(\alpha,\beta)}_{xx})^{1/2}$ of the individual transitions.
For $(\alpha,\beta)=(4,2)$, these amplitudes are the adiabaticity parameters of Eq.~\eqref{eqn: faquad eta} per unit speed, $\eta_{kn}/|\dot{x}|$, so that Eq.~\eqref{eq: hypergeo protocol} keeps the $\ell^2$ norm of $(\eta_{kn})_{n\neq k}$ constant, not their sum as in Eq.~\eqref{eqn: faquad}.
To cover the alternatives discussed below Eq.~\eqref{eqn: faquad eta}, \texttt{geodesiq} combines the transitions with an $\ell^p$ norm of order $p\in[1,\infty]$,
\begin{align}
    \label{eqn: transition norm}
    G^{(\alpha,\beta)}_{xx,p}=\Big[\sum_{n\neq k}\big(G^{kn,(\alpha,\beta)}_{xx}\big)^{p/2}\Big]^{2/p},
    \qquad
    G^{(\alpha,\beta)}_{xx,\infty}=\max_{n\neq k}G^{kn,(\alpha,\beta)}_{xx},
\end{align}
which replaces $G^{(\alpha,\beta)}_{xx}$ in Eqs.~\eqref{eq: hypergeo protocol} and~\eqref{eqn: path length}.
We call this scalar the single-control metric and write it as $G(x)\equiv G^{(\alpha,\beta)}_{xx,p}(x)$; for $p\neq2$, it is no longer the component of a tensor.
The default $p=2$ is the hypermetric tensor itself, $p=1$ with $(\alpha,\beta)=(4,2)$ gives the multilevel FAQUAD condition, Eq.~\eqref{eqn: faquad}, and $p=\infty$ follows only the transition with the largest weight at each control value, which for $(\alpha,\beta)=(4,2)$ is the generalized FAQUAD of Ref.~\cite{garcia-ripollQuantumControlFrequencyTunable2020} for a single followed state.
All orders agree when a single transition contributes.
For $K$ contributing transitions, they obey $G_{xx,\infty}\le G_{xx,2}\le G_{xx,1}\le K\,G_{xx,2}$, and since these ratios vary along the sweep, the orders differ in where the pulse slows down: $p=1$ adds up the weights of all transitions, whereas $p=\infty$ responds only to the dominant one.

We choose $p=2$ as the default for three reasons.
First, it is the only order for which the combined weight $\sum_{n\neq k}G^{kn}_{\mu\nu}dx^\mu dx^\nu$ is a quadratic form in the parameter displacements, as required for a metric tensor; for $(\alpha,\beta)=(2,2)$, it is the quantum metric tensor $g_{\mu\nu}$, with its connection to the local infidelity, Eq.~\eqref{eqn: fidelity susceptibility}, and to the energy variance.
Second, it has a direct perturbative meaning: for $(\alpha,\beta)=(4,2)$, $\sum_{n\neq k}\eta_{kn}^2$ is the squared norm of the first-order correction to the adiabatically followed state~\cite{degrandiAdiabaticPerturbationTheory2010}, i.e., the local population outside it, which the sum of Eq.~\eqref{eqn: faquad} only bounds, $\|\eta\|_2\le\|\eta\|_1\le\sqrt{K}\,\|\eta\|_2$.
Third, the metric remains smooth wherever its contributions are, whereas the maximum ($p=\infty$) has kinks where the dominant transition changes, and these carry over to the pulse.
None of these local conditions fixes the final infidelity, which depends on the interference of the transition amplitudes accumulated along the path, as the oscillations of the FAQUAD infidelity with the pulse time show (Fig.~\ref{fig: landau_zener}).
The order $p$ is therefore exposed as an option, \code{norm}, that can be tuned together with the exponents.
Appendix~\ref{app: metric comparison} compares the exponents and orders for systems with different spectra, including a qubit register in which the choice of $p$ changes the infidelity by orders of magnitude.

\subsubsection{Beyond adiabatic evolution: Diabatic--adiabatic control}
The resulting adiabatic protocols achieve high-fidelity state transfers by staying in a particular eigenstate of the Hamiltonian.
They are therefore not suited to transfers that require selected diabatic passages.
Motivated by the adiabatic-impulse approximation~\cite{tomkaAccuracyAdiabaticimpulseApproximation2018, suzukiGeneralizedAdiabaticImpulse2022}, we reduce the weight $G_{\mu\nu}^{\,kn}$ of the transitions that should be passed diabatically, so that the pulse traverses these anticrossings rapidly (impulse regime), e.g., by choosing a diabatic gap exponent $\tilde{\alpha}<0$.
Despite the non-perturbative nature of diabatic evolution, we interpolate between adiabatic and diabatic weighting with an ansatz that is linear in the exponents of each transition, and define the \textit{diabatic--adiabatic (di--ad) tensor}
\begin{subequations}
    \begin{align}
        \label{eqn: di-ad tensor}
        \left(\mathcal{G}_{\mathrm{di\text{--}ad}}\right)_{\mu\nu}^{(\alpha,\beta;\;\tilde{\alpha},\tilde{\beta})}(x) &=\sum_{n\neq k(x)} G_{\mu\nu}^{\,kn,(\alpha_{kn},\,\beta_{kn})},\\
        \alpha_{kn}(x)&=\xi_{kn}(x)\,\alpha+[1-\xi_{kn}(x)]\,\tilde{\alpha}, \\
        \beta_{kn}(x)&=\xi_{kn}(x)\,\beta+[1-\xi_{kn}(x)]\,\tilde{\beta},
    \end{align}
\end{subequations}
where $k(x)$ labels the instantaneous eigenstate that is occupied at the control value $x$.
The new parameters $(\tilde{\alpha}, \tilde{\beta})$ are free parameters that can be chosen to favor diabatic passages
\footnote{The adiabatic and diabatic contributions carry different dimensions, $E^{\beta-\alpha}X^{-\beta}$ and $E^{\tilde\beta-\tilde\alpha}X^{-\tilde\beta}$, where $E$ and $X$ are the units of energy and of the control $x$. A change of units therefore rescales the diabatic weights relative to the adiabatic ones by a constant factor, which changes the di--ad pulse whenever $(\alpha,\beta)\neq(\tilde\alpha,\tilde\beta)$. This only affects how the pulse $x_\text{gq}(t)$ is generated; hence, it might obstruct an interpretation of the diabatic and adiabatic components.}
As in the adiabatic case, only transitions out of the occupied state contribute.
For $p\neq2$, they are combined with the $\ell^p$ norm of Eq.~\eqref{eqn: transition norm} instead of the sum, each with its blended exponents, and the single-control metric $G(x)$ is then this combination for $\mu=\nu=x$.
The components of the \textit{di--ad transition matrix} $\xi_{kn}(x)\in[0,1]$ are defined in the eigenbasis of the Hamiltonian as
\begin{align}
    \xi_{kn}(x)=\begin{cases}
        0 & \text{diabatic: } \ket{\psi_k}{\to}\ket{\psi_n}\\
        1 & \text{adiabatic: } \ket{\psi_k}{\to}\ket{\psi_k} \text{ and }\ket{\psi_k}{\not\to}\ket{\psi_n}
    \end{cases},
\end{align}
with intermediate values only where a diabatic passage is smoothly switched off (see below).
Wherever $\xi_{kn}\in\{0,1\}$, the di--ad tensor splits into an adiabatic component, built from the transitions with $\xi_{kn}=1$, and a diabatic component, built from those with $\xi_{kn}=0$.

The transition matrix is fixed by the initial and target states, $\ket{\psi_{k_\mathrm{i}}}$ and $\ket{\psi_{k_\mathrm{f}}}$, together with the spectrum along the sweep $x_\mathrm{i}\to x_\mathrm{f}$.
Starting from $k(x_\mathrm{i})=k_\mathrm{i}$, the occupied state is followed along the sweep, with $s=\operatorname{sgn}(k_\mathrm{f}-k_\mathrm{i})$ the direction towards the target:
\begin{enumerate}
    \item Each anticrossing between the occupied state $\ket{\psi_k}$ and its neighbor $\ket{\psi_{k+s}}$, i.e., each local minimum $x^*$ of the gap $|E_{k+s}-E_k|$ inside the sweep, is passed diabatically, and the occupied state becomes $\ket{\psi_{k+s}}$ at $x^*$.
    A minimum counts only if it is resolved, i.e., if the gap rises above it on both sides by more than $10^{-12}$ times the spectral bandwidth, so that rounding errors of a flat gap are not mistaken for anticrossings.
    This is repeated until the target state $\ket{\psi_{k_\mathrm{f}}}$ is reached.
    \item All other anticrossings are passed adiabatically: anticrossings with the neighbor away from the target, and all anticrossings after the target has been reached, including a second anticrossing of the same pair of levels.
    Hence, two levels are never crossed diabatically back and forth.
    \item A diabatic passage acts only within its anticrossing, i.e., between the closest maxima $x^*_-<x^*$ and $x^*_+>x^*$ of the gap on either side of $x^*$.
    It is fully diabatic over the first half of the way from $x^*$ to each maximum.
    Over the second half, $\xi_{kn}$ increases smoothly to one as $3u^2-2u^3$, with $u=2|x-x^*|/|x^*_\pm-x^*|-1\in[0,1]$, so that the metric does not jump where the diabatic passage ends.
    The sweep endpoints are not treated as gap maxima, so a passage is not switched off towards them.
    \item Overlapping anticrossings, for instance a fan of levels crossing at a single point, are passed together.
\end{enumerate}
A true level crossing, at which $\mel{\psi_k}{\partial_x\hat{H}}{\psi_n}=0$ for the crossing level $\ket{\psi_n}$ (for instance, because the two states belong to different symmetry sectors), is not an anticrossing.
There, the occupied state keeps its identity, and only its energy index $k(x)$ changes; \texttt{geodesiq} follows it through the crossing by the overlap of the eigenvectors at neighboring grid points.
The same holds for an adiabatic transfer, whose target index $k_\mathrm{f}$ is then the index in which the followed state ends.
If the target state cannot be reached through anticrossings towards it, the transfer is not defined (and \texttt{geodesiq} raises an \code{InvalidControlParameterError}).
This includes a sweep that stops before the anticrossing: a gap that is smallest at an endpoint of the sweep is not an anticrossing.
If an anticrossing is to be traversed, extending the sweep to bracket it is usually the clearest physical setup.
For a deliberately truncated passage, \code{allow_endpoint_passages=True} also counts such an endpoint minimum as an anticrossing; a passage at $x_\mathrm{i}$ then relabels the initial state before any evolution, so that its transitions never enter the metric.
For a diabatic transfer ($k_\mathrm{f}\neq k_\mathrm{i}$) in a two-level system with a single anticrossing, such as the Landau--Zener problem, the gap has no maximum inside the sweep, and the di--ad tensor reduces to $G^{(\tilde{\alpha},\tilde{\beta})}_{\mu\nu}$ along the whole sweep.

This construction allows for state transfers even if there is no direct adiabatic path between the initial and target states.
The exponents $(\alpha,\beta;\; \tilde{\alpha}, \tilde{\beta})$ determine the degree of adiabaticity and diabaticity, respectively, and were shown to allow for high-fidelity state transfer between eigenstates~\cite{ventura-meinersenMultilevelSpectralNavigation2026a}.
As for the adiabatic protocols, we replace the hypermetric tensor by the di--ad tensor in Eq.~\eqref{eq: hypergeo protocol} and solve this first-order equation for the pulse.
For a derivation of the di--ad construction, we refer the reader to Ref.~\cite{ventura-meinersenMultilevelSpectralNavigation2026a}; Appendix~\ref{app: formal details} collects formal details for the two-level case.

The di--ad tensor contains all of the above protocols as special cases, so that a single implementation covers them.
Section~\ref{sec: geodesiq} shows how this is used in \texttt{geodesiq}.

\section{\texorpdfstring{\texttt{geodesiq}}{geodesiq}: methods and usage}
\label{sec: geodesiq}
\begin{figure}[!ht]
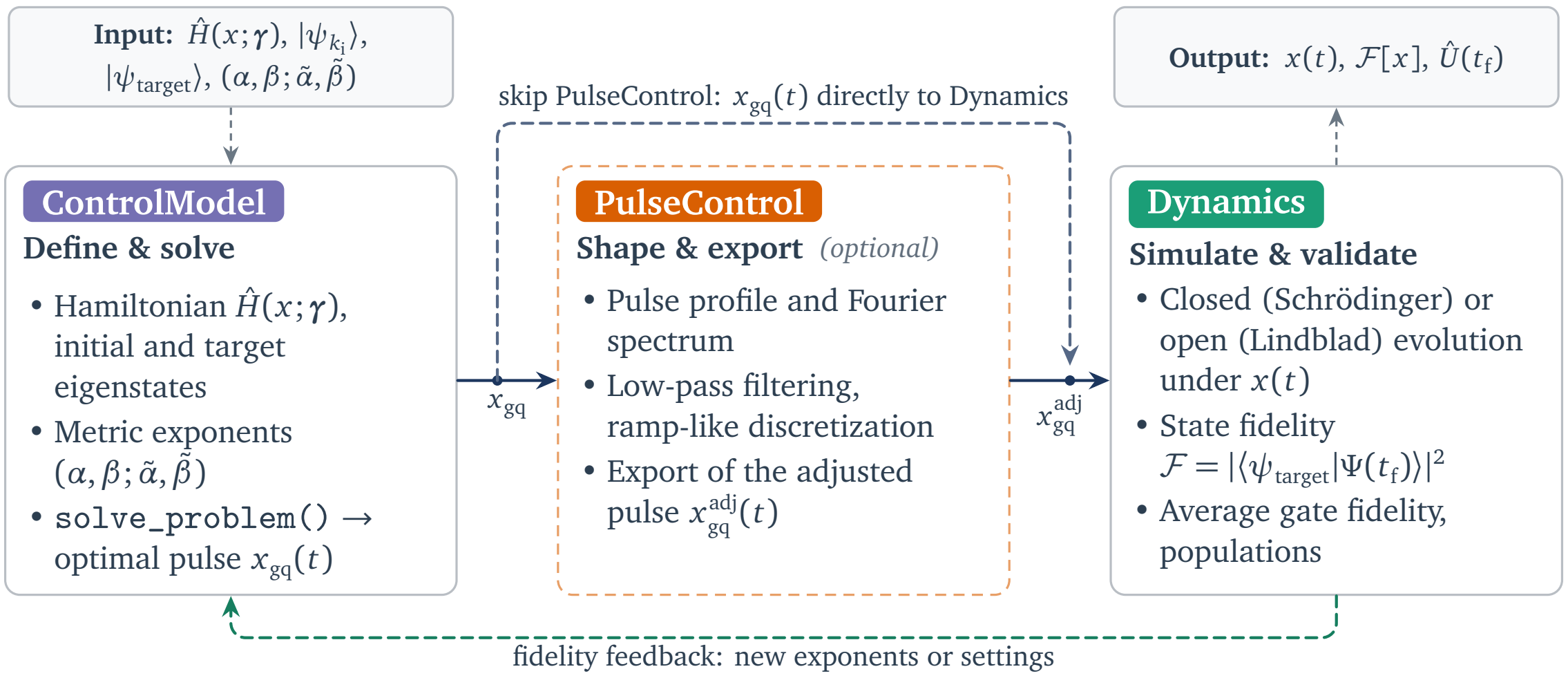

    \centering
    \includestandalone{Extras/summary_fig_v4}
    \caption{Overview of the \texttt{geodesiq} Python package contents and the recommended workflow.
    The package is set up around the main class \code{geodesiq.ControlModel}, which allows the user to define the Hamiltonian and the associated control problem to be solved.
    The user sets the Hamiltonian parameters $\boldsymbol{\gamma}$, the initial and target eigenstates, and the exponents of the metric.
    \code{solve_problem()} computes the optimal pulse $x_\mathrm{gq}(t)$.
    The \code{PulseControl} and \code{Dynamics} classes post-process and simulate the pulse, so that it can be checked before use in theory or experiment; $x^\mathrm{adj}_\mathrm{gq}(t)$ denotes the pulse adjusted with \code{PulseControl}.
    The dashed border marks \code{PulseControl} as optional: the dashed arrow above it passes $x_\mathrm{gq}(t)$ directly to \code{Dynamics}, and the dashed arrow below the cards indicates that the simulated fidelity can guide a new choice of the \code{ControlModel} settings, such as the exponents.}
    \label{fig: geodesiq summary}
\end{figure}

As in Sec.~\ref{sec: theory}, $x$ denotes a generic single control; in the examples, it is called $z$ (Landau--Zener sweep), $\varepsilon$ (detuning), and $\lambda$ (interpolation parameter of the Ising chain).
To showcase the use of the \texttt{geodesiq} package (see overview in Fig.~\ref{fig: geodesiq summary}), we use the Landau--Zener problem~\cite{ivakhnenkoNonadiabaticLandauZener2023} as an example, which is described by the following Hamiltonian:
\begin{align}
\label{eqn: Landau Zener model}
    \hat{H}(z(t))=z(t)\sigma^z+\Delta\sigma^x=\begin{pmatrix}
        z(t) & \Delta \\
        \Delta & -z(t)
    \end{pmatrix},
\end{align}
where $2|\Delta|$ is the minimum gap, reached at $z=0$, and $z(t)$ is the control.
Here, $z(t)$ need not be a linear ramp: its shape is what we optimize.
We restate the Landau--Zener problem: starting from the ground state $\ket{\psi_0(z_\mathrm{i})}$ for $z_\mathrm{i}=z(0)<0$, we want to change the control $z(t)$ to the final value $z(t_\mathrm{f})=z_\mathrm{f}>0$, such that we maximize the state-transfer fidelity
\begin{align}
    \mathcal{F}(t_\mathrm{f})=|\braket{\psi_\mathrm{target}}{\hat{U}(t_\mathrm{f})\psi_0(z_\mathrm{i})}|^2,
\end{align}
where $\hat{U}(t_\mathrm{f})$ is the time-evolution operator generated by $\hat{H}(z(t))$ from $t=0$ to $t_\mathrm{f}$, and the target state can be either the ground state $\ket{\psi_\mathrm{target}}=\ket{\psi_0(z_\mathrm{f})}$ (adiabatic evolution) or the excited state $\ket{\psi_\mathrm{target}}=\ket{\psi_1(z_\mathrm{f})}$ (diabatic evolution).

Here, we will focus on the fully adiabatic evolution for simplicity.
The following subsections cover solving the problem (Sec.~\ref{sec: controlmodel}), post-processing the pulse (Sec.~\ref{sec: pulsecontrol}), and simulating the dynamics to obtain figures of merit such as the state fidelity (Sec.~\ref{sec: dynamics}).
This example is also available as the notebook \href{https://geodesiq-control.github.io/geodesiq-control/notebooks/index.html?path=landau_zener.ipynb}{examples/landau\_zener.ipynb}, which, like the other example notebooks, runs directly in the web browser through \texttt{JupyterLite}, without any installation.

\subsection{The \texorpdfstring{\texttt{ControlModel}}{ControlModel} class: solving for the optimal pulse}
\label{sec: controlmodel}
The \code{ControlModel} class supports two representations of a Hamiltonian depending on a single real control parameter.
For a general dependence, the Hamiltonian is supplied as a callable, together with any parameters that remain constant during the evolution.
Its derivative with respect to the control parameter can optionally be supplied as a second callable; otherwise, it is computed by high-order finite differences of the sampled Hamiltonians on the control grid.
Only the traceless part of the Hamiltonian is differentiated, since a multiple of the identity does not change the metric.
Points that fail an error estimate are recomputed using \code{scipy.differentiate.jacobian} with an adaptive step.
The error estimate and the adaptive step both account for the rounding of large offsets in the control values and in the energies, and where the adaptive step does not converge the finite-difference value is kept with a warning.
Since the derivative is taken of $\hat{H}$ and not of its eigenstates, a narrow anticrossing does not require a smaller differentiation step; it does, however, require a control grid fine enough to resolve the resulting peak of the metric.

Alternatively, a Hamiltonian that depends affinely on the control can be specified directly through two constant matrices,
\begin{equation}
    \hat{H}(z(t)) = \hat{H}_{\mathrm{d}} + z(t)\hat{H}_{\mathrm{c}},
    \qquad
    \partial_z \hat{H} = \hat{H}_{\mathrm{c}},
    \label{eqn: affine Hamiltonian}
\end{equation}
where $\hat{H}_{\mathrm{d}}$ is the drift Hamiltonian and $\hat{H}_{\mathrm{c}}$ is the control operator.
Both matrices must be square, of equal shape, Hermitian, and have finite entries.
This representation allows the Hamiltonians over the control grid to be constructed by array broadcasting and supplies the derivative exactly, without numerical differentiation.
The geometric construction and pulse equation remain unchanged.
Although $\hat{H}_{\mathrm{c}}$ is constant in the original basis, its matrix elements in the instantaneous eigenbasis generally depend on $z$.

A state-transfer problem is specified by the following ingredients:
\begin{enumerate}
    \item A \code{ControlModel} initialized either from a Hamiltonian callable, with an optional derivative callable, or from the constant matrices \code{H_d} and \code{H_c}.
    These two interfaces cannot be combined.
    \item The pulse range $(z_\mathrm{i},z_{\mathrm{f}})$, specified by \code{pulse_initial} and \code{pulse_final}.
    \item The initial and target eigenstate indices, specified by \code{initial_state} and \code{final_state}.
    These zero-based indices refer to the ascending energy spectrum at the corresponding pulse endpoints.
    For adiabatic ground-state preparation, both are set to zero.
    \item The control name, specified by \code{control_name}.
    For a callable Hamiltonian, this must match its control argument.
    For an affine Hamiltonian, it only labels the scalar coefficient multiplying \code{H_c} and may be omitted.
    \item The parameters of the di--ad tensor $(\alpha,\beta;\tilde{\alpha},\tilde{\beta})$, specified by \code{alpha}, \code{beta}, \code{dia_alpha}, and \code{dia_beta}.
    The diabatic exponents are required for a diabatic transfer, i.e., when the final index differs from the index in which the initial state, followed adiabatically, ends.
    Unless the state passes a true level crossing (Sec.~\ref{sec: theory}), that index is the initial one.
    Each diabatic passage needs an anticrossing inside the sweep (Sec.~\ref{sec: theory}), unless \code{allow_endpoint_passages} is set.
    The order $p$ of the norm that combines the transitions, Eq.~\eqref{eqn: transition norm}, is set by \code{norm} (default $p=2$).
\end{enumerate}
\begin{table}[!ht]
    \centering
    \begin{tabular}{l c l}
        \hline
        Parameter & Type & Description \\
        \hline
        \code{control_name} & \texttt{str} & Name of the scalar control \\
        \code{pulse_initial} & \texttt{float} & Initial value $z_\mathrm{i}=z(0)$ of the control \\
        \code{pulse_final} & \texttt{float} & Final value $z_\mathrm{f}=z(t_\mathrm{f})$ of the control \\
        \code{initial_state} & \texttt{int} & Index $k_\mathrm{i}$ of the initial eigenstate $|\psi_{k_\mathrm{i}}\rangle$ \\
        \code{final_state} & \texttt{int} & Index $k_\mathrm{f}$ of the target eigenstate $|\psi_{k_\mathrm{f}}\rangle$ \\
        \code{alpha} & \texttt{float} & Adiabatic exponent $\alpha$ of the di--ad tensor $\mathcal{G}_{\mathrm{di\text{--}ad}}^{(\alpha,\beta;\tilde{\alpha},\tilde{\beta})}$  \\
        \code{beta} & \texttt{float} & Adiabatic exponent $\beta$ of the di--ad tensor $\mathcal{G}_{\mathrm{di\text{--}ad}}^{(\alpha,\beta;\tilde{\alpha},\tilde{\beta})}$ \\
        \code{dia_alpha} & \texttt{float} & Diabatic exponent $\tilde{\alpha}$ of the di--ad tensor $\mathcal{G}_{\mathrm{di\text{--}ad}}^{(\alpha,\beta;\tilde{\alpha},\tilde{\beta})}$  \\
        \code{dia_beta} & \texttt{float} & Diabatic exponent $\tilde{\beta}$ of the di--ad tensor $\mathcal{G}_{\mathrm{di\text{--}ad}}^{(\alpha,\beta;\tilde{\alpha},\tilde{\beta})}$  \\
        \code{norm} & \texttt{float} & Order $p\geq1$ of the norm combining the transitions, \\
        & & Eq.~\eqref{eqn: transition norm} (default $2$; \code{np.inf} allowed) \\
        \code{num_steps} & \texttt{int} & Number $N$ of control-grid points (default $2^{10}+1$); \\
        & & the default Romberg integrator requires $N=2^r+1$ \\
        \code{allow_endpoint_passages} & \texttt{bool} & Count a gap minimum at an endpoint \\
        & & as an anticrossing (default \code{False}) \\
        \hline
    \end{tabular}
    \caption{\code{set_control()} method parameters and their respective descriptions.
    The exponents $(\alpha, \beta)$ and $(\tilde{\alpha}, \tilde{\beta})$ determine the adiabatic and diabatic components of the di--ad tensor; see Table~\ref{tab: adiabatic protocols and hypergeo}.}
    \label{tab: set control}
\end{table}

The above elements are summarized in Table~\ref{tab: set control}.
For the Landau--Zener Hamiltonian in Eq.~\eqref{eqn: Landau Zener model}, the affine representation is obtained by choosing $\hat{H}_{\mathrm{d}}=\Delta\sigma^x$ and $\hat{H}_{\mathrm{c}}=\sigma^z$.
The corresponding implementation is
\begin{minted}[bgcolor=LightGray, fontsize=\footnotesize]{python}
import numpy as np
import geodesiq as gq

# ----- Set system parameters -----
delta = 1.0
z0, zf = -10.0, 10.0
alpha, beta = 2.0, 2.0

# ----- Define the affine Hamiltonian -----
H_d = np.array([[0.0, delta], [delta, 0.0]])
H_c = np.array([[1.0, 0.0], [0.0, -1.0]])
model = gq.ControlModel(H_d=H_d, H_c=H_c)

# ----- Configure and solve the control problem -----
model.set_control(control_name="z", pulse_initial=z0, pulse_final=zf, initial_state=0, alpha=alpha, beta=beta)
model.solve_problem(pulse_accuracy=1000)
\end{minted}
All fixed system parameters are already incorporated into \code{H_d} and \code{H_c}.
Consequently, affine models do not support \code{set_parameters()}.
To change a parameter absorbed into these matrices, a new \code{ControlModel} must be constructed.
The control configuration, including the pulse endpoints and metric exponents, can still be changed using \code{set_control()}.

With the callable interface, the same Hamiltonian reads:
\begin{minted}[bgcolor=LightGray, fontsize=\footnotesize]{python}
def ham(z, delta):
    return np.array([[z, delta], [delta, -z]])

def partial_ham(z, delta):
    return np.array([[1.0, 0.0], [0.0, -1.0]])

model = gq.ControlModel(ham, partial_H_func=partial_ham)
model.set_parameters(delta=delta)
model.set_control(control_name="z", pulse_initial=z0, pulse_final=zf, initial_state=0, alpha=alpha, beta=beta)
model.solve_problem(pulse_accuracy=1000)
\end{minted}
Omitting \code{partial_H_func} would select numerical differentiation.
The explicit affine interface avoids both repeated Hamiltonian-function evaluations over the control grid and the need to provide a derivative callable.

As long as \code{final_state} has not been set explicitly, it follows \code{initial_state}, which selects an adiabatic protocol; after a true level crossing, it is the index in which the followed state ends.
The argument \code{pulse_accuracy} specifies the number of samples of the optimized waveform on the normalized-time grid, whereas \code{num_steps} determines the control grid used to evaluate the eigensystem and metric.
The model configuration and computation status can be inspected with
\begin{minted}[bgcolor=LightGray, fontsize=\footnotesize]{python}
model.print_summary()
\end{minted}
For an affine model, the summary identifies the Hamiltonian as \code{H_d + control * H_c} and its derivative as the exact constant matrix \code{H_c}.

Two methods help inspect the problem: \code{model.plot_eigenvalues()} plots the spectrum and \code{model.plot_metric_tensor()} the single-control metric $G$ of Sec.~\ref{sec: transition norm}, both as functions of the control.
The solver integrates the pulse equation
\begin{equation*}
    \left|\dv{z}{\tau}\right|=\frac{\mathcal{L}^{(\alpha,\beta)}}{\sqrt{G(z)}},
\end{equation*}
where $G(z)$ is the plotted metric at the control value $z$ and $\mathcal{L}^{(\alpha,\beta)}$ is the length of the path, Eq.~\eqref{eqn: path length}, i.e., the integral of $\sqrt{G}$ from $z_\mathrm{i}$ to $z_\mathrm{f}$.
Hence, the pulse slows down where $G$ is large, i.e., near small gaps.
Table~\ref{tab: controlmodel table} lists the public methods of \code{ControlModel}.
\begin{table}[hbt]
    \centering
    \begin{tabular}{l l} 
    \hline 
        Method & Description \\ \hline
        \code{set_parameters()} & Set fixed parameters of a callable Hamiltonian \\
        \code{set_control()} & Set the control variables that determine the di--ad protocol\\
        \code{plot_eigenvalues()} & Plot the eigenvalues $E_n$  \\
        \code{plot_metric_tensor()} & Plot the metric $G(x)$ \\
        \code{solve_problem()} & Generate the geometric control pulse \\
        \code{synthesize_pulse()} & Create a \code{PulseControl} object with a fixed duration \\
        \code{print_summary()} & Print the configuration and computation status \\
     \hline
    \end{tabular}
    \caption{\code{ControlModel} class methods and their respective descriptions.}
    \label{tab: controlmodel table}
\end{table}

Appendix~\ref{app: numerical benchmarking} compares \texttt{geodesiq} with two widely used numerical methods, gradient-ascent pulse engineering (GRAPE)~\cite{khanejaOptimalControlCoupled2005} and chopped random-basis optimization (CRAB)~\cite{doriaOptimalControlTechnique2011, canevaChoppedRandombasisQuantum2011}.
In summary, we can draw two main conclusions from our analysis: (1) \texttt{geodesiq} generates pulses one to more than three orders of magnitude faster, and the pulse shape does not need to be recomputed for each total pulse time $t_\mathrm{f}$, and (2) under low-pass filtering, as in experimental electronic equipment, \texttt{geodesiq} pulses keep a low infidelity at lower cutoff frequencies, whereas GRAPE reaches the smallest infidelities when the full bandwidth is available.
These conclusions hold for the systems studied in the appendix; both GRAPE and CRAB have many hyperparameters (e.g., target fidelity, number of time slots) that affect their performance.

\subsection{The \texorpdfstring{\texttt{PulseControl}}{PulseControl} class: additional pulse methods}
\label{sec: pulsecontrol}
The \code{ControlModel} class is the core class for defining, parametrizing, and solving the optimal control problem.
For further theoretical analysis or for direct implementation on experimental platforms, we provide additional methods.
These methods are included in the class \code{PulseControl}, which is instantiated by calling:
\begin{minted}
    [bgcolor=LightGray, fontsize=\footnotesize]{python}
total_time = 10  # For Landau–Zener it is measured in units of 1 / delta
pulse = model.synthesize_pulse(duration=total_time)
\end{minted}
Here \code{pulse} is an instance of the \code{PulseControl} class, which has methods to plot, discretize, and export the pulse, among others.
The discretization of the pulse may be useful for direct experimental application, such as the piecewise-linear voltage ramps of an arbitrary waveform generator, as used in Ref.~\cite{farinaSiteresolvedMagnonTriplon2025}.
For instance, the pulse is plotted with (Fig.~\ref{fig: LZ pulse}):
\begin{minted}
    [bgcolor=LightGray, fontsize=\footnotesize]{python}
fig, ax = pulse.plot_pulse()
\end{minted}
\begin{figure}[!ht]
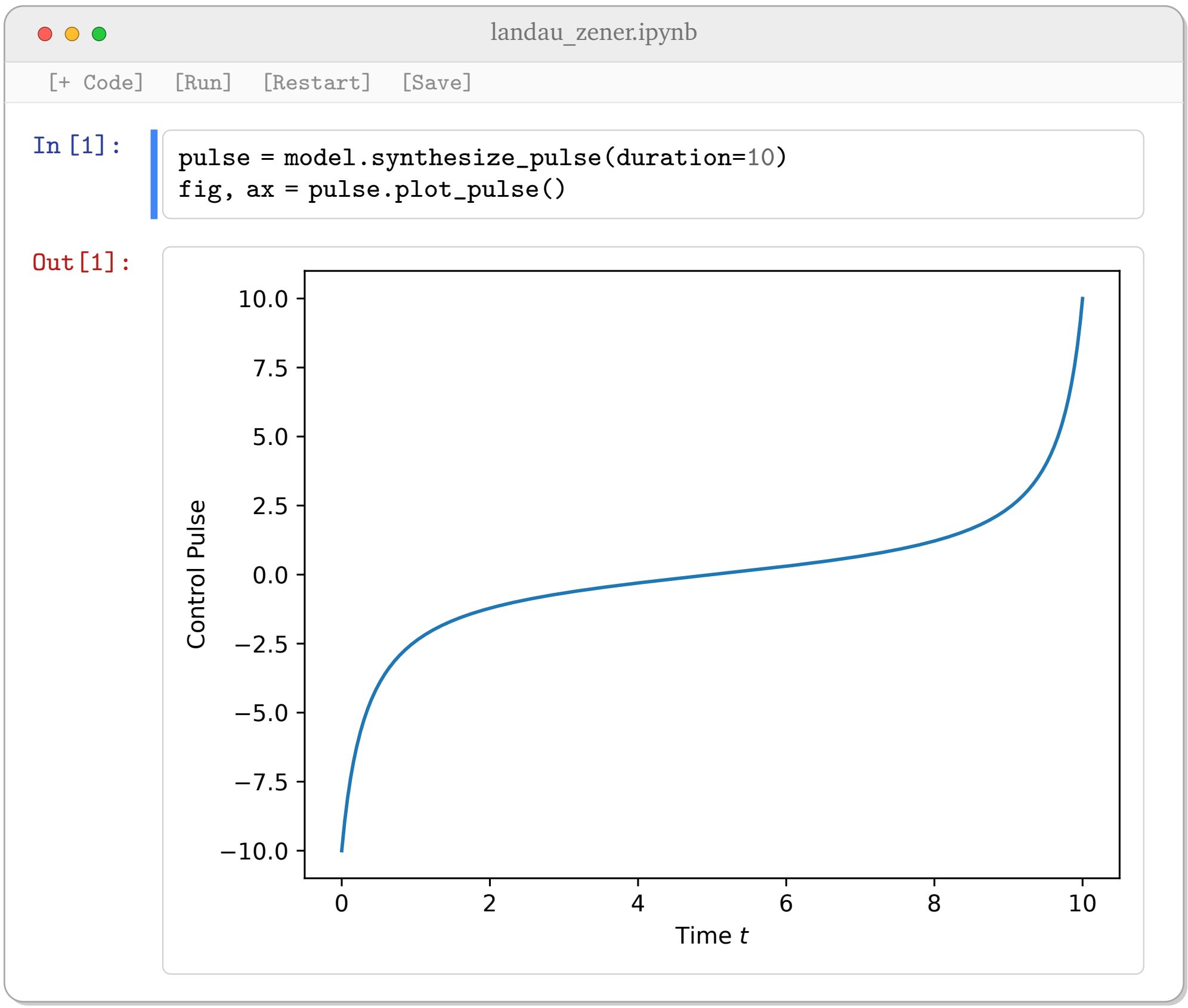

    \centering
    \includestandalone{Extras/lz_output}
    \caption{Solved control pulse $z(t)$ for the Landau--Zener problem.
    The exponents $(\alpha,\beta)=(2,2)$ enforce adiabatic dynamics, and the figure shows the raw output of \code{pulse.plot_pulse()}.
    The time $t$ is in units of $1/\Delta$, with $t_\mathrm{f}=10/\Delta$.}
    \label{fig: LZ pulse}
\end{figure}
To account for restrictions of the control electronics, the pulse can be discretized:
\begin{minted}
    [bgcolor=LightGray, fontsize=\footnotesize]{python}
discretized_pulse = pulse.discretized_pulse(linear_steps=5)  # 5 points, i.e., 4 linear ramps
\end{minted}
The main demands a pulse places on the control electronics are its bandwidth and slew rate.
The bandwidth requirement can be read from the short-time Fourier transform (STFT) of the pulse (Fig.~\ref{fig: LZ freq spectrum}):
\begin{minted}
    [bgcolor=LightGray, fontsize=\footnotesize]{python}
freqs, times, magnitude = pulse.fourier_spectrum(
    window_len=256, hop=1, detrend=True, padding="hold", mfft=4096
)
\end{minted}
For the pulse of Fig.~\ref{fig: LZ pulse}, the largest STFT amplitude at any time falls below $10\%$ of its maximum above $f\approx1.6\,\Delta$ and below $1\%$ above $f\approx5.3\,\Delta$.
A control line with a bandwidth of a few $\Delta$ therefore reproduces the pulse well.

\begin{figure}[!ht]
    \centering
    \includegraphics[width=\linewidth]{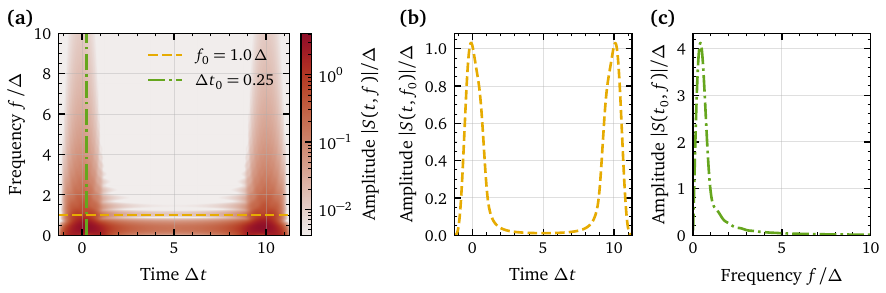}
    \caption{Short-time Fourier spectrum of the Landau--Zener pulse in Fig.~\ref{fig: LZ pulse}, with total pulse time $t_\mathrm{f}=10/\Delta$.
    \textbf{(a)} STFT amplitude $|S(t,f)|$ as a function of time and frequency, on a logarithmic color scale spanning three decades.
    It is the physical amplitude of the oscillation at each frequency: a sinusoid of amplitude $A$ within a window gives $|S|=A$ at its frequency.
    The pulse has 1000 samples, and each window is a periodic Hann window of 256 samples ($2.56/\Delta$, i.e., a frequency resolution of about $0.39\,\Delta$), advanced by one sample.
    Each window is zero-padded to an FFT of 4096 points (\code{mfft=4096}), which only interpolates the frequency axis.
    The mean of each window is removed (\code{detrend=True}), so that the DC part of the sweep does not dominate.
    Windows that overhang the pulse see its first and last values held (\code{padding="hold"}), as an arbitrary waveform generator holds its output, so the window centers extend $1.27/\Delta$ beyond $[0,t_\mathrm{f}]$.
    The amplitude is largest near the start and the end, where the geodesic pulse is steepest and leaves or joins the held values; zero padding would add a spurious jump to zero there.
    \textbf{(b)} Cut of panel \textbf{(a)} at the fixed frequency $f_0=1.0\,\Delta$ (yellow dashed line), as a function of time.
    \textbf{(c)} Cut of panel \textbf{(a)} at the fixed time $\Delta t_0=0.25$ (light-green dash-dotted line), as a function of frequency.
    Time is measured in units of $1/\Delta$, the frequency $f$, in cycles per unit time, in units of $\Delta$, and the amplitude, like the control $z$, in units of $\Delta$.}
    \label{fig: LZ freq spectrum}
\end{figure}

Once the pulse satisfies the experimental constraints and the fidelity target, it can be exported:
\begin{minted}
    [bgcolor=LightGray, fontsize=\footnotesize]{python}
pulse.export_pulse(filename="pulse_data", file_extension="csv", overwrite=True)
\end{minted}
Supported formats are \texttt{.txt}, \texttt{.npz}, and \texttt{.csv}.
The public methods of \code{PulseControl} are listed in Table~\ref{tab: pulsecontrol table}.
\begin{table}[hbt]
    \centering
    \begin{tabular}{l p{0.55\linewidth}}
    \hline
        Method & Description \\ \hline
        \code{plot_pulse()} & Plot the pulse\\
        \code{discretized_pulse()} & Resample as linear ramps through \code{linear_steps} equally spaced points\\
        \code{filtered_pulse()} & Return the low-pass-filtered pulse (Butterworth, default order 3) \\
        \code{fourier_spectrum()} & Return the short-time Fourier spectrum of the pulse \\
        \code{export_pulse()} & Export the pulse to a file (\texttt{.txt}, \texttt{.npz}, \texttt{.csv}) \\
        
     \hline 
    \end{tabular}
    \caption{\code{PulseControl} class methods and their respective descriptions.}
    \label{tab: pulsecontrol table}
\end{table}

\subsection{The \texorpdfstring{\texttt{Dynamics}}{Dynamics} class: time-dependent simulators and methods}
\label{sec: dynamics}

Another important ingredient in the analysis of the generated pulses is the simulation of the dynamics\footnote{The \code{Dynamics} class is built on \texttt{QuTiP}~\cite{johanssonQuTiPOpensourcePython2012, johanssonQuTiP2Python2013, lambertQuTiP5Quantum2026}.}.
The \code{Dynamics} class takes a configured \code{ControlModel} and a physical pulse duration:
\begin{minted}[bgcolor=LightGray, fontsize=\footnotesize]{python}
total_time = 100
dynamics = gq.Dynamics(duration=total_time, model=model)
\end{minted}
During construction, the normalized pulse is mapped to the physical-time interval $[0,t_{\mathrm{f}}]$ and a \texttt{QuTiP} \code{QobjEvo} representation of $\hat{H}(t)/\hbar$ is prepared.
For affine models, the constant operators $\hat{H}_{\mathrm{d}}$ and $\hat{H}_{\mathrm{c}}$ are supplied directly, with the sampled control waveform as the coefficient of $\hat{H}_{\mathrm{c}}$.
The optimal pulse is densely sampled and smooth, and its coefficient is interpolated with cubic splines.

For callable models, the Hamiltonian is evaluated at the pulse sample times and decomposed into constant Hermitian operators with time-dependent coefficients using a singular-value decomposition.
The drift is chosen as the mean sampled Hamiltonian, and singular values exceeding $10^{-10}$ times the largest singular value are retained.
The resulting coefficient arrays are interpolated in the same way, that is, entry-wise in $\hat{H}$ rather than through the control.
This decomposition is a numerical representation of the sampled Hamiltonian and does not introduce additional independently optimized controls.

The same prepared Hamiltonian is used for propagators and state evolution.
Without collapse operators, state evolution uses the Schr\"odinger-equation solver; with collapse operators, it uses the Lindblad master-equation solver.
A new \code{Dynamics} object should be constructed after changing the model parameters, control configuration, or pulse solution.
The keyword argument \code{options} configures the \texttt{QuTiP} solver for all propagation methods, for example \code{options={"atol": 1e-10, "rtol": 1e-8, "nsteps": 10000}}.
Omitting it preserves the solver's numerical defaults; precision studies should tighten the tolerances and reduce \code{max_step} independently of pulse sampling and interpolation.
Instead of a duration, \code{Dynamics} also accepts any custom pulse sampled at strictly increasing times, for instance the output of \code{filtered_pulse()} or \code{discretized_pulse()}:
\begin{minted}[bgcolor=LightGray, fontsize=\footnotesize]{python}
times, values = pulse.filtered_pulse(cutoff_freq=1.0)
filtered_dynamics = gq.Dynamics(model=model, times=times, pulse=values)
\end{minted}
In this case, the model only supplies the Hamiltonian and the endpoint eigenstates.
These eigenstates are still taken at \code{pulse_initial} and \code{pulse_final}, so a warning is emitted when the custom pulse starts or ends further than $1\%$ of the control range from them, as a strongly filtered pulse does.
A custom pulse is interpolated linearly between its samples, so that, e.g., the piecewise-linear output of \code{discretized_pulse()} is simulated as the waveform it describes; a cubic spline through a few samples can deviate strongly between them and overshoot the sampled range.
The argument \code{order} selects the interpolation explicitly: \code{0} holds each sample until the next one, as an arbitrary waveform generator does, \code{1} is linear, and \code{3} is a cubic spline.
For a callable Hamiltonian that is nonlinear in the control, the interpolated decomposition deviates from $\hat{H}(x(t))$ between the samples, except with \code{order=0}.
This deviation decreases with the number of samples and is negligible for the densely sampled optimal pulse, but it can matter for a custom pulse with few samples.
The attribute \code{interpolation_error} estimates the resulting error of the propagator, and a warning is emitted when it exceeds $10^{-3}$.
The argument \code{exact_hamiltonian=True} then evaluates $\hat{H}(x(t))$ at every solver step instead, which is exact but slower.

\begin{figure}[!ht]
    \centering
    \includegraphics[width=0.67\linewidth]{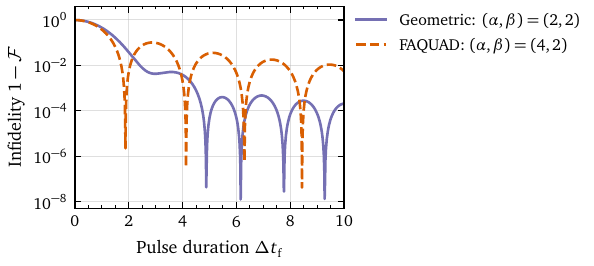}
    \caption{State-transfer infidelity $1-\mathcal{F}$ of the Landau--Zener model, Eq.~\eqref{eqn: Landau Zener model}, versus total pulse time $\Delta t_\mathrm{f}$ for the geometric [$(\alpha,\beta)=(2,2)$] and FAQUAD [$(\alpha,\beta)=(4,2)$] pulses, with $\Delta=1$, $z_\mathrm{i}=-10\Delta$ and $z_\mathrm{f}=10\Delta$.
    The sharp dips are interference minima; their depth is limited by the sampling of $t_\mathrm{f}$ (1000 points in $[0.01, 10]/\Delta$).}
    \label{fig: landau_zener}
\end{figure}

The main methods of the class compute the full time-evolution operator, the state-transfer fidelity, and the average gate fidelity, which can be called as
\begin{minted}[bgcolor=LightGray, fontsize=\footnotesize]{python}
import qutip as qt

# Average gate fidelity is always with respect to a target gate
target_gate = qt.sigmax()
gate_fid = dynamics.average_gate_fidelity(target_gate=target_gate)

# State fidelity between the evolved initial state and the target state
fidelity = dynamics.state_fidelity()

# state_fidelity optionally accepts Lindblad collapse operators,
# which include the square root of their rate
gamma = 1e-3  # dephasing rate
dephasing_op = np.sqrt(gamma) * qt.sigmaz()
fidelity_noisy = dynamics.state_fidelity(c_ops=[dephasing_op])
\end{minted}
These methods can be combined in user-defined functions, for instance to sweep the total pulse time:
\begin{minted}[bgcolor=LightGray, fontsize=\footnotesize]{python}
# We may define a custom function to do a total pulse time sweep
def fidelity_vs_time(durations, model, alpha, beta):

    model.set_control(alpha=alpha, beta=beta)
    model.solve_problem(pulse_accuracy=1000)

    fidelities = []
    # Durations must be strictly positive
    for duration in durations:
        dynamics = gq.Dynamics(duration=duration, 
                               model=model)
        fidelities.append(dynamics.state_fidelity())

    return np.array(fidelities)

# And then we run it as follows
durations = np.linspace(0.01, 10, 1000)
fidelities_geometric = fidelity_vs_time(durations, model, alpha=2, beta=2)
fidelities_faquad = fidelity_vs_time(durations, model, alpha=4, beta=2)
\end{minted}
Figure~\ref{fig: landau_zener} compares the infidelities $1-\mathcal{F}$ of the geometric and FAQUAD protocols.
Both reach infidelities below $10^{-4}$ within a few units of $1/\Delta$, but they differ in how sensitive the result is to the exact value of $t_\mathrm{f}$: for $\Delta t_\mathrm{f}\gtrsim 5$, the geometric pulse keeps the infidelity below $5\times10^{-4}$, whereas the FAQUAD infidelity oscillates with $t_\mathrm{f}$, with maxima between $10^{-2}$ and $10^{-1}$, and drops below $10^{-4}$ only near isolated pulse times.
These oscillations come from the endpoints of the sweep: since the pulse speed scales as $|dz/dt|\propto G^{-1/2}\propto|z|^{(\alpha+\beta)/2}$, FAQUAD leaves $z_\mathrm{i}=-10\Delta$ and reaches $z_\mathrm{f}=10\Delta$ about seven times faster than the geometric pulse, so the non-adiabatic amplitudes generated at the two endpoints~\cite{degrandiAdiabaticPerturbationTheory2010} are larger and interfere as $t_\mathrm{f}$ varies.
Figure~\ref{fig: landau_zener} is therefore not a general ranking of the two protocols: FAQUAD first reaches $1-\mathcal{F}\leq10^{-2}$ at a shorter pulse time than the geometric pulse ($\Delta t_\mathrm{f}\approx1.7$ against $2.4$; App.~\ref{app: metric comparison}), but its infidelity exceeds this threshold again at longer times.
As mentioned above, we optimize state-transfer fidelities by engineering an appropriate unitary evolution through the control parameter.
The resulting time-evolution operator can therefore also be used to implement a gate.
For the Landau--Zener model, adiabatic transfer implements an $X$ gate up to a relative phase, which we check with:
\begin{minted}
[bgcolor=LightGray, fontsize=\footnotesize]{python}
gate_dyn = gq.Dynamics(duration=50, model=model)
gate_fid = gate_dyn.average_gate_fidelity(target_gate=qt.sigmax())
print("Average gate fidelity (at t=tf): ", gate_fid[-1])
\end{minted}
For the FAQUAD pulse solved last in the example above, the resulting average gate fidelity is roughly $99\%$ after a total pulse time of $t_\mathrm{f}=50/\Delta$.
Unlike the state fidelity, the average gate fidelity is sensitive to the relative phase acquired by the two adiabatically transported eigenstates, which includes the dynamical phase $\int dt\,(E_1-E_0)$.
This phase is neither optimized nor removed here.
Hence, the protocol implements an $X$ gate up to a phase gate, and the gate fidelity oscillates with $t_\mathrm{f}$ (for instance, it drops to about $41\%$ at $t_\mathrm{f}=55/\Delta$), while the state fidelity remains close to one.
For $\hat{U}(t_\mathrm{f})=\sigma^x\,\mathrm{diag}(e^{i\varphi_0},e^{i\varphi_1})$, the average gate fidelity is $\mathcal{F}_\mathrm{gate}=[2+|\mathrm{Tr}(\sigma^{x\dagger}\hat{U}(t_\mathrm{f}))|^2]/6=[2+\cos(\varphi_1-\varphi_0)]/3$, which ranges from $1/3$ to $1$.
The phase can be compensated, e.g., by a virtual $Z$ rotation or by tuning $t_\mathrm{f}$.
Table~\ref{tab: dynamics table} lists the public methods of \code{Dynamics}.
\begin{table}[hbt]
    \centering
    \begin{tabular}{l p{0.55\linewidth}}
    \hline
        Method & Description \\ \hline
        \code{time_evolution_operator()} & Return the time-evolution operator $\hat{U}(t)$ of the pulse\\
        \code{state_fidelity()} & Return the state-transfer fidelity \\
        \code{average_gate_fidelity()} & Return the average gate fidelity $\mathcal{F}_\mathrm{gate}$ with respect to a target gate $\hat{\mathcal{O}}$ \\
        \code{populations()} & Return the populations of the instantaneous (adiabatic) or basis (diabatic) states \\
        \code{plot_populations()} & Plot the state populations versus time \\

     \hline 
    \end{tabular}
    \caption{\code{Dynamics} class methods and their respective descriptions.}
    \label{tab: dynamics table}
\end{table}
Appendix~\ref{app: api} summarizes the API; the complete, versioned documentation is available online at \url{https://geodesiq.readthedocs.io}.
\section{Examples}
\label{sec: examples}
Next, we demonstrate the features of the \texttt{geodesiq} package by exploring three examples: spin-qubit initialization, many-body ground-state preparation, and diabatic--adiabatic state transfer in a four-level system.
The three examples are accompanied by the notebooks \href{https://geodesiq-control.github.io/geodesiq-control/notebooks/index.html?path=dqd_model.ipynb}{examples/dqd\_model.ipynb}, \href{https://geodesiq-control.github.io/geodesiq-control/notebooks/index.html?path=many_body.ipynb}{examples/many\_body.ipynb}, and \href{https://geodesiq-control.github.io/geodesiq-control/notebooks/index.html?path=four_level.ipynb}{examples/four\_level.ipynb}, respectively.

\subsection{Spin-qubit initialization}
\begin{figure}[!ht]
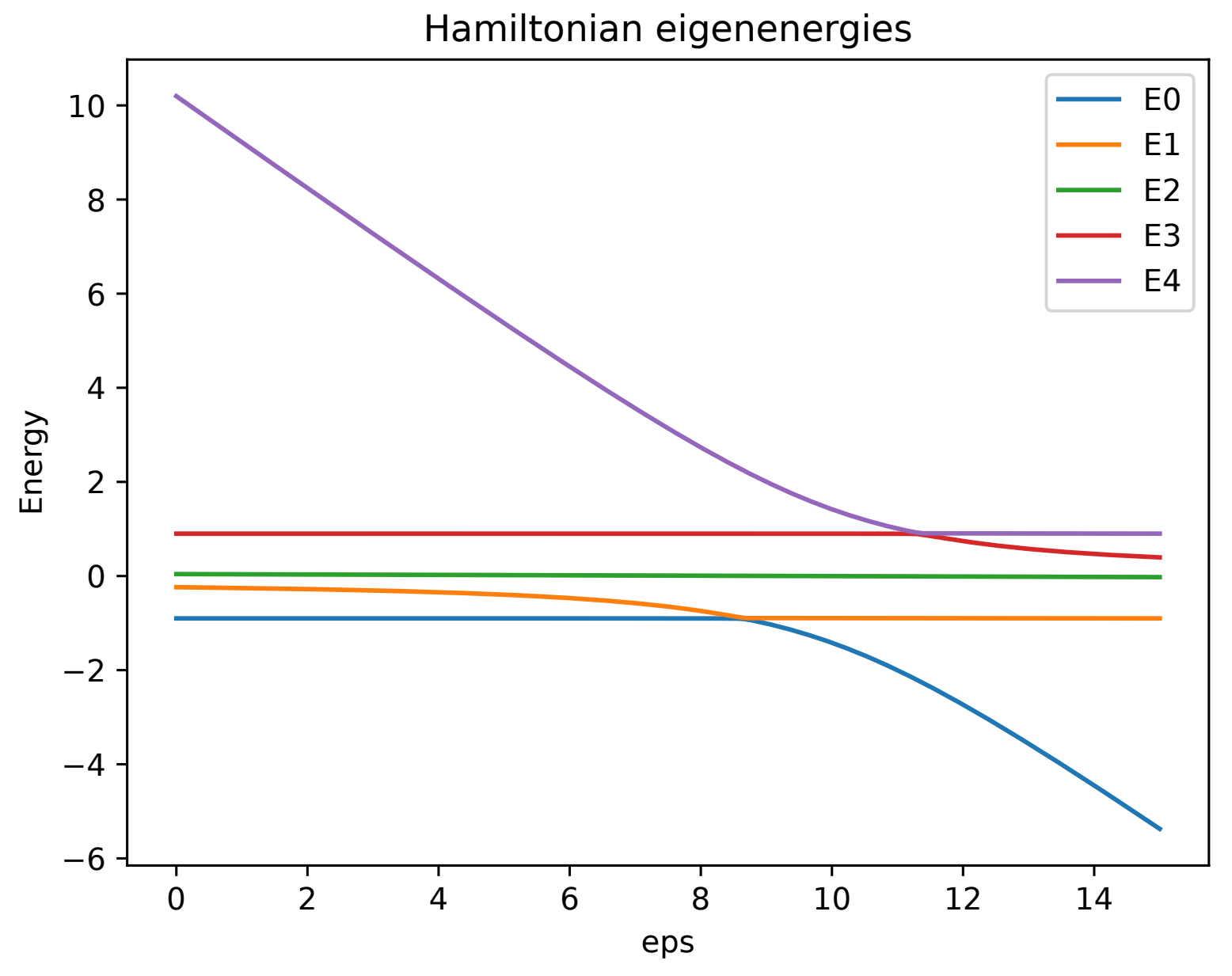

    \centering
    \includestandalone{Extras/dqd_energies}
    \caption{Energy level diagram for the DQD model in Eq.~\eqref{eqn: 5x5 Hamiltonian}.
    The parameters are those of the example below.
    We identify a very small anticrossing between the ground and first excited states, which causes non-adiabatic errors when the qubit is initialized by sweeping the detuning from $\varepsilon(0)/t_\mathrm{c}=15$ to $\varepsilon(t_\mathrm{f})/t_\mathrm{c}=0$.
    Its minimum gap, $E_1-E_0\approx 0.024\,t_\mathrm{c}$ at $\varepsilon^*\approx 8.65\,t_\mathrm{c}$, is set by $\Delta\!E_X$ and is too small to be resolved on this scale.
    The figure is the unmodified output of \code{dqd_model.plot_eigenvalues()}.}
    \label{fig: DQD energies}
\end{figure}
Semiconductor quantum dots are a promising platform for quantum processors~\cite{burkardSemiconductorSpinQubits2023} and support a wide range of operation schemes~\cite{kolokProtocolsMeasureNonAbelian2024, nguyenSuppressingSpinQubit2026, meinekeSpinQubitLeapfrogging2026, jakobFastReadoutQuantum2025}.
As a practical example, we show how our \texttt{geodesiq} package allows for a high-fidelity initialization of a semiconductor spin qubit within a double quantum dot (DQD) by reversing spin-to-charge conversion\footnote{Spin-to-charge conversion maps the spin state onto a charge configuration, here $(2,0)\leftrightarrow(1,1)$, which can then be read out electronically.
Run in reverse, the same mechanism initializes the qubit: the detuning is swept from the $(2,0)$ singlet to the $(1,1)$ regime.}~\cite{seedhousePauliBlockadeSilicon2021, niegemannParitySingletTripletHighFidelity2022, jirovecSinglettripletHoleSpin2021, krzywdaDecoherenceElectronSpin2025, jirovecDynamicsHoleSingletTriplet2022}.
A common bottleneck is the narrow anticrossing opened by weak spin-flip terms, i.e., by small differences between the Zeeman vectors of the two dots~\cite{farinaSiteresolvedMagnonTriplon2025}.
The effective model can be obtained from a Fermi--Hubbard model, which we can project onto the basis states $\{\ket{\uparrow\downarrow,\cdot},\allowbreak\ket{\uparrow,\uparrow},\allowbreak\ket{\uparrow,\downarrow},\allowbreak\ket{\downarrow,\uparrow},\allowbreak\ket{\downarrow,\downarrow}\}$\footnote{The other doubly-occupied state $\ket{\cdot, \uparrow\downarrow}$ that hosts two spins in the right dot can be neglected as it is far apart in energy.}, resulting in the following matrix representation~\cite{ungererStrongCouplingMicrowave2024, geyerAnisotropicExchangeInteraction2024, kellyIdentifyingMitigatingErrors2025}
\begin{align}
\label{eqn: 5x5 Hamiltonian}
    \hat{H}_{\mathrm{DQD}}(\varepsilon(t)) = \begin{pmatrix}
 U{-}\varepsilon(t) & 0 & -t_\mathrm{c} & t_\mathrm{c} & 0 \\
 0 & E_Z & -\Delta\!E_X & \Delta\!E_X & 0 \\
 -t_\mathrm{c} & -\Delta\!E_X & \Delta\!E_Z & 0 & \Delta\!E_X \\
 t_\mathrm{c} & \Delta\!E_X & 0 & -\Delta\!E_Z & -\Delta\!E_X \\
 0 & 0 & \Delta\!E_X & -\Delta\!E_X & -E_Z
\end{pmatrix}.
\end{align}
Here, $t_\mathrm{c}$ represents the tunnel coupling, and $\sum_{d,a}E_{a,d}\,\sigma^a_d$ is the Zeeman term of the $(1,1)$ states, where $E_{a,d}$ is the Zeeman energy of dot $d=\mathrm{L},\mathrm{R}$ along the axis $a=X,Y,Z$, i.e., half the Zeeman splitting of that dot, and the difference between the dots arises from a spatially varying g-factor~\cite{vanriggelen-doelmanCoherentSpinQubit2024, vanriggelenPhaseFlipCode2022, boscoFullyTunableLongitudinal2022, geyerAnisotropicExchangeInteraction2024, saez-mollejoExchangeAnisotropiesMicrowavedriven2025, nguyenSingleStepHighFidelityThreeQubit2025, jakobFastReadoutQuantum2025, valvoElectricallyTuneableVariability2025, valvoProbingElectrostaticDisorder2026} or a magnetic field gradient~\cite{takedaQuantumErrorCorrection2022, xueQuantumLogicSpin2022, desmetHighfidelitySinglespinShuttling2025}.
We choose the quantization axis $Z$ along the total Zeeman field, so that $E_{X,\mathrm{L}}+E_{X,\mathrm{R}}=E_{Y,\mathrm{L}}+E_{Y,\mathrm{R}}=0$, and the axis $X$ along the transverse part of the Zeeman difference, so that $E_{Y,\mathrm{L}}=E_{Y,\mathrm{R}}=0$.
Three Zeeman parameters then remain in Eq.~\eqref{eqn: 5x5 Hamiltonian}: the total Zeeman energy $E_Z=E_{Z,\mathrm{L}}+E_{Z,\mathrm{R}}$, the Zeeman energy difference $\Delta\!E_Z=E_{Z,\mathrm{L}}-E_{Z,\mathrm{R}}$, and the transverse term $\Delta\!E_X=E_{X,\mathrm{L}}=-E_{X,\mathrm{R}}$.
Finally, $U$ is the effective Coulomb repulsion and $\varepsilon\geq 0$ the detuning between the chemical potentials of the two dots.

\begin{figure}[!ht]
    \centering
    \includegraphics[width=\linewidth]{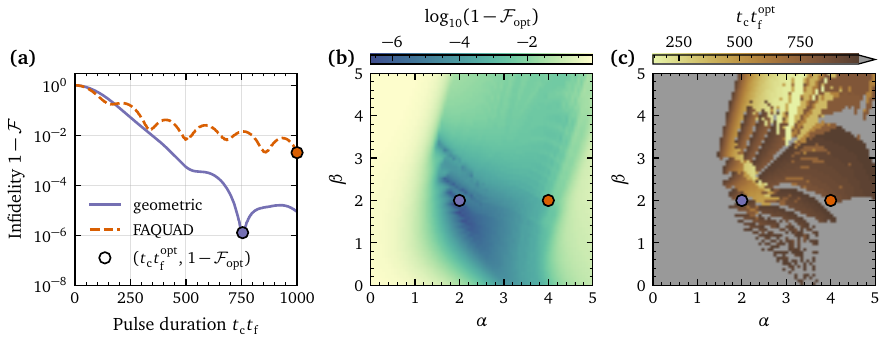}
    \caption{Time-dependent simulations of the adiabatic spin qubit initialization using \texttt{geodesiq}.
    \textbf{(a)} Infidelity of the final state using the geometric and FAQUAD protocols as a function of (rescaled) total time $t_\mathrm{c}t_\mathrm{f}$.
    The dot on each curve marks its minimum, $(t_\mathrm{c}t^\mathrm{opt}_\mathrm{f},\,1-\mathcal{F}_\mathrm{opt})$.
    \textbf{(b)} Lowest infidelity $1-\mathcal{F}_\mathrm{opt}$, minimized over the scanned pulse times $t_\mathrm{c}t_\mathrm{f} \in (0,10^3]$, as a function of $(\alpha,\beta)$.
    \textbf{(c)} Pulse time $t_\mathrm{c}t^\mathrm{opt}_\mathrm{f}$ at which this minimum is reached.
    In (b) and (c), the purple and orange dots mark the geometric and FAQUAD protocols.
    Gray regions (the gray triangle of the color bar) mark exponents for which the best fidelity is found at the longest scanned pulse time, $t_\mathrm{c}t^\mathrm{opt}_\mathrm{f}=10^3$, i.e., at the boundary of the scan.}
    \label{fig: dqd}
\end{figure}

Initializing the qubit by reverse spin-to-charge conversion requires shaping the detuning pulse $\varepsilon(t)$ from a far-detuned position, with $\varepsilon(0)-U$ well above $|t_\mathrm{c}|$ and $|E_Z|$, where the ground state has two charges in one quantum dot (i.e., $\ket{\uparrow\downarrow, \cdot}$), to the symmetric point $\varepsilon(t_\mathrm{f})=0$, transferring one charge to the other dot, $(2,0)\to(1,1)$, where $(n_\mathrm{L},n_\mathrm{R})$ denotes the charge occupation of the left and right dots.
Thereby, we initialize a computational state.
The finite spin-flip term $\Delta\!E_X$ opens a small gap between the ground and first excited states (Fig.~\ref{fig: DQD energies}).
Starting from the $\ket{\uparrow\downarrow, \cdot}$ state, we have two initialization schemes, resulting in different parity sectors:
\begin{itemize}
    \item Adiabatic evolution: $\ket{\uparrow\downarrow,\cdot}\mapsto \ket{\downarrow,\downarrow}$ 
    \item Diabatic--adiabatic evolution: $\ket{\uparrow\downarrow,\cdot}\mapsto \ket{\downarrow,\uparrow}$
\end{itemize}
For the diabatic--adiabatic case, we set \code{initial_state=0} and \code{final_state=1}.
Along the sweep, the gap between the ground and first excited state has a single minimum, at $\varepsilon^*\approx 8.65\,t_\mathrm{c}$, which is the small anticrossing in Fig.~\ref{fig: DQD energies}, and no interior maximum.
Hence, $\xi_{01}=0$ over the whole sweep, and the anticrossing is passed diabatically: the occupied state is $\ket{\psi_0}$ for $\varepsilon>\varepsilon^*$ and $\ket{\psi_1}$ for $\varepsilon<\varepsilon^*$.
The resulting di--ad transition matrix (in the eigenbasis of the Hamiltonian) is
\begin{align}
    \xi_{kn}(\varepsilon>\varepsilon^*)=\begin{pmatrix}
\cdot & 0 & 1 & 1 & 1 \\
\cdot & \cdot & \cdot & \cdot & \cdot \\
\cdot & \cdot & \cdot & \cdot & \cdot \\
\cdot & \cdot & \cdot & \cdot & \cdot \\
\cdot & \cdot & \cdot & \cdot & \cdot
\end{pmatrix},\qquad
    \xi_{kn}(\varepsilon<\varepsilon^*)=\begin{pmatrix}
\cdot & \cdot & \cdot & \cdot & \cdot \\
0 & \cdot & 1 & 1 & 1 \\
\cdot & \cdot & \cdot & \cdot & \cdot \\
\cdot & \cdot & \cdot & \cdot & \cdot \\
\cdot & \cdot & \cdot & \cdot & \cdot
\end{pmatrix}.
\end{align}
The dots represent the excluded transitions, which do not involve the occupied state.
The transitions from the occupied state to the three higher-energy states are passed adiabatically along the whole sweep.
The diabatic--adiabatic case also requires \code{final_state=1} along with the diabatic exponents \code{dia_alpha} and \code{dia_beta}; Sec.~\ref{sec: diad example} demonstrates a diabatic--adiabatic transfer.
Here we will focus on the purely adiabatic evolution, which initializes the $\ket{\downarrow,\downarrow}$ computational state.
The adiabatic problem is solved within \texttt{geodesiq} as follows:
\begin{minted}[bgcolor=LightGray, fontsize=\footnotesize]{python}
import numpy as np
import geodesiq as gq

def dqd_hamiltonian(eps, U, tc, Ez, dEz, dEx):
    ham = np.array([
        [U-eps, 0, -tc, tc, 0],
        [0, Ez, -dEx, dEx, 0],
        [-tc, -dEx, dEz, 0, dEx],
        [tc, dEx, 0, -dEz, -dEx],
        [0, 0, dEx, -dEx, -Ez]
    ])
    return ham

dqd_model = gq.ControlModel(dqd_hamiltonian)

# ----- Set system and control parameters -----
alpha, beta = 2, 2

U, tc, Ez, dEz, dEx = 10, 1, 0.9, 0.1, 0.01
eps0, epsf = 15, 0

dqd_model.set_parameters(U=U, tc=tc, Ez=Ez, dEz=dEz, dEx=dEx)
dqd_model.set_control(control_name="eps",
                      pulse_initial=eps0, 
                      pulse_final=epsf,
                      initial_state=0, 
                      alpha=alpha, 
                      beta=beta)

# ----- Solve for optimal pulse -----
dqd_model.solve_problem(pulse_accuracy=1000)
\end{minted}
This code is also available as a Jupyter notebook: \href{https://geodesiq-control.github.io/geodesiq-control/notebooks/index.html?path=dqd_model.ipynb}{examples/dqd\_model.ipynb}.
The results are shown in Fig.~\ref{fig: dqd}.
The infidelity as a function of pulse duration $t_\mathrm{c}t_\mathrm{f}$ for the FAQUAD (orange, dashed) and geometric (purple) methods is shown in Fig.~\ref{fig: dqd}(a), alongside the optimal (minimal) infidelity as a function of the adiabatic parameters $(\alpha,\beta)$ [See Fig.~\ref{fig: dqd}(b)], and the corresponding optimal pulse duration [See Fig.~\ref{fig: dqd}(c)].
Note that for adiabatic state transfer, the minimal infidelity is usually accompanied by longer pulse durations unless the pulse duration is tuned to an interference minimum, as seen for the geometric pulse.
As can be seen by the infidelity maps, the minimal infidelity regions are found in non-trivial locations in the $(\alpha,\beta)$ space.

\subsection{Many-body ground-state preparation}
\label{sec: many-body example}
\begin{figure}[!ht]
    \centering
    \includegraphics[width=\linewidth]{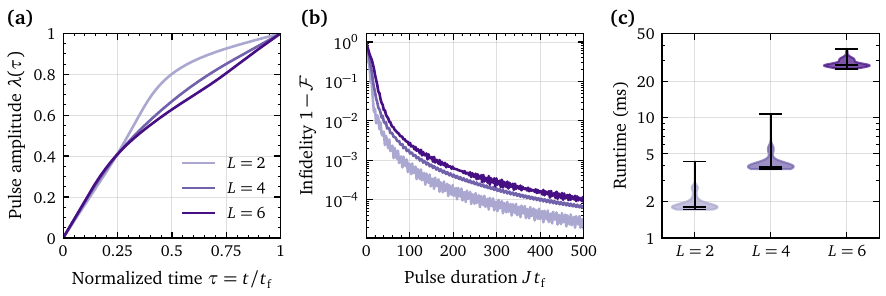}
    \caption{Preparing the ground state of the Ising model.
    We initialize the non-interacting ground state $\ket{-}^{\otimes L}$ and sweep the global control parameter $\lambda$ to prepare the interacting ground state of $\hat{H}_\mathrm{target}$.
    Panel \textbf{(a)} shows the resulting optimal pulse shapes as functions of the normalized time $\tau=t/t_\mathrm{f}$, with the corresponding infidelities in panel \textbf{(b)} as a function of total pulse time $Jt_\mathrm{f}$.
    Panel \textbf{(c)} shows the distribution of runtimes for generating the optimal pulse.
    The parameters chosen are $h_x/J=1$, $h_z/J=0.8$, and $(\alpha,\beta)=(2,2)$, with the Hamiltonian in the reflection-even sector of dimension $d=3$, $10$, and $36$ for $L=2$, $4$, and $6$.
    All pulses use \code{num_steps=257}, with \code{pulse_accuracy=1000} for panels (a) and (c), and \code{pulse_accuracy=100} for the infidelities in panel (b), evaluated at 200 pulse times in $[0.1, 500]/J$.
    For the runtime simulation, we repeat each measurement 200 times to account for CPU performance fluctuations.
    Each sample measures the wall-clock time for constructing the Hamiltonian with \texttt{QuSpin}, constructing the \code{ControlModel}, and calling \code{solve_problem()}; the horizontal bars mark the minimum, median, and maximum.
    The runtimes were measured on an Intel Core i7-13700K processor (16 cores / 24 threads, base clock 3.4~GHz) with 64~GB of RAM under Windows~11.}
    \label{fig: many body}
\end{figure}

To illustrate a many-body application of the \texttt{geodesiq} package, we prepare the ground state of a many-body Hamiltonian, namely an Ising chain with local fields along the $x$ and $z$ directions on each site $j$.
The model (with open boundary conditions) is given by
\begin{align}
    \hat{H}&=\lambda(t)\hat{H}_\mathrm{target}+(1-\lambda(t))\hat{H}_\mathrm{init}\nonumber\\
    &=\lambda(t)\left[J\sum_{j=1}^{L-1} \,\sigma_j^z\sigma_{j+1}^z +h_z\sum_{j=1}^L\sigma_j^z\right]+ \Big(1-\lambda(t)\Big)h_x\sum_{j=1}^L\sigma_j^x,
    \label{eqn: ising hamiltonian}
\end{align}
where $(h_x,h_z)$ parametrize the strengths of the local fields, $\lambda(t)$ is the control field, and $J=1$.
For $J>0$, open boundaries, even $L$, and $h_z<J$ (here $h_z=0.8\,J$), $\hat{H}_\mathrm{target}$ has two degenerate N\'eel ground states, which are mapped onto each other by the reflection $j\to L+1-j$.
Since $\hat{H}$ is reflection symmetric for all $\lambda$, we work in the reflection-even sector, where the target ground state is unique: the even superposition of the two N\'eel states, i.e., a cat-like state.
This sector has dimension $d=3$, $10$, and $36$ for $L=2$, $4$, and $6$, compared with $2^L$ for the full Hilbert space.
This interpolation has the affine form of Eq.~\eqref{eqn: affine Hamiltonian}, with
\begin{subequations}
    \begin{align}
        \hat{H}_{\mathrm{d}} &= \hat{H}_{\mathrm{init}}, \\
        \hat{H}_{\mathrm{c}} &= \hat{H}_{\mathrm{target}}-\hat{H}_{\mathrm{init}}.
    \end{align}
\end{subequations}
In particular, the control operator includes the negative of the initial transverse-field term:
\begin{equation}
    \hat{H}_{\mathrm{c}}
    = J\sum_{j=1}^{L-1}\sigma_j^z\sigma_{j+1}^z
    + h_z\sum_{j=1}^{L}\sigma_j^z
    - h_x\sum_{j=1}^{L}\sigma_j^x.
\end{equation}
Once the two endpoint Hamiltonians have been constructed as arrays in the same basis, the model is initialized as follows:
\begin{minted}[bgcolor=LightGray, fontsize=\footnotesize]{python}
model = gq.ControlModel(H_d=H_init, H_c=H_target - H_init)
model.set_control(control_name="lam", pulse_initial=0.0, pulse_final=1.0, initial_state=0, alpha=2.0, beta=2.0)
model.solve_problem(pulse_accuracy=1000)
\end{minted}

We aim to prepare the interacting ground state of the Ising model by initializing the ground state of $\hat{H}_\mathrm{init}$, which is the state $\ket{\psi_0(\lambda_\mathrm{i})}=\ket{-}^{\otimes L}$, with $\sigma^x\ket{-}=-\ket{-}$, and then sweeping the control field from $\lambda_\mathrm{i}=\lambda(0)=0$ to $\lambda_\mathrm{f}=\lambda(t_\mathrm{f})=1$.
The code for this example is in \href{https://geodesiq-control.github.io/geodesiq-control/notebooks/index.html?path=many_body.ipynb}{examples/many\_body.ipynb} and uses the \texttt{QuSpin} package~\cite{weinbergQuSpinPythonPackage2017, weinbergQuSpinPythonPackage2019} to construct the spin-chain Hamiltonian in the reflection-even sector (\code{spin_basis_1d(L, pblock=1)}).
Where \texttt{QuSpin} is not installed, as in the web browser, the notebook builds the same matrix with \texttt{NumPy}.
In Fig.~\ref{fig: many body}, we plot the state-transfer infidelity as a function of the rescaled total pulse time $Jt_\mathrm{f}$ for different system sizes.
As expected, the pulse time needed for a given infidelity increases with $L$ [Fig.~\ref{fig: many body}(b)].
Constructing the Hamiltonian and generating the optimal pulse takes a median of $1.8$, $3.9$, and $27$~ms for $L=2$, $4$, and $6$ [Fig.~\ref{fig: many body}(c)].

\subsection{Diabatic--adiabatic state transfer}
\label{sec: diad example}

Finally, we demonstrate how \texttt{geodesiq} can combine desired diabatic transitions with adiabatic passages within the same control sweep.
We consider a four-level system consisting of two coupled sites, left ($\mathrm{L}$) and right ($\mathrm{R}$), each with a local ground state $\ket{g}$ and excited state $\ket{e}$.
In the basis $\{\ket{\mathrm{L},e},\ket{\mathrm{L},g},\ket{\mathrm{R},e},\ket{\mathrm{R},g}\}$, the Hamiltonian reads
\begin{align}
    \label{eqn: four level Hamiltonian}
    \hat{H}(\varepsilon(t)) =
    \begin{pmatrix}
        \varepsilon(t)/2+\Delta_\mathrm{L} & 0 & t_{\mathrm{ee}} & t_{\mathrm{eg}} \\
        0 & \varepsilon(t)/2-\Delta_\mathrm{L} & t_{\mathrm{ge}} & t_{\mathrm{gg}} \\
        t_{\mathrm{ee}}^* & t_{\mathrm{ge}}^* & -\varepsilon(t)/2+\Delta_\mathrm{R} & 0 \\
        t_{\mathrm{eg}}^* & t_{\mathrm{gg}}^* & 0 & -\varepsilon(t)/2-\Delta_\mathrm{R}
    \end{pmatrix},
\end{align}
where $2\Delta_\mathrm{L}$ and $2\Delta_\mathrm{R}$ are the local energy splittings and $\varepsilon(t)$ is the detuning between the sites.
The tunnel matrix elements are parametrized as
\begin{align}
    t_{\mathrm{ee}} &= \frac{t_\mathrm{c}}{2}\left[1+e^{i(\varphi_\mathrm{L}-\varphi_\mathrm{R})}\right],
    & t_{\mathrm{eg}} &= \frac{t_\mathrm{c}}{2}\left(e^{i\varphi_\mathrm{L}}-e^{i\varphi_\mathrm{R}}\right), \\
    t_{\mathrm{gg}} &= t_{\mathrm{ee}}^*,
    & t_{\mathrm{ge}} &= -t_{\mathrm{eg}}^*.
\end{align}
We choose $\Delta_\mathrm{R}=2\Delta_\mathrm{L}$, $t_\mathrm{c}=0.2\Delta_\mathrm{L}$, $\varphi_\mathrm{L}=0.1$, and $\varphi_\mathrm{R}=\pi/2$.

\begin{figure}[!ht]
    \centering
    \includegraphics[width=\linewidth]{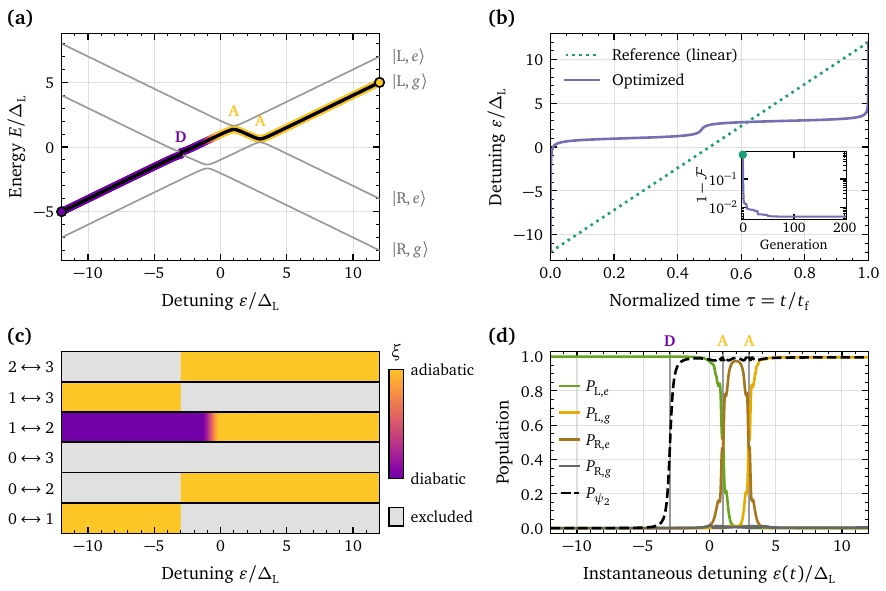}
    \caption{Diabatic--adiabatic state transfer in a four-level system using \texttt{geodesiq}, for $\Delta_\mathrm{R}=2\Delta_\mathrm{L}$, $t_\mathrm{c}=0.2\Delta_\mathrm{L}$, $\varphi_\mathrm{L}=0.1$, and $\varphi_\mathrm{R}=\pi/2$.
    The detuning is swept from $\varepsilon_\mathrm{i}=-12\Delta_\mathrm{L}$ to $\varepsilon_\mathrm{f}=12\Delta_\mathrm{L}$, connecting the initial eigenstate $\ket{\psi_1}$ to the target eigenstate $\ket{\psi_2}$, which approximately correspond to $\ket{\mathrm{L},e}$ and $\ket{\mathrm{L},g}$ at the respective endpoints.
    Panel \textbf{(a)} shows the instantaneous eigenenergies (gray lines), labeled on the right by the diabatic states they approach at large detuning.
    The occupied eigenstate (bold black line, highlighted with the colors of $\xi$ in panel \textbf{(c)}) indicates the prescribed diabatic passage (D), followed by two adiabatic passages (A).
    Panel \textbf{(b)} compares the reference linear ramp (green dotted line) with the optimized pulse (purple), as functions of the normalized time $\tau=t/t_\mathrm{f}$.
    The four exponents $(\alpha,\beta;\tilde{\alpha},\tilde{\beta})$ and the total pulse time are optimized using differential evolution; the inset shows the corresponding infidelity as a function of optimization generation; the green marker at generation $-1$ is the reference linear ramp (infidelity $0.72$), which is not part of the search, and generation $0$ is the best member of the random initial population.
    Panel \textbf{(c)} shows the di--ad transition matrix $\xi_{kn}(\varepsilon)$, with $\xi=0$ and $\xi=1$ corresponding to diabatic and adiabatic passages, respectively.
    Each row is a pair of levels, labeled with the lower index first (e.g., $1\leftrightarrow2$), and shows $\xi_{kn}$ wherever one of the two levels is the occupied state $\ket{\psi_k}$.
    Gray regions indicate excluded transitions that do not involve the prescribed occupied eigenstate.
    Panel \textbf{(d)} shows the resulting diabatic-state populations $P_{\mathrm{L},e},\dots,P_{\mathrm{R},g}$ as a function of instantaneous detuning.
    The black dashed line gives the population $P_{\psi_2}$ of $\ket{\psi_2(\varepsilon)}$, with a final state-transfer infidelity $1-\mathcal{F}\approx4.9\times10^{-3}$.
    The gray vertical lines mark the crossings D and A of panel (a).}
    \label{fig:diabatic_transition}
\end{figure}

Labeling the instantaneous eigenstates by increasing energy, starting from zero, the control task is to connect $\ket{\psi_1(\varepsilon_\mathrm{i})}$ to $\ket{\psi_2(\varepsilon_\mathrm{f})}$ while the detuning is swept from $\varepsilon_\mathrm{i}=-12\Delta_\mathrm{L}$ to $\varepsilon_\mathrm{f}=12\Delta_\mathrm{L}$.
We chose these values of detuning endpoints to ensure these states approximately correspond to $\ket{\mathrm{L},e}$ and $\ket{\mathrm{L},g}$, respectively.
Purely adiabatic evolution would preserve the eigenstate index and therefore would not reach the desired target.
Instead, the prescribed path in Fig.~\ref{fig:diabatic_transition}(a) passes the anticrossing near $\varepsilon=-3\Delta_\mathrm{L}$ diabatically, changing the occupied eigenstate from $\ket{\psi_1}$ to $\ket{\psi_2}$ while approximately preserving the diabatic character $\ket{\mathrm{L},e}$.
The subsequent anticrossings near $\varepsilon=\Delta_\mathrm{L}$ and $\varepsilon=3\Delta_\mathrm{L}$ are passed adiabatically, transferring the population through $\ket{\mathrm{R},e}$ to $\ket{\mathrm{L},g}$.

Within \texttt{geodesiq}, this path is specified by setting \code{initial_state=1} and \code{final_state=2}.
The di--ad transition matrix is then constructed from the spectrum, as described in Sec.~\ref{sec: theory}.
Figure~\ref{fig:diabatic_transition}(c) shows that the element $\xi$ between $\ket{\psi_1}$ and $\ket{\psi_2}$ (row $1\leftrightarrow2$) selects the first passage diabatically and subsequently switches smoothly to the adiabatic prescription.
In particular, the later anticrossing of the same pair of eigenstates is passed adiabatically, since the target eigenstate index has already been reached.
All other transitions involving the prescribed occupied eigenstate are given adiabatic weights, while transitions between the remaining states are excluded from the di--ad tensor.

To determine the pulse shape, we minimize the final state-transfer infidelity over the four exponents $(\alpha,\beta;\tilde{\alpha},\tilde{\beta})$ and the total pulse time $t_\mathrm{f}$ using the differential evolution of SciPy~\cite{virtanenSciPy10Fundamental2020}.
Each exponent is varied within $[-10,10]$, and the duration is restricted to $\Delta_\mathrm{L}t_\mathrm{f}\in[100,1000]$.
For each trial, \texttt{geodesiq} constructs the pulse from the di--ad tensor, and the resulting fidelity is evaluated through time-dependent propagation.
The search uses a population of 80 (\code{popsize=16}) and runs for 200 generations with \code{seed=0}, without a final local polishing step.
With both $\hat{H}$ and the control $\varepsilon$ in units of $\Delta_\mathrm{L}$, it finds the optimum at $\alpha=4.07$, $\beta=2.31$, $\tilde{\alpha}=-7.33$, $\tilde{\beta}=9.96$, and $\Delta_\mathrm{L}t_\mathrm{f}=123.9$.
As a reference, we use a linear ramp with all four exponents set to zero and $\Delta_\mathrm{L}t_\mathrm{f}=500$.
The notebook \href{https://geodesiq-control.github.io/geodesiq-control/notebooks/index.html?path=four_level.ipynb}{examples/four\_level.ipynb} builds this model and reproduces the optimized pulse and both infidelities from the optimum given above, without repeating the search.

The optimized pulse and the convergence of the search are shown in Fig.~\ref{fig:diabatic_transition}(b).
The pulse traverses the first anticrossing rapidly and allocates more time to the subsequent adiabatic passages.
The simulated populations in Fig.~\ref{fig:diabatic_transition}(d) follow the intended sequence, with the population of $\ket{\psi_2(\varepsilon)}$ increasing after the diabatic passage and remaining close to unity thereafter.
The final infidelity decreases from approximately $0.72$ for the reference linear ramp to $4.9\times10^{-3}$ for the optimized pulse.
This example illustrates how the di--ad tensor provides a restricted family of pulses that can be further optimized to combine selective excitations with adiabatic state transfer.

\section{Conclusion}
\label{sec: conclusion}
Geometric protocols and methods have been gaining attention in the pursuit of an interpretable and optimal approach to quantum control.
Here, optimality refers to a predefined heuristic cost whose extremal solutions are geodesics of a (generalized) quantum metric.
In this work, we have shown how different heuristics, for instance, those that define the geometric and the FAQUAD protocols, may be generalized to include many adiabatic and diabatic protocols, allowing for a flexible toolkit to explore optimality for a given system.

We have presented the \texttt{geodesiq} Python package, which provides an easy-to-use and flexible framework for geometric pulse design for Hamiltonians with a single control parameter, as long as the occupied instantaneous eigenstate remains non-degenerate along the sweep, apart from symmetry-protected crossings, through which the state is followed.
The package also provides methods for plotting, discretizing, filtering, and exporting pulses, and simulators of the time evolution for quick prototyping and testing.

We have applied the package to spin-qubit initialization in a double quantum dot, to ground-state preparation in an Ising chain, and to a diabatic--adiabatic transfer in a four-level system.
In the benchmarks of App.~\ref{app: numerical benchmarking}, \texttt{geodesiq} generates pulses one to more than three orders of magnitude faster than GRAPE and CRAB and keeps a low infidelity at lower filter cutoffs, whereas GRAPE reaches the smallest infidelities when the full bandwidth is available.
For a fixed choice of exponents, the pulse shape follows from a single first-order ordinary differential equation.
For the four-level transfer in Sec.~\ref{sec: diad example}, however, the exponents $(\alpha,\beta;\tilde{\alpha},\tilde{\beta})$ and $t_\mathrm{f}$ were tuned by an outer differential-evolution search that propagates the dynamics for every trial.
The exponents thus act as a low-dimensional family of pulses that may still need to be tuned against the simulated dynamics in multilevel diabatic--adiabatic transfers.

In the future, we plan to extend the code in several directions that are active areas of research.
First, we aim to provide full multi-parameter control, which requires the solution of the coupled, non-linear geodesic equations.
Second, we aim to reduce the main computational costs, which are the dense diagonalization of the Hamiltonian, the construction of the di--ad tensor, and the solution of the first-order ordinary differential equation.
For many-body systems, the dense diagonalization at every grid point dominates the cost, and the memory for the $N\times d\times d$ arrays limits the present implementation to Hilbert-space dimensions of order $10^3$ (e.g., spin chains with $L\lesssim 10$ without symmetry reduction)\footnote{A chain of $L$ spins-$1/2$ has the Hilbert-space dimension $d=2^L$, e.g., $d=1024$ for $L=10$, for which each real $N\times d\times d$ array occupies about $8$~GB at the default $N=1025$ grid points.
Symmetries of the Hamiltonian or the restriction to a smaller set of basis states reduce the effective dimension, as in Sec.~\ref{sec: many-body example}, and a coarser control grid (\code{num_steps}) reduces $N$.}.
Iterative methods to solve for the eigenvalues and eigenvectors could significantly improve the computation of the di--ad tensor, as one is typically interested only in a small subset of the many-body spectrum (for instance, adiabatic ground-state preparation), provided that the contributions of the omitted states to the metric are negligible\footnote{One may also envision the use of Schrieffer--Wolff perturbation theory methods provided by existing Python libraries like SymPT~\cite{diotalleviSymPTComprehensiveTool2024} and Pymablock~\cite{dayPymablockAlgorithmPackage2025}.}.
Finally, Krylov-based schemes~\cite{finzgarCounterdiabaticDrivingPerformance2025, morawetzUniversalCounterdiabaticDriving2025} could be used to accelerate the approximate construction of the quantum metric tensor and its generalizations.

\section*{Acknowledgements}
We thank all members of the Rimbach-Russ and Bosco groups for providing valuable insights.

\paragraph{Author contributions}
D.F.-F. and C.V.-M. contributed equally to the development of the Python package and to the writing of the manuscript.
G.P. and M.R.-R. jointly supervised the project.
All authors discussed the results and reviewed the manuscript.

\paragraph{Funding information}
G.P. and D.F.-F. acknowledge support from the Spanish Ministry of Science through grant PID2023-149072NB-I00 and from the CSIC Research Platform PTI-001.
M.R.-R. and C.V.-M. acknowledge support from the EU through the Horizon Europe QLSI2 project and from the Army Research Office under Award Number W911NF-23-1-0110.
M.R.-R. acknowledges support from the Dutch Research Council (NWO) under Award Number Vidi TTW 22204.
The views and conclusions contained in this document are those of the authors and should not be interpreted as representing the official policies, either expressed or implied, of the Army Research Office or the U.S.
Government.
The U.S.
Government is authorized to reproduce and distribute reprints for Government purposes, notwithstanding any copyright notation herein.

\paragraph{Data and code availability}
The repository for the \texttt{geodesiq} Python package is available at \url{https://github.com/geodesiq-control/geodesiq-control}.
Its example notebooks can be run in the web browser, without installation, at \url{https://geodesiq-control.github.io/geodesiq-control/}.
The software version described in this work is \texttt{geodesiq} v0.5.0, archived on Zenodo with the DOI \href{https://doi.org/10.5281/zenodo.24321210}{\nolinkurl{10.5281/zenodo.24321210}}.
All versions are collected under the concept DOI \href{https://doi.org/10.5281/zenodo.23985948}{\nolinkurl{10.5281/zenodo.23985948}}, which resolves to the latest release.
This version is also available on PyPI at \url{https://pypi.org/project/geodesiq/0.5.0/}.
The archived release contains the notebook and data that produce the figures, \nolinkurl{dev/manuscript_figures.ipynb} and \nolinkurl{dev/data/}, and the benchmark scripts and results of App.~\ref{app: numerical benchmarking}, in \nolinkurl{dev/other_qoctrl/}.

The package is built on \texttt{NumPy}~\cite{harrisArrayProgrammingNumPy2020}, \texttt{SciPy}~\cite{virtanenSciPy10Fundamental2020}, \texttt{Matplotlib}~\cite{hunterMatplotlib2DGraphics2007}, and \texttt{QuTiP}~\cite{johanssonQuTiPOpensourcePython2012, johanssonQuTiP2Python2013, lambertQuTiP5Quantum2026}.
The many-body example uses \texttt{QuSpin}~\cite{weinbergQuSpinPythonPackage2017, weinbergQuSpinPythonPackage2019}, and the GRAPE and CRAB benchmarks in App.~\ref{app: numerical benchmarking} use \texttt{qutip-qtrl}, the optimal control package of \texttt{QuTiP}~\cite{lambertQuTiP5Quantum2026}.
The figures and benchmarks were produced with Python 3.14, \texttt{NumPy} 2.5.4, \texttt{SciPy} 1.18.1, \texttt{Matplotlib} 3.11.2, \texttt{QuTiP} 5.3.1, \texttt{qutip-qtrl} 0.2.0, \texttt{QuSpin} 1.0.1, and \texttt{geodesiq} v0.5.0.

\paragraph{Use of AI tools}
Claude (Anthropic), through Claude Code, was used to assist with code development, testing, and language editing of this manuscript.
All AI-generated output was checked by the authors, who take full responsibility for the scientific content, interpretations, and conclusions.

\begin{appendix}
\numberwithin{equation}{section}

\section{Note on open-source license}
\label{app: LGPL}
The Python package presented in this work is licensed under the GNU Lesser General Public License, version 2.1 only (LGPL-2.1-only), a weak-copyleft license.
It allows the package to be used in academic and commercial pipelines, including closed-source software, without imposing the license on the code that uses it.
The practical implications for users and developers working with this package include:
\begin{itemize}
    \item \textbf{Integration:} Researchers and developers can import and use the unmodified package within closed-source or commercial applications without copyleft obligations for their own application code.
    \item \textbf{Modification:} Modifications may be kept private, but if a modified version of the package is distributed, its source code must be made available to the recipients under the terms of the LGPL.
    \item \textbf{Redistribution:} Anyone redistributing the package---whether standalone or bundled within a larger software product---must include the original license terms and ensure that the library remains distinct enough for users to modify or update it independently.
\end{itemize}
More information about the license can be found at \url{https://opensource.org/license/lgpl-2-1}.

\section{Formal details and derivations}
\label{app: formal details}

\subsection{Quantum metric and energy variance}
As defined in Eq.~\eqref{eqn: fidelity susceptibility}, the quantum metric tensor measures the distance between two infinitesimally close states $\ket{\psi_k(x)}$ and $\ket{\psi_k(x+dx)}$.
The quantum metric can also be understood in light of the energy variance.
Consider a state $\ket{\Psi(t)}$ evolving under the Schrödinger equation.
To second order in $dt$,
\begin{align}
    \ket{\Psi(t+dt)}&\approx \ket{\Psi(t)}+\dv{}{t}\ket{\Psi(t)}dt +\frac{1}{2}\dv[2]{}{t}\ket{\Psi(t)}dt^2\nonumber\\
    &= \ket{\Psi(t)}+(-i\hat{H})\ket{\Psi(t)}dt+\frac{1}{2}\left(-i \dv{\hat{H}}{t}-\hat{H}^2\right)\ket{\Psi(t)}dt^2.
\end{align}
The resulting overlap with the original wave function yields
\begin{align}
    |\braket{\Psi(t)}{\Psi(t+dt)}|^2\approx 1- \Big(\expval{\hat{H}^2}-\expval{\hat{H}}^2\Big)\,dt^2\equiv 1- \sigma_E^2 dt^2.
\end{align}
Hence, over a time $dt$, the quantum metric assigns to the evolution the length $d\ell=\sigma_E\,dt$, which connects the metric to the energy variance discussed in Sec.~\ref{sec: geometric control}.

\subsection{Analytic \texorpdfstring{$(\alpha,\beta)$}{(alpha, beta)}-hypermetric of the Bloch sphere}

The Bloch sphere is the space of pure states of a two-level system, parametrized by the angles $(\theta,\varphi)$.
We may investigate the generalized geometry of the same model.
We consider the single-qubit Hamiltonian with unit field strength in spherical coordinates,
\begin{align}
    \label{eqn: qubit Hamiltonian}
    \hat{H}_\mathrm{qubit}(\theta,\varphi)=\vec{n}(\theta,\varphi)\cdot \vec{\sigma}=\begin{pmatrix}
        \cos \theta & e^{-i\varphi}\sin \theta \\
        e^{i\varphi}\sin \theta  & -\cos \theta
    \end{pmatrix},
\end{align}
where $\vec{n}(\theta,\varphi)$ is the unit vector on the 2-sphere and $\vec{\sigma}=(\sigma^x,\sigma^y,\sigma^z)$ is the Pauli vector.
The angles $x^\mu=(\theta,\varphi)$ represent the parameters of the Hamiltonian.
The eigenstates (with eigenvalues $E_{0,1}=\mp 1$) of the qubit Hamiltonian in Eq.~\eqref{eqn: qubit Hamiltonian} are
\begin{align}
    \ket{\psi_0} & = \begin{pmatrix}
        \sin(\theta/2) \\
        -e^{i\varphi} \cos(\theta/2)
    \end{pmatrix}, & \ket{\psi_1} & = \begin{pmatrix}
        \cos(\theta/2) \\
        e^{i\varphi} \sin(\theta/2)
    \end{pmatrix}.
\end{align}
To compute the $(\alpha,\beta)$-hypergeometric tensor $\mathcal{Q}_{\mu \nu}^{(\alpha,\beta)}$ [Eq.~\eqref{eqn: generalized qgt}] of the ground state, $k=0$, we need the matrix elements, which do not depend on $(\alpha,\beta)$:
\begin{subequations}
    \begin{align}
        \mel{\psi_0}{\partial_\theta\hat{H}_\mathrm{qubit}}{\psi_1}&=-1,\\
        \mel{\psi_0}{\partial_\varphi\hat{H}_\mathrm{qubit}}{\psi_1}&=-i\sin \theta,\\
        E_1-E_0&=2.
    \end{align}
\end{subequations}
As we are dealing with a two-level system, we can either be purely adiabatic or purely diabatic.
Hence, we only need a single set of $(\alpha,\beta)$.
Therefore, we find that the $(\alpha,\beta)$-hypergeometric tensor (in matrix representation in parameter space $\mu,\nu$) for a single-qubit system is given by
\begin{align}
   [\mathcal{Q}_{\mu \nu}^{(\alpha,\beta)}(\theta,\varphi)]= \frac{1}{2^\alpha}\begin{pmatrix}
       1 & -i \sin^{\beta/2} \theta\\
       i \sin^{\beta/2} \theta & \sin^\beta \theta
   \end{pmatrix}.
\end{align}
Here we used $P^{(01)}_{\theta\theta}=1$, $P^{(01)}_{\varphi\varphi}=\sin^2\theta$, and $P^{(01)}_{\theta\varphi}=(P^{(01)}_{\varphi\theta})^*=-i\sin\theta$, with $\sin\theta\geq 0$ for $\theta\in[0,\pi]$.
Since the off-diagonal elements are purely imaginary, they do not contribute to the hypermetric tensor.
Hence, the single-qubit hypermetric tensor gives the line element
\begin{align}
    G_{\mu \nu}^{(\alpha,\beta)}dx^\mu dx^\nu =\frac{1}{2^\alpha}\Big(d\theta^2+\sin^\beta \theta \,d\varphi^2\Big),
\end{align}
which we call the hyper-Bloch sphere.
For the geometric case $(\alpha,\beta)=(2,2)$, the prefactor is $1/2^\alpha=1/4$, and we find the line element of the 2-sphere of radius $1/2$, the standard Bloch sphere.

\subsection{Analytic pulses for the Landau--Zener problem}
For the Landau--Zener problem, we take the Hamiltonian in Eq.~\eqref{eqn: Landau Zener model}, where $z(t)$ is the control parameter.
With $|\mel{\psi_0}{\sigma^z}{\psi_1}|=|\Delta|/\sqrt{\Delta^2+z^2}$ and $E_1-E_0=2\sqrt{\Delta^2+z^2}$, this yields the following metric
\begin{align}
    G_{zz}^{(\alpha,\beta)}=\frac{1}{2^\alpha} \frac{|\Delta|^\beta}{(\Delta^2+z(t)^2)^{(\alpha+\beta)/2}}.
\end{align}
Since the pulse equation, Eq.~\eqref{eq: hypergeo protocol}, reads $\sqrt{G_{zz}}\,|dz/d\tau|=\mathcal{L}^{(\alpha,\beta)}$, the antiderivative $F$ of $\sqrt{G_{zz}}$ is linear in $\tau$ along a monotonic pulse,
\begin{align}
    F(z)=\int^z dz'\,\sqrt{G_{zz}^{(\alpha,\beta)}(z')},\qquad F(z(\tau))=F(z_\mathrm{i})+\tau\,[F(z_\mathrm{f})-F(z_\mathrm{i})].
\end{align}
For general exponents, the antiderivative is a Gauss hypergeometric function,
\begin{align}
    F(z)=\frac{z}{2^{\alpha/2}|\Delta|^{\alpha/2}}\;{}_2F_1\!\left(\frac{1}{2},\frac{\alpha+\beta}{4};\frac{3}{2};-\frac{z^2}{\Delta^2}\right),
\end{align}
so that the pulse follows from inverting $F$ for any $(\alpha,\beta)$; for instance, $(\alpha,\beta)=(1,1)$ gives $F(z)=\sqrt{\Delta/2}\,\operatorname{arcsinh}(z/\Delta)$ for $\Delta>0$.
For $(2,2)$ and $(4,2)$, $F$ reduces to elementary functions that can be inverted in closed form.
For the geometric protocol, $(\alpha,\beta)=(2,2)$, we have $\sqrt{G_{zz}}=|\Delta|/[2(\Delta^2+z^2)]$ and, for $\Delta>0$, $F(z)=\arctan(z/\Delta)/2$.
Hence, $\arctan(z/\Delta)$ is linear in $\tau$, and the pulse follows in closed form,
\begin{align}
    z(\tau)=\Delta\tan[\vartheta_\mathrm{i}+\tau(\vartheta_\mathrm{f}-\vartheta_\mathrm{i})],\qquad \vartheta_{\mathrm{i},\mathrm{f}}=\arctan(z_{\mathrm{i},\mathrm{f}}/\Delta).
\end{align}
For the FAQUAD protocol, $(\alpha,\beta)=(4,2)$, we have
\begin{align}
    \sqrt{G_{zz}}=\frac{|\Delta|}{4(\Delta^2+z^2)^{3/2}},\qquad F(z)=\frac{z}{4|\Delta|\sqrt{\Delta^2+z^2}}.
\end{align}
Hence, $w=z/\sqrt{\Delta^2+z^2}$ is linear in $\tau$, and
\begin{align}
    z(\tau)=\frac{|\Delta|\,w(\tau)}{\sqrt{1-w(\tau)^2}},\qquad w(\tau)=w_\mathrm{i}+\tau(w_\mathrm{f}-w_\mathrm{i}),\qquad w_{\mathrm{i},\mathrm{f}}=\frac{z_{\mathrm{i},\mathrm{f}}}{\sqrt{\Delta^2+z_{\mathrm{i},\mathrm{f}}^2}}.
\end{align}
For the parameters of Sec.~\ref{sec: geodesiq} ($\Delta=1$, $z_\mathrm{i}=-10$, $z_\mathrm{f}=10$), both analytic pulses agree with the output of \code{solve_problem()} to within $5\times10^{-7}$, which provides a check of the numerical solution.

\section{Numerical benchmarking}
\label{app: numerical benchmarking}

In this appendix, we benchmark the pulses generated by \texttt{geodesiq} against two widely used numerical optimal-control methods: gradient-ascent pulse engineering (GRAPE)~\cite{khanejaOptimalControlCoupled2005} and the chopped-random-basis method (CRAB)~\cite{doriaOptimalControlTechnique2011, canevaChoppedRandombasisQuantum2011}.
The comparison is intended to characterize complementary practical aspects of the protocols rather than to establish a universal ranking.
In particular, we compare the structure of the optimized pulses, the fidelity retained under a finite control bandwidth, and the computational cost of constructing the pulse.

For GRAPE and CRAB, the control amplitude is constrained to the interval spanned by the \texttt{geodesiq} endpoints, $[z_\mathrm{i},z_\mathrm{f}]$ or $[\lambda_\mathrm{i},\lambda_\mathrm{f}]$, and the iterations are stopped once the target infidelity is reached.
GRAPE pulses are obtained with the \texttt{qutip-qtrl} implementation~\cite{lambertQuTiP5Quantum2026}, using a piecewise-constant discretization of the control over $N$ time slots and a maximum of 2000 optimization iterations.
The first and last control amplitudes of the initial pulse are set to the lower and upper endpoints, respectively, but the optimizer can change them within the amplitude bounds.
To assess the dependence on initialization, GRAPE is run using both a linear ramp and a random initial pulse.
CRAB is also implemented with \texttt{qutip-qtrl} and uses a truncated random Fourier basis with 10 frequency components (20 coefficients) and $N$ time slots.
We employ a sequence similar to dressed CRAB (dCRAB)~\cite{rachDressingChoppedrandombasisOptimization2015} with up to 10 super-iterations: the first optimization is performed around a linear ramp, while each subsequent super-iteration uses the best pulse obtained so far as its reference and adds a new random-basis correction with relative scale $0.1$.
Each CRAB optimization is limited to 2000 iterations, and the sequence is terminated early when the target infidelity is reached.
The \texttt{geodesiq} pulse is instead constructed by defining the same initial and final control values and solving the geometric control equation described in Sec.~\ref{sec: theory} with \code{num_steps=N} and \code{pulse_accuracy=N}; the resulting solution is subsequently rescaled to the desired total evolution time $t_{\mathrm f}$.
The state-transfer fidelity used to determine successful pulse generation is evaluated by propagating the common initial state under the corresponding time-dependent Hamiltonian and computing its overlap with the target state.

To assess sensitivity to finite control bandwidth, we apply a third-order low-pass Butterworth filter (the default of \code{filtered_pulse()}) to each generated pulse and recompute the final-state fidelity.
The filter uses the sampling frequency $f_\mathrm{s}=1/\Delta t$, where $\Delta t$ is the uniform spacing of the pulse time grid.
The cutoff frequency $f_\mathrm{c}$ is expressed in cycles per unit time and satisfies $0<f_\mathrm{c}<f_\mathrm{s}/2$.
Filtering is performed in both the forward and backward directions, yielding zero phase shift and an overall magnitude response equal to the square of the single-pass response (gain $1/2$ at $f_\mathrm{c}$).
Before filtering, each pulse is extended by holding its initial and final values until the impulse response of the filter has decayed, as an arbitrary waveform generator holds its output before and after a pulse.
The filtered pulses retain their original time grid and duration.

To benchmark the computational cost, we measure the wall-clock time of each pulse-generation routine using Python's \code{time.perf_counter()}.
For each $N$, we perform 50 sequential runs with identical inputs, except that no fixed seed is set, so each run of GRAPE with a random initial pulse starts from a different pulse, and each run of CRAB uses a different random basis.
Each timing covers the full pulse-generation call, including the fidelity evaluation that decides success: the internal fidelity evaluations of GRAPE and CRAB, and, for \texttt{geodesiq}, the time propagation of the generated pulse.
Runs that do not reach the target infidelity count as unsuccessful.
We test $N=2^r+1$ with $r=2,\dots,11$, i.e., from $N=5$ to $N=2049$.
For \texttt{geodesiq}, $N$ sets both \code{num_steps} and \code{pulse_accuracy}, whereas for GRAPE and CRAB it is the number of time slots.
The runtime panels show the mean over the successful runs, and values of $N$ with no successful run are omitted.
All runtimes in this paper, in this appendix and in Fig.~\ref{fig: many body}, were measured on the same machine: an Intel Core i7-13700K processor (16 cores / 24 threads, base clock 3.4~GHz) with 64~GB of RAM under Windows~11.
The software versions are listed under Data and code availability.
The BLAS libraries (OpenBLAS, bundled with \texttt{NumPy} and \texttt{SciPy}) were left at their default of 24 threads.
\texttt{geodesiq} limits them to a single thread around its dense linear algebra for matrices of dimension up to 256, which includes every system benchmarked here, whereas GRAPE and CRAB in \texttt{qutip-qtrl} run with the default setting.

\subsection{Landau--Zener benchmark}
\label{app: LZ benchmark}

\begin{figure*}[!ht]
    \centering
    \includegraphics[width=\textwidth]{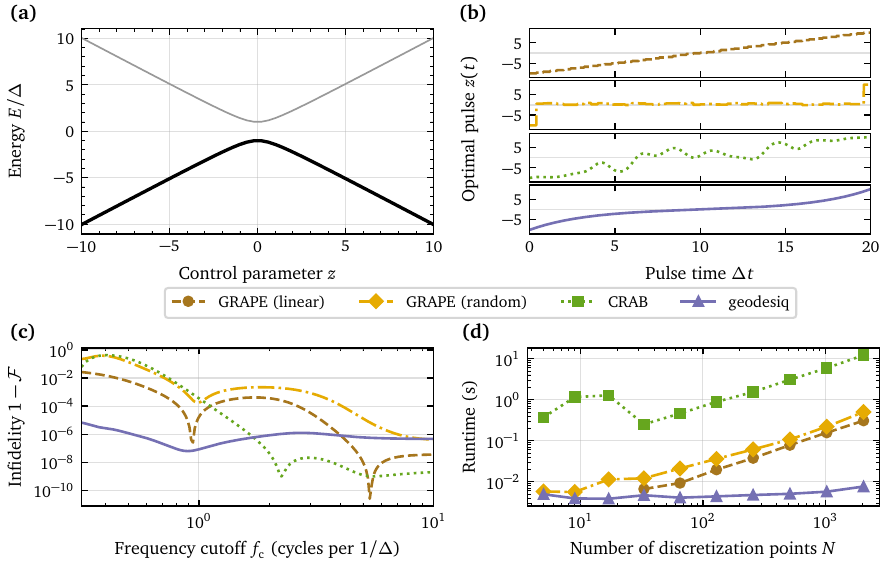}
    \caption{Benchmark of state-transfer protocols for the Landau--Zener problem.
    \textbf{(a)} Instantaneous eigenenergies (gray) as a function of the control parameter $z$, with the eigenenergy branch to follow in bold black.
    \textbf{(b)} Control pulses generated using GRAPE initialized with a linear (dashed brown) or a random pulse (dash-dotted yellow), CRAB (dotted green), and \texttt{geodesiq} (solid purple), for $t_\mathrm{f}=20/\Delta$.
    Each pulse is shown in its own strip with a common vertical scale; the thin gray line marks $z=0$.
    \textbf{(c)} Final-state infidelity after low-pass filtering the control pulses, plotted as a function of the filter cutoff frequency.
    The filter is a zero-phase third-order Butterworth filter with the initial and final values held before and after the pulse (App.~\ref{app: numerical benchmarking}).
    Frequencies are expressed in cycles per unit time.
    \textbf{(d)} Wall-clock time for pulse generation as a function of the number of discretization points, averaged over the successful runs; values of $N$ without a successful run are omitted.
    The common legend applies to panels \textbf{(b)--(d)}.
    In all panels, $\Delta=1$ and $(\alpha,\beta)=(1,1)$ for the \texttt{geodesiq} pulse.
    The target infidelity is $4\times10^{-7}$ in panels \textbf{(b)} and \textbf{(c)} and $2\times10^{-6}$ in panel \textbf{(d)}, where $t_\mathrm{f}=50/\Delta$.
    In panels \textbf{(b)} and \textbf{(c)}, it only stops the iterative methods; the \texttt{geodesiq} pulse reaches $1-\mathcal{F}\approx4.5\times10^{-7}$.}
    \label{fig: LZ benchmarks}
\end{figure*}

We first consider the Landau--Zener Hamiltonian in Eq.~\eqref{eqn: Landau Zener model}, with $\Delta=1$ and a sweep of the control
parameter from $z_\mathrm{i}=-10$ to $z_\mathrm{f}=10$.
The initial state is the ground state at $z_\mathrm{i}$, and the target is the instantaneous ground state at
$z_\mathrm{f}$, as shown in Fig.~\ref{fig: LZ benchmarks}(a).
All methods address the same transfer between the same states; unlike \texttt{geodesiq}, GRAPE and CRAB may change the endpoint control values.
The common fidelity is
\begin{align}
    \mathcal{F}=\left|\braket{\psi_0(z_\mathrm{f})}
    {\hat{U}(t_\mathrm{f})\psi_0(z_\mathrm{i})}\right|^2,
\end{align}
with the time-evolution operator $\hat{U}(t_\mathrm{f})$ generated by the respective pulse.
For \texttt{geodesiq}, we use $(\alpha,\beta)=(1,1)$, whose pulse has the closed form given in App.~\ref{app: formal details}.
For this sweep, it gives the lowest infidelity of the exponent pairs $(0,0)$, $(1,1)$, $(2,2)$, and $(4,2)$ at both pulse times used below, and it is the only one that reaches the target infidelity of the runtime benchmark.
Its infidelity decreases with the pulse time, staying below $10^{-6}$ for all $t_\mathrm{f}\gtrsim29/\Delta$, whereas the geometric choice $(2,2)$, faster for moderate targets such as $10^{-3}$, reaches only $8\times10^{-6}$ at $t_\mathrm{f}=50/\Delta$.
Panels (b) and (c) of Fig.~\ref{fig: LZ benchmarks} use $t_\mathrm{f}=20/\Delta$ and a target infidelity of $4\times10^{-7}$, and panel (d) uses $t_\mathrm{f}=50/\Delta$ and $2\times10^{-6}$.
The target only stops the iterative GRAPE and CRAB optimizations; the \texttt{geodesiq} pulse is not iterated and reaches $1-\mathcal{F}\approx4.5\times10^{-7}$ at $t_\mathrm{f}=20/\Delta$.

Figure~\ref{fig: LZ benchmarks}(b) illustrates the qualitatively different control solutions obtained for a total duration $t_\mathrm{f}=20/\Delta$.
The \texttt{geodesiq} protocol produces a smooth monotonic sweep, while the numerically optimized GRAPE and CRAB solutions contain oscillatory or step-like features.
To quantify the practical relevance of these features, we apply the same low-pass filter to each pulse and recompute the state-transfer fidelity.
As shown in Fig.~\ref{fig: LZ benchmarks}(c), the iterative optimal control methods can reach very small infidelities when the cutoff is high enough to retain their high-frequency components, but their response to the cutoff is non-monotonic and depends on the optimized solution.
By contrast, the smooth \texttt{geodesiq} pulse retains a comparatively small infidelity already at low cutoff frequencies and varies only weakly once the relevant pulse bandwidth is resolved.
Thus, the benchmark exposes a trade-off between the lowest attainable numerical infidelity and robustness to bandwidth limitations.

Figure~\ref{fig: LZ benchmarks}(d) compares the wall-clock time required to construct each pulse as the number of time slots $N$ is increased.
The geometric construction is the least expensive method throughout the tested
range and displays the weakest increase in runtime.
The cost of GRAPE grows faster with $N$, and CRAB is the most expensive method throughout.
At the largest discretization shown, the \texttt{geodesiq} runtime is approximately $0.008$~s, compared with $0.31$--$0.50$~s for GRAPE and approximately $12$~s for CRAB, i.e., between one and two orders of magnitude faster than GRAPE and more than three orders of magnitude faster than CRAB.
In this benchmark, \texttt{geodesiq} reaches the target infidelity for all tested $N$, down to $N=5$, whereas GRAPE initialized with a linear ramp requires $N\geq 33$.
GRAPE with a random initialization and CRAB miss the target in some runs for $N\leq 17$ and $N\leq 9$, respectively.

\subsection{Many-body benchmark}
\label{app: many-body benchmark}

We next benchmark the protocols for the Ising chain of Eq.~\eqref{eqn: ising hamiltonian} (Sec.~\ref{sec: many-body example}), using the global control parameter $\lambda(t)$.
Here, we take a chain of $L=4$ spins and adiabatically follow the sixth excited state of the reflection-even sector (index 6, \code{initial_state=final_state=6}) as an example of excited-state transfer.
The sweep starts at $\lambda_\mathrm{i}=0.1$, because at $\lambda=0$ this state is degenerate with the fifth excited state, and, symmetrically, ends at $\lambda_\mathrm{f}=0.9$.
Along the sweep, the state remains non-degenerate, approaches its upper neighbor towards $\lambda_\mathrm{f}$, and has a narrow anticrossing with the fifth excited state (gap $\approx 0.08\,J$ near $\lambda\approx 0.67$).

\begin{figure*}[!ht]
    \centering
    \includegraphics[width=\textwidth]{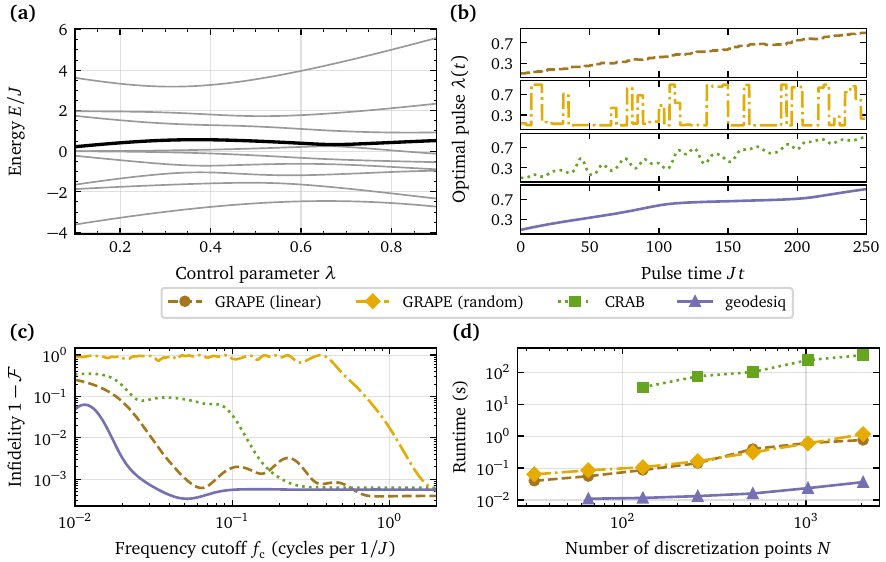}
    \caption{Benchmark of state-transfer protocols for the many-body Ising model.
    \textbf{(a)} Instantaneous eigenenergies (gray) as a function of the control parameter $\lambda$, with the eigenenergy branch to follow in bold black.
    \textbf{(b)} Control pulses generated using GRAPE initialized with a linear (dashed brown) or a random pulse (dash-dotted yellow), CRAB (dotted green), and \texttt{geodesiq} (solid purple), for $t_\mathrm{f}=250/J$.
    Each pulse is shown in its own strip with a common vertical scale.
    \textbf{(c)} Final-state infidelity after low-pass filtering the control pulses, plotted as a function of the filter cutoff frequency.
    The filter is a zero-phase third-order Butterworth filter with the initial and final values held before and after the pulse (App.~\ref{app: numerical benchmarking}).
    Frequencies are expressed in cycles per unit time.
    \textbf{(d)} Wall-clock time for pulse generation as a function of the number of discretization points, averaged over the successful runs; values of $N$ without a successful run are omitted.
    The common legend applies to panels \textbf{(b)--(d)}.
    In all panels, $L=4$, $h_x / J = 1$, $h_z / J = 0.8$, $(\alpha,\beta)=(2,2)$ for the \texttt{geodesiq} pulse, and the target infidelity is $6\times10^{-4}$.
    The followed state is the sixth excited state of the reflection-even sector (index 6, of dimension $d=10$), and $\lambda$ is swept from $\lambda_\mathrm{i}=0.1$ to $\lambda_\mathrm{f}=0.9$; panel \textbf{(d)} also uses $t_\mathrm{f}=250/J$.}
    \label{fig: many-body benchmarks}
\end{figure*}

The pulses in Fig.~\ref{fig: many-body benchmarks}(b) exhibit substantial differences despite addressing the same transfer task.
GRAPE initialized with a linear ramp produces a predominantly increasing pulse with local modulations, whereas random initialization leads to pronounced fluctuations.
The CRAB solution also contains oscillatory features.
In contrast, \texttt{geodesiq} yields a smooth, monotonic protocol with a non-uniform sweep rate.

These differences are reflected in the filtered-pulse infidelities shown in Fig.~\ref{fig: many-body benchmarks}(c).
The geometric protocol reaches infidelities below $10^{-3}$ at lower cutoff frequencies than the other protocols in this benchmark.
Its error is non-monotonic, exhibiting a minimum before approaching a plateau of approximately
$6\times10^{-4}$.
CRAB and the randomly initialized GRAPE approach a comparable plateau, while GRAPE initialized with a linear ramp attains a somewhat smaller error at the largest cutoffs shown.
However, GRAPE initialized randomly is particularly sensitive to filtering, requiring a substantially larger bandwidth to recover
high-fidelity transfer.
These results illustrate that the performance after filtering depends on both the optimization method and its initialization.

Figure~\ref{fig: many-body benchmarks}(d) shows that \texttt{geodesiq} has the lowest pulse-generation runtime at every sampled $N$.
The curves start at different numbers of steps because, for fewer steps, the target infidelity is not reached (in any run, for GRAPE and CRAB): GRAPE requires $N\geq 33$ for both initializations, \texttt{geodesiq} $N\geq 65$, and CRAB $N\geq 129$.
CRAB also misses the target in some runs at larger $N$.
At the largest discretization shown, the \texttt{geodesiq} runtime is approximately $0.04$~s, compared with $0.76$--$1.2$~s for GRAPE and approximately $340$~s for CRAB.
Thus, for this example, the geometric construction provides high-fidelity transfer with substantially lower computational cost, although it does not yield the smallest infidelity at large bandwidth.

\section{Comparison of the metrics}
\label{app: metric comparison}
In this appendix, we compare the pulses obtained with different exponents $(\alpha,\beta)$ and transition norms $p$ (Sec.~\ref{sec: transition norm}) for systems with different spectra.
We consider the linear ramp $(0,0)$, the geometric protocol $(2,2)$, and FAQUAD $(4,2)$, the latter with the three orders $p=1,2,\infty$ of Eq.~\eqref{eqn: transition norm}; unless stated otherwise, $p=2$.
Since $\sqrt{G}\,|dx/d\tau|=\mathcal{L}^{(\alpha,\beta)}$ [Eq.~\eqref{eq: hypergeo protocol}], the time that the pulse spends per unit of the control is proportional to $\sqrt{G(x)}$.
We quantify it by the local slowdown
\begin{align}
    \label{eqn: slowdown}
    s(x)=|x_\mathrm{f}-x_\mathrm{i}|\left|\dv{\tau}{x}\right|=\frac{|x_\mathrm{f}-x_\mathrm{i}|\sqrt{G(x)}}{\mathcal{L}^{(\alpha,\beta)}},
\end{align}
which equals one for the linear ramp and averages to one over the sweep for every metric.
The infidelity $1-\mathcal{F}$ is computed with \code{Dynamics.state_fidelity()}, and $t_\mathrm{f}^*$ denotes the shortest pulse time for which it does not exceed a given threshold.

\subsection{Systems with different spectra}
We use four systems, which are shown in the left column of Fig.~\ref{fig: metric profiles}.
\begin{enumerate}
    \item The Landau--Zener problem, Eq.~\eqref{eqn: Landau Zener model}, swept from $z_\mathrm{i}=-10$ to $z_\mathrm{f}=10$, with a single anticrossing of gap $2\Delta$.
    \item A level swept through a band of eight equally spaced levels,
    \begin{align}
        \label{eqn: level through band}
        \hat{H}(x)=x\ketbra{0}{0}+\sum_{j=1}^{8}\epsilon_j\ketbra{j}{j}+g\sum_{j=1}^{8}\big(\ketbra{0}{j}+\ketbra{j}{0}\big),
    \end{align}
    with $\epsilon_j$ equally spaced in $[-2,2]$ and $x$ swept from $x_\mathrm{i}=-6$ to $x_\mathrm{f}=6$.
    We follow the fifth level from the bottom (index $4$), which starts and ends inside the band.
    While the level $\ket{0}$ passes through it, the occupied state approaches its lower and upper neighbors at once, and the coupling $g$ sets the minimum gap.
    \item The adiabatic initialization of the DQD, Eq.~\eqref{eqn: 5x5 Hamiltonian}, with $U=10\,t_\mathrm{c}$, $E_Z=0.9\,t_\mathrm{c}$, and $\Delta\!E_Z=0.1\,t_\mathrm{c}$ as in Fig.~\ref{fig: DQD energies}, swept from $\varepsilon_\mathrm{i}=15\,t_\mathrm{c}$ to $\varepsilon_\mathrm{f}=0$.
    The occupied state first changes its charge character at the anticrossing set by $t_\mathrm{c}$ and then passes the narrow anticrossing near $\varepsilon\approx8.7\,t_\mathrm{c}$, whose gap is set by $\Delta\!E_X$, which we vary.
    \item The Ising chain, Eq.~\eqref{eqn: ising hamiltonian}, with $L=6$ and $h_x/J=1$ in the reflection-even sector ($d=36$), swept from $\lambda_\mathrm{i}=0$ to $\lambda_\mathrm{f}=1$.
    For small $h_z$, the gap at the transition hardly depends on $h_z$ ($\approx0.69\,J$ for $L=6$), whereas the gap at the end of the sweep, $2(J-h_z)$, closes as $h_z\to J$.
    We therefore vary $h_z\in[0.2,0.95]\,J$, which moves the minimum gap to the end of the sweep.
\end{enumerate}
In all four cases, the transfer is adiabatic, and the target is the followed eigenstate at $x_\mathrm{f}$.

\subsection{Pulse profiles}
\begin{figure*}[!ht]
    \centering
    \includegraphics[width=\linewidth]{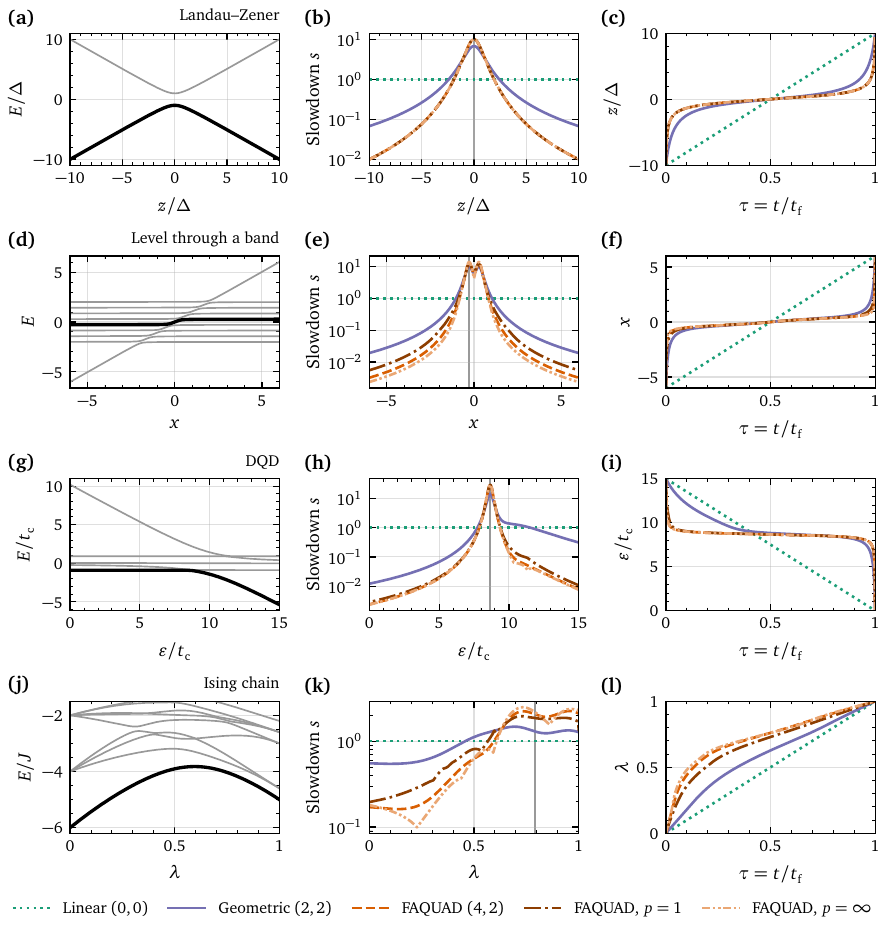}
    \caption{Pulse profiles of the compared metrics for four systems (rows).
    \textbf{Left column:} instantaneous eigenenergies, with the followed level in black and the other levels in gray; for the Ising chain, only the lowest levels are shown.
    \textbf{Middle column:} local slowdown $s(x)$, Eq.~\eqref{eqn: slowdown}; the gray vertical line marks the minimum gap of the followed level.
    \textbf{Right column:} pulses $x(\tau)$.
    \textbf{(a)--(c)} Landau--Zener problem, Eq.~\eqref{eqn: Landau Zener model}, with $\Delta=1$.
    \textbf{(d)--(f)} Level through a band, Eq.~\eqref{eqn: level through band}, with $g=0.12$ (minimum gap $0.23$).
    \textbf{(g)--(i)} DQD, Eq.~\eqref{eqn: 5x5 Hamiltonian}, with $\Delta\!E_X=0.03\,t_\mathrm{c}$ (minimum gap $0.072\,t_\mathrm{c}$) and the other parameters of Fig.~\ref{fig: DQD energies}.
    \textbf{(j)--(l)} Ising chain, Eq.~\eqref{eqn: ising hamiltonian}, with $L=6$, $h_x/J=1$, and $h_z/J=0.8$ (minimum gap $0.35\,J$).
    The orders $p=1$ and $p=\infty$ of FAQUAD are drawn in a darker and a lighter shade of its color; for a single transition (top row), the three FAQUAD curves coincide.}
    \label{fig: metric profiles}
\end{figure*}

Figure~\ref{fig: metric profiles} shows the slowdown and the pulses for one gap of each system.
For the Landau--Zener problem, only one transition contributes, so that the three orders coincide.
Its metric is $G\propto(\Delta^2+z^2)^{-(\alpha+\beta)/2}$ (App.~\ref{app: formal details}), so that $s$ decays as $|z|^{-2}$ for the geometric protocol and as $|z|^{-3}$ for FAQUAD.
FAQUAD therefore concentrates the pulse time more strongly at the anticrossing and passes the far-detuned regions faster.
For the level in the band, the slowdown has two maxima, one for each neighbor that the occupied state approaches.
Away from them, the occupied state couples weakly to several levels of the band, and the orders separate: $p=1$ adds these weights and slows down more than $p=2$, whereas $p=\infty$ follows only the largest one.
For the DQD, FAQUAD spends almost the whole pulse time at the narrow anticrossing, while the geometric protocol keeps a shoulder for $\varepsilon\gtrsim10\,t_\mathrm{c}$, where the occupied state changes its charge character at the larger gap.
The additional factor $1/|E_n-E_k|$ of FAQUAD makes it pass this region about an order of magnitude faster.
For the Ising chain, many transitions contribute along the whole sweep, so that the metrics differ everywhere, and $p=\infty$ shows the kinks discussed in Sec.~\ref{sec: transition norm} where the dominant transition changes.

\subsection{Dependence on the gap}
\begin{figure*}[!ht]
    \centering
    \includegraphics[width=\linewidth]{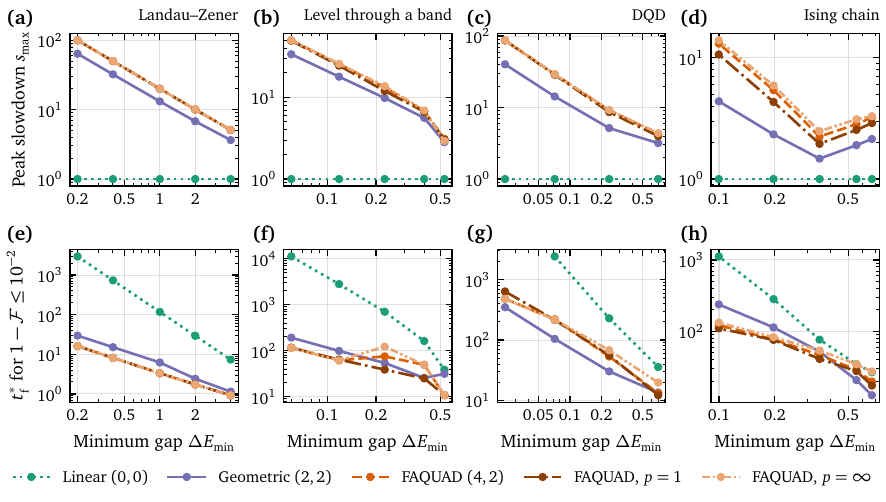}
    \caption{Dependence of the compared metrics on the minimum gap $\Delta E_\mathrm{min}$ of the followed level, for the four systems of Fig.~\ref{fig: metric profiles} (columns).
    \textbf{(a)--(d)} Largest local slowdown $s_\mathrm{max}=\max_x s(x)$.
    \textbf{(e)--(h)} Shortest pulse time $t_\mathrm{f}^*$ for $1-\mathcal{F}\leq10^{-2}$; the linear ramp of the DQD does not reach the threshold within $t_\mathrm{f}\,t_\mathrm{c}\leq2\times10^4$ at the smallest gap.
    The gap is varied through $\Delta\in[0.1,2]$ (Landau--Zener), $g\in[0.03,0.5]$ (level through a band), $\Delta\!E_X\in[0.01,0.3]\,t_\mathrm{c}$ (DQD), and $h_z\in[0.2,0.95]\,J$ (Ising chain), and is given in the energy unit of each system ($\Delta E_\mathrm{min}=2\Delta$ for Landau--Zener).}
    \label{fig: metric gap scan}
\end{figure*}

Figure~\ref{fig: metric gap scan} shows, as a function of the minimum gap, the largest slowdown and the shortest pulse time for $1-\mathcal{F}\leq10^{-2}$.
For $t_\mathrm{f}^*$, the pulse time is increased by factors of $1.25$ until the threshold is reached and then refined by bisection.
For the Landau--Zener problem, $t_\mathrm{f}^*$ scales as $\Delta E_\mathrm{min}^{-2}$ for the linear ramp, as expected from the Landau--Zener formula, but only as $\Delta E_\mathrm{min}^{-1}$ for the geometric protocol and FAQUAD, and FAQUAD needs about half the time of the geometric protocol.
For the level in the band, the order $p$ matters at intermediate couplings: for $g=0.12$, FAQUAD reaches the threshold at $t_\mathrm{f}^*=39$ with $p=1$, $75$ with $p=2$, and $123$ with $p=\infty$, against $54$ for the geometric protocol, whereas the orders nearly coincide for the weakest and the strongest coupling.
For the DQD, the geometric protocol is the fastest at small gaps ($t_\mathrm{f}^*\,t_\mathrm{c}=105$ against $216$--$220$ for FAQUAD at $\Delta\!E_X=0.03\,t_\mathrm{c}$), because FAQUAD passes the charge anticrossing too fast.
The linear ramp needs $2.4\times10^3/t_\mathrm{c}$ there and does not reach the threshold within $2\times10^4/t_\mathrm{c}$ for $\Delta\!E_X=0.01\,t_\mathrm{c}$.
For the Ising chain, the geometric protocol is the fastest while the gap at the transition is the smallest one ($t_\mathrm{f}^*J=12.7$ against $17$--$27$ for FAQUAD at $h_z=0.2\,J$).
Once the gap closes at the end of the sweep, FAQUAD becomes faster ($110$--$133$ against $238$ at $h_z=0.95\,J$, with $1.1\times10^3$ for the linear ramp).
No metric is therefore the fastest for all spectra: FAQUAD wins when a single anticrossing dominates, while the geometric protocol wins when the occupied state also changes at larger gaps.
These durations refer to $1-\mathcal{F}\leq10^{-2}$, where the loss at the anticrossings dominates.
For lower thresholds, the ranking can change, because the error is then set increasingly by the non-adiabatic terms at the endpoints of the sweep~\cite{degrandiAdiabaticPerturbationTheory2010}, whose size depends on the metric through the speed $|dx/dt|$ there.

\subsection{Transition norm: a frequency-crowded qubit register}
The order $p$ changes the pulse where the number of contributing transitions changes along the sweep.
A simple system in which this number changes strongly is a register of $N_\mathrm{q}+1$ independent qubits that are inverted by a single chirped drive, as in global control of qubits with nearly equal frequencies.
In the frame rotating with the drive frequency $x$, the Hamiltonian is
\begin{align}
    \label{eqn: qubit register}
    \hat{H}(x)=\sum_{q=1}^{N_\mathrm{q}+1}\left[\frac{\omega_q-x}{2}\,\sigma^z_q+\frac{\Omega}{2}\,\sigma^x_q\right],
\end{align}
which is affine in $x$.
We take $\Omega=0.2$, $N_\mathrm{q}$ qubits with frequencies $\omega_q$ equally spaced in $[-0.05,0.05]$, and one isolated qubit with $\omega_{N_\mathrm{q}+1}=10$, and sweep $x$ from $x_\mathrm{i}=-10$ to $x_\mathrm{f}=20$.
Starting in the ground state, the target is the instantaneous ground state at $x_\mathrm{f}$, i.e., every qubit is inverted.
Since the qubits are independent, $\mathcal{F}=\prod_q\mathcal{F}_q$, and we compute each $\mathcal{F}_q$ with the same pulse.

\begin{figure*}[!ht]
    \centering
    \includegraphics[width=\linewidth]{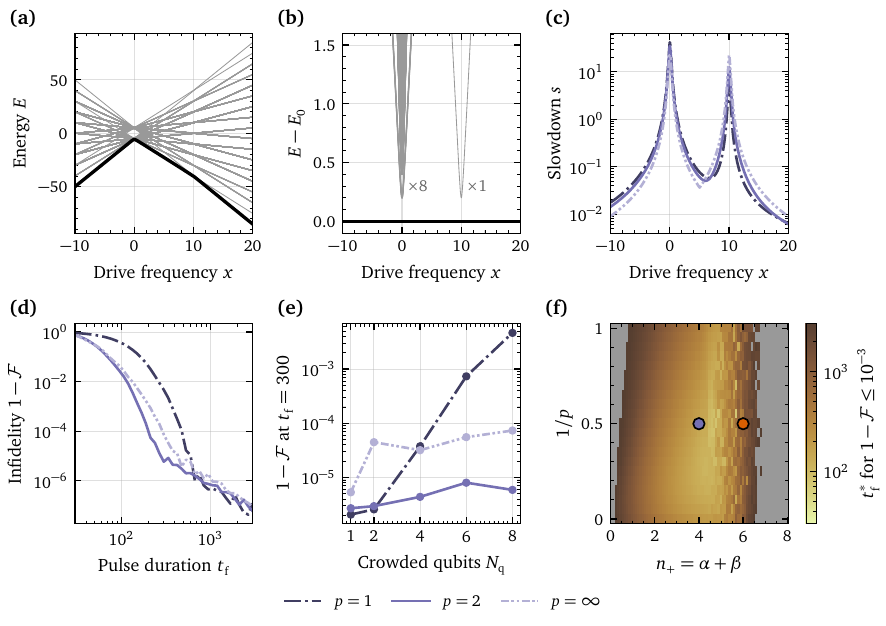}
    \caption{Effect of the transition norm for the qubit register of Eq.~\eqref{eqn: qubit register}, with $N_\mathrm{q}$ crowded qubits and one isolated qubit.
    \textbf{(a)} All $2^{N_\mathrm{q}+1}=512$ levels (gray) as a function of the drive frequency $x$, for $N_\mathrm{q}=8$, with the followed ground state in black.
    \textbf{(b)} Excitation energies $E-E_0$ above the followed state: the lowest are the gaps $\Delta E_q$ to the states with one flipped qubit, eight of which close together at $x\approx0$ ($\times8$) and one at $x=10$ ($\times1$).
    \textbf{(c)} Local slowdown, Eq.~\eqref{eqn: slowdown}, for $(\alpha,\beta)=(2,2)$ and $p=1,2,\infty$ (darker, solid, and lighter).
    \textbf{(d)} Infidelity as a function of the pulse time for the same pulses.
    \textbf{(e)} Infidelity at $t_\mathrm{f}=300$ as a function of the number $N_\mathrm{q}$ of crowded qubits.
    \textbf{(f)} Shortest pulse time $t_\mathrm{f}^*$ for $1-\mathcal{F}\leq10^{-3}$ as a function of $n_+=\alpha+\beta$ (with $\alpha=\beta$) and $1/p$, for $N_\mathrm{q}=8$, from 41 pulse times between $30$ and $3000$; gray cells do not reach the threshold.
    The purple and orange dots mark the geometric protocol and FAQUAD with $p=2$.
    Energies and times are in units of an arbitrary energy scale ($\hbar=1$), with $\Omega=0.2$.}
    \label{fig: metric crowded qubits}
\end{figure*}

The ground state couples through $\partial_x\hat{H}=-\frac{1}{2}\sum_q\sigma^z_q$ only to the $N_\mathrm{q}+1$ states with a single flipped qubit, with the gaps $\Delta E_q=[(\omega_q-x)^2+\Omega^2]^{1/2}$ [Fig.~\ref{fig: metric crowded qubits}(a,b)] and the matrix elements $\Omega/(2\Delta E_q)$.
The weight of each transition is therefore $(\Omega/2)^\beta\,\Delta E_q^{-(\alpha+\beta)}$, and for $\beta>0$ the pulse depends on the exponents only through $n_+=\alpha+\beta$.
Near $x=0$, the $N_\mathrm{q}$ crowded qubits contribute with nearly equal weights $r$, so that Eq.~\eqref{eqn: transition norm} gives $G=N_\mathrm{q}^{2/p}\,r$.
Relative to the isolated qubit, the order $p$ thus weights the cluster by $N_\mathrm{q}^{1/p}$, and the isolated qubit receives approximately the fraction $1/(1+N_\mathrm{q}^{1/p})$ of the pulse time [Fig.~\ref{fig: metric crowded qubits}(c)].
The infidelity is approximately the sum of the leakage probabilities of the qubits, $1-\mathcal{F}\approx\sum_q(1-\mathcal{F}_q)$, which is the combination that $p=2$ weights.
For $p=1$, the isolated qubit receives too little time, and its loss grows exponentially as its share shrinks; for $p=\infty$, every anticrossing receives the same time, and the cluster loses about $N_\mathrm{q}$ times more than a single qubit.

For $N_\mathrm{q}=8$ and $(\alpha,\beta)=(2,2)$, Fig.~\ref{fig: metric crowded qubits}(d), the infidelity at $t_\mathrm{f}=300$ is $4.7\times10^{-3}$ for $p=1$, $5.9\times10^{-6}$ for $p=2$, and $7.4\times10^{-5}$ for $p=\infty$.
At long pulse times, the curves merge, because the endpoints of the sweep then dominate the error.
The loss for $p=1$ grows exponentially with the number of crowded qubits [Fig.~\ref{fig: metric crowded qubits}(e)], from about the value for $p=2$ at $N_\mathrm{q}\leq2$ to $800$ times larger at $N_\mathrm{q}=8$.
Figure~\ref{fig: metric crowded qubits}(f) shows that this asymmetry holds for all $n_+$ at which the anticrossings limit the fidelity.
At $n_+=4$, $t_\mathrm{f}^*$ for $1-\mathcal{F}\leq10^{-3}$ is $369$ for $p=1$, $157$ for $p=2$, $124$ for $p\approx3.3$, and $207$ for $p=\infty$: lowering $p$ below $2$ is costly, raising it to $3$--$5$ is harmless or slightly favorable, and only $p=\infty$ costs again.
For $n_+\gtrsim5$, the endpoints of the sweep limit the fidelity, so that $p$ matters little, and $t_\mathrm{f}^*$ varies between neighboring exponents because the contributions of the two endpoints interfere.
Together with the examples above, this supports $p=2$ as the default: it adds the losses of independent transitions as probabilities and was never far from the best order in these examples.

\section{API documentation}
\label{app: api}
\begingroup
\setlength{\emergencystretch}{1em}
This appendix documents the public interface of \texttt{geodesiq} version 0.5.0 (DOI \href{https://doi.org/10.5281/zenodo.24321210}{\nolinkurl{10.5281/zenodo.24321210}}).
Signatures and behavior follow the implementation.
Private helpers and caches
are excluded.
The notation \code{np} denotes \texttt{NumPy}, \code{plt} denotes \code{matplotlib.pyplot}, and \code{qt} denotes \texttt{QuTiP}; typing names in signatures have their usual Python meanings.
\code{ArrayLike} is \code{numpy.typing.ArrayLike}, i.e., any input that \code{np.asarray()} accepts, such as a nested list.
\code{NDArray} denotes \code{numpy.typing.NDArray}; \code{Figure} and \code{SubFigure} are from \code{matplotlib.figure}, \code{Axes} is from \code{matplotlib.axes}, and \code{Path} is \code{pathlib.Path}.
Types are written as in the source annotations: \code{NDArray[np.float64]} for real arrays and \code{np.ndarray} for matrices.
Each parameter entry gives the type, a description with the constraints, and, if the parameter is optional, ends with its default value.
\par\endgroup

\subsection{Package interface and conventions}
\begin{apisignature}
from geodesiq import ControlModel, PulseControl, Dynamics
\end{apisignature}

\code{ControlModel} constructs a pulse for a Hamiltonian depending on one
real control parameter.
\code{PulseControl} stores samples on a physical-time
grid and provides processing, plotting, and export methods.
\code{Dynamics} evaluates the resulting quantum evolution and fidelities.

\apiheading{Additional package exports}
The following names are also available directly from \code{geodesiq}.
\apiitem{decompose_hamiltonian}{\code{Callable}}{Sample a time-dependent Hermitian Hamiltonian and approximate it as a constant drift plus a sum of constant Hermitian control operators with real sampled coefficients, using a singular-value decomposition.
The required arguments are \code{H_func(t)} and sample \code{times}; optional arguments select the drift, retained rank, singular-value cutoffs, and Hermiticity tolerance.
The result is a \code{HamiltonianDecomposition}.}

\apiitem{HamiltonianDecomposition}{\code{dataclass}}{Stores \code{times}, \code{H_d}, \code{H_controls}, \code{coefficients}, \code{singular_values}, \code{rank}, \code{relative_residual_error}, and \code{relative_total_error}.
The coefficient array has shape \code{(rank, len(times))}.
The method \code{qobjevo(order=3)} builds a \texttt{QuTiP} \code{QobjEvo}; \code{reconstruct_sample(index)} returns the reconstructed Hamiltonian at one sample, using the retained terms.}

\apiitem{about()}{\code{None}}{Print the package and dependency versions and runtime environment information.}
\apiitem{__version__}{\code{str}}{Installed package version.}

\apiheading{Errors and warnings}
All package error and warning classes below are exported from \code{geodesiq}; messages carry a \code{[geodesiq]} prefix.
\apiitem{GeodesiQError}{\code{Exception}}{Base class for package errors.}
\apiitem{ValidationError}{\code{GeodesiQError}, \code{ValueError}}{Invalid input values or shapes.
Its subclasses \code{InvalidControlParameterError} and \code{MissingArgsError} identify invalid control settings and missing required arguments.}
\apiitem{ConfigurationError}{\code{GeodesiQError}}{Incomplete or inconsistent configuration.
Its subclasses are \code{MissingControlParameterError} and \code{ImmutableConfigurationError}.}
\apiitem{ComputationError}{\code{GeodesiQError}}{Numerical computation failure.
Its subclasses are \code{SolverError} and \code{MetricComputationError}.}
\apiitem{IOErrorGeodesiQ}{\code{GeodesiQError}}{Input/output, serialization, or export failure.}
\apiitem{GeodesiQWarning}{\code{Warning}}{Base class for package warnings.
Its subclasses \code{NumericalStabilityWarning}, \code{PerformanceWarning}, and \code{ExperimentalFeatureWarning} indicate numerical stability concerns, expensive configurations, and experimental features, respectively.}

\apiheading{Grids and units}
Let $N=\text{\code{num_steps}}$, $M=\text{\code{pulse_accuracy}}$, and let $d$ be
the Hilbert-space dimension.
In this appendix, $x$ is the control, $H$ (without a hat) is the Hamiltonian matrix that the model evaluates, and $\langle k|\partial_x H|n\rangle$ is short for $\mel{\psi_k(x)}{\partial_x\hat{H}}{\psi_n(x)}$, with $k$ the occupied state.
State indices are zero-based and refer to the ascending instantaneous energy spectrum, for \code{initial_state} at \code{pulse_initial} and for \code{final_state} at \code{pulse_final}.
Along the grid, the occupied state is followed by the overlap of the eigenvectors at neighboring points.
At a true level crossing, where $\partial_x H$ does not couple the two states (matrix element at most $10^{-8}$ times the largest one), it keeps its identity and changes its index, also when the crossing falls on a grid point.
An eigenvector swap between coupled states is an anticrossing that the grid does not resolve, and keeps the energy order.

The control grid \code{control_pulse} holds $N$ uniformly spaced values from \code{pulse_initial} to \code{pulse_final}, on which the eigensystem and the metric are computed; it is not the optimized waveform.
The optimized control values \code{control_sol} are given at the $M$ points of the uniform normalized-time grid \code{tau}, $\tau=t/t_\mathrm{f}\in[0,1]$, also for a custom solver.
A pulse synthesized for a duration $t_\mathrm{f}$ has the uniform physical-time grid \code{times}, $t\in[0,t_\mathrm{f}]$, and pulse processing methods return physical times, not normalized times.
Types and access behavior are given in the \code{ControlModel} and \code{PulseControl} entries.

Use mutually consistent energy, time, and reduced-Planck-constant units.
\code{Dynamics} evolves with $H/\hbar$; its default is $\hbar=1$.
Spectral and filter frequencies are in cycles per unit time, not angular frequency.

\apiheading{Minimal workflow}
\begin{minted}[bgcolor=LightGray, fontsize=\footnotesize]{python}
import numpy as np
from geodesiq import ControlModel, Dynamics

def H(eps, delta):
    return np.array([[eps / 2, delta], [delta, -eps / 2]])

def dH(eps, delta):
    return np.array([[0.5, 0.0], [0.0, -0.5]])

model = ControlModel(H, partial_H_func=dH)
model.set_parameters(delta=1.0)
model.set_control(control_name="eps", pulse_initial=-5.0, pulse_final=5.0, initial_state=0, alpha=4.0, beta=2.0, num_steps=1025)
model.solve_problem(pulse_accuracy=1000)
pulse = model.synthesize_pulse(duration=10.0)
dynamics = Dynamics(duration=pulse.duration, model=model)
fidelity = dynamics.state_fidelity()
\end{minted}

The exponents in this example are an explicit choice, not package defaults.
Generating a metric-based pulse does not guarantee a prescribed dynamical
fidelity; evaluate the dynamics at the intended duration.

\subsection{\code{ControlModel}}
\begin{apisignature}
class ControlModel(H_func: Callable[..., ArrayLike] | None=None,
                   partial_H_func: Callable[..., ArrayLike] | None=None,
                   *, H_d: ArrayLike | None=None, H_c: ArrayLike | None=None)
\end{apisignature}

Represents a single-control Hamiltonian, constructs the adiabatic metric or the di--ad tensor of Eq.~\eqref{eqn: di-ad tensor}, and solves for a normalized control trajectory.
Supply either a Hamiltonian callable, optionally with its derivative, or two constant matrices defining $H(x)=H_{\mathrm{d}}+xH_{\mathrm{c}}$.
Configuration changes invalidate dependent cached calculations.
Properties may perform numerical work when accessed.
Private diagnostic arguments are omitted from this signature.

\apiheading{Parameters}
\apiitem{H_func}{\code{Callable[..., ArrayLike] | None}}{Hamiltonian callable.
Required in callable mode; must be omitted in affine mode.
It must accept the control variable and fixed parameters as keyword arguments and return a finite Hermitian square array of shape \code{(d, d)}.
Its dimension must remain fixed after the first validated evaluation.
Default: \code{None}.}

\apiitem{partial_H_func}{\code{Callable[..., ArrayLike] | None}}{Analytic derivative with respect to the control variable, accepting the same keyword arguments and returning the same matrix shape as \code{H_func}.
If omitted in callable mode, numerical differentiation is used.
Must be omitted in affine mode, where the derivative is exactly \code{H_c}.
Default: \code{None}.}

\apiitem{H_d}{\code{ArrayLike | None}}{Constant drift matrix for an affine Hamiltonian.
Must be supplied together with \code{H_c}, without either callable argument.
The input is copied and must be numeric, finite, Hermitian, and square.
Keyword-only.
Default: \code{None}.}

\apiitem{H_c}{\code{ArrayLike | None}}{Constant matrix multiplying the scalar control parameter.
Must satisfy the same requirements as \code{H_d} and have the same shape.
It also supplies the exact Hamiltonian derivative.
Keyword-only.
Default: \code{None}.}

\apiheading{Attributes}
\apiitem{H_func}{\code{Callable[..., ArrayLike] | None}}{Hamiltonian function in callable mode; \code{None} in affine mode.
A configured function cannot be replaced.
Assignment in affine mode raises \code{ImmutableConfigurationError}.}

\apiitem{partial_H_func}{\code{Callable[..., ArrayLike] | None}}{Derivative function in callable mode.
May be assigned once if initially omitted; assignment invalidates the eigensystem and dependent results.
A configured callable cannot be replaced.
In affine mode this property is \code{None}, although the exact derivative is available internally as \code{H_c}; assignment raises \code{ImmutableConfigurationError}.}

\apiitem{H_d}{\code{np.ndarray | None}}{Read-only property returning a copy of the drift matrix in affine mode, or \code{None} in callable mode.}

\apiitem{H_c}{\code{np.ndarray | None}}{Read-only property returning a copy of the control matrix in affine mode, or \code{None} in callable mode.}

\apiitem{affine_hamiltonian}{\code{bool}}{Read-only flag indicating whether the model was constructed from \code{H_d} and \code{H_c}.}

\apiitem{parameters}{\code{Mapping[str, Any]}}{Read-only mapping of a deep copy of fixed callable-Hamiltonian parameters.
Initially empty.
Updated through \code{set_parameters()} in callable mode; remains empty in affine mode.}

\apiitem{hamiltonian_dimension}{\code{int | None}}{Read-only matrix dimension $d$.
Determined during construction in affine mode; initially \code{None} in callable mode until a validated Hamiltonian evaluation determines it.}

\apiitem{control_name}{\code{str | None}}{Writable control name.
Must be a non-empty string distinct from stored fixed-parameter names.
In callable mode it identifies the Hamiltonian keyword; in affine mode it labels the scalar coefficient multiplying \code{H_c}.
Required for pulse construction in callable mode.
Initially \code{None}.}

\apiitem{pulse_initial}{\code{float | None}}{Writable finite initial control value.
Must differ from the final value.
Initially \code{None}.}

\apiitem{pulse_final}{\code{float | None}}{Writable finite final control value.
Both increasing and decreasing sweeps are supported.
Initially \code{None}.}

\apiitem{initial_state}{\code{int | None}}{Writable non-negative initial energy-level index.
A full solve checks that it is smaller than $d$.
As long as \code{final_state} has never been set explicitly, it follows \code{initial_state}, also when \code{initial_state} changes later.
Initially \code{None}.}

\apiitem{final_state}{\code{int | None}}{Writable non-negative final energy-level index.
Until it is set explicitly, a solve resolves it to the index in which the state followed from \code{initial_state} ends, which differs from \code{initial_state} only after a true level crossing.
Once set explicitly, it keeps its value when \code{initial_state} changes.
A value different from that index selects the di--ad tensor, in which the anticrossings that are passed diabatically are determined from the spectrum along the sweep (see the box ``\code{solve_problem()}: metric and solver conventions'').
Initially \code{None}.}

\apiitem{alpha}{\code{float | None}}{Writable finite real exponent $\alpha$ of the energy gap $|E_n-E_k|$.
Required for a full solve; there is no automatic default.
Initially \code{None}.}

\apiitem{beta}{\code{float | None}}{Writable finite real exponent $\beta$ of the matrix element $|\langle k|\partial_x H|n\rangle|$.
Required for a full solve; there is no automatic default.
Initially \code{None}.}

\apiitem{dia_alpha}{\code{float | None}}{Writable finite real diabatic exponent $\tilde\alpha$ of the energy gap $|E_n-E_k|$, used for the transitions with $\xi_{kn}<1$ in the di--ad tensor.
Required for a diabatic transfer (see \code{final_state}).
Initially \code{None}.}

\apiitem{dia_beta}{\code{float | None}}{Writable finite real diabatic exponent $\tilde\beta$ of the matrix element $|\langle k|\partial_x H|n\rangle|$, used for the transitions with $\xi_{kn}<1$ in the di--ad tensor.
Required for a diabatic transfer (see \code{final_state}).
Initially \code{None}.}

\apiitem{norm}{\code{float}}{Writable order $p\geq1$, or \code{np.inf}, of the norm that combines the transitions out of the occupied state into the metric, Eq.~\eqref{eqn: transition norm}.
Initially \code{2.0}, the sum of the hypermetric tensor.}

\apiitem{allow_endpoint_passages}{\code{bool}}{Writable; whether a gap minimum at an end of the control grid also counts as an anticrossing to pass diabatically (see \code{set_control()}).
Initially \code{False}.}

\apiitem{num_steps}{\code{int | None}}{Writable number $N$ of control-grid points, at least 3.
The default Romberg integrator requires $N=2^r+1$.
Initially \code{None}; the first \code{set_control()} call sets it to \code{1025} if omitted.}

\apiitem{control_pulse}{\code{NDArray[np.float64]}, shape \code{(N,)}}{Read-only property returning a copy of the uniform control grid.
Access computes the eigensystem if needed and therefore requires its configuration.}

\apiitem{eigenenergies}{\code{NDArray[np.float64]}, shape \code{(N, d)}}{Read-only property returning real (\code{float64}) eigenenergies, ordered ascending within each grid point, also for complex Hermitian Hamiltonians.
Access computes the eigensystem if needed.
The trace part $\operatorname{tr}(H)/d$, which is removed internally before diagonalization, is added back.}

\apiitem{control_sol}{\code{NDArray[np.float64]}, shape \code{(M,)}}{Read-only property returning a copy of the optimized control samples.
Access calls \code{solve_problem()} if the ODE result is invalid or unavailable, reusing the \code{pulse_accuracy} of the previous solve (\code{1000} if there was none).}

\apiitem{tau}{\code{NDArray[np.float64]}, shape \code{(M,)}}{Read-only property returning a copy of the normalized time grid $\tau=t/t_\mathrm{f}\in[0,1]$ corresponding to \code{control_sol}.
Access calls \code{solve_problem()} if the ODE result is invalid or unavailable, reusing the previous \code{pulse_accuracy} in the same way.}

The constructor raises \code{ValidationError} for a missing or non-callable Hamiltonian in the callable mode, a non-callable derivative, incomplete or invalid affine matrices, or simultaneous use of the callable and affine interfaces.
For the writable control settings, assigning \code{None} preserves the current value; it does not clear the setting.
Use \code{set_control()} for coordinated updates, particularly when changing both endpoints.
Computed arrays and affine matrix properties are returned as copies.

\apiheading{Methods}
\apiitem{__call__(*args, **kwargs)}{\code{np.ndarray}}{Evaluate the Hamiltonian with explicit call arguments.}

\apiitem{evaluate_hamiltonian(control_value)}{\code{np.ndarray}}{Evaluate the configured Hamiltonian at one control value.}

\apiitem{set_parameters(**parameters)}{\code{None}}{Merge fixed Hamiltonian parameters in callable mode.
Unavailable in affine mode.}

\apiitem{set_control(...)}{\code{None}}{Validate and update the control configuration atomically.}

\apiitem{solve_problem(...)}{\code{None}}{Compute the eigensystem, metric, and normalized pulse.}

\apiitem{synthesize_pulse(duration)}{\code{PulseControl}}{Create a pulse on a physical-time grid.}

\apiitem{plot_eigenvalues(...)}{\code{tuple[Figure | SubFigure, Axes]}}{Plot energy branches versus the control parameter.}

\apiitem{plot_metric_tensor(...)}{\code{tuple[Figure | SubFigure, Axes]}}{Plot the metric versus the control parameter.}

\apiitem{print_summary()}{\code{None}}{Print configuration and computation status.}

\apiheading{Detailed methods}
\begin{methodbox}{\code{__call__()}}
\begin{minted}{python}
ControlModel.__call__(*args: Any, **kwargs: Any) -> np.ndarray
\end{minted}

Evaluate the Hamiltonian directly.
In callable mode, arguments are forwarded to \code{H_func}, with runtime keyword arguments overriding stored fixed parameters for this evaluation only.
In affine mode, supply exactly one positional control value, or the configured \code{control_name} as the only keyword argument.

\apiheading{Parameters}
\apiitem{args}{\code{Any}}{Arguments forwarded to the Hamiltonian callable, or one scalar control value in affine mode.}

\apiitem{kwargs}{\code{Any}}{Keywords merged over stored parameters in callable mode.
In affine mode, only the configured control keyword is accepted, without positional arguments.}

\apiheading{Returns}
\apiitem{hamiltonian}{\code{np.ndarray}, shape \code{(d, d)}}{Hamiltonian matrix at the supplied arguments.}

\apiheading{Notes}
A direct callable evaluation does not require \code{control_name} to be configured.
An affine evaluation with one positional value also does not require it; keyword evaluation does.
Unsupported affine calling patterns raise \code{InvalidControlParameterError}.

\end{methodbox}

\begin{methodbox}{\code{evaluate_hamiltonian()}}
\begin{minted}{python}
ControlModel.evaluate_hamiltonian(control_value: float) -> np.ndarray
\end{minted}

Evaluate the Hamiltonian at one scalar control value.
In callable mode, the configured \code{control_name} and the stored fixed parameters are used.
In affine mode, return \code{H_d + control_value * H_c}; this evaluation does not require \code{control_name}.

\apiheading{Parameters}
\apiitem{control_value}{\code{float}}{Finite real control value.
It need not lie between the configured pulse endpoints.}

\apiheading{Returns}
\apiitem{hamiltonian}{\code{np.ndarray}, shape \code{(d, d)}}{Hamiltonian matrix at the supplied control value.}

\end{methodbox}

\begin{methodbox}{\code{set_parameters()}}
\begin{minted}{python}
ControlModel.set_parameters(**parameters: Any) -> None
\end{minted}

In callable mode, merge the supplied fixed parameters with previously stored values.
Unspecified parameters are retained.
Values are stored as deep copies, so an array changed in place takes effect only when it is passed again.
Changed values invalidate the eigensystem and dependent calculations.

\apiheading{Parameters}
\apiitem{parameters}{\code{Any}}{Keyword arguments passed to \code{H_func} and \code{partial_H_func}.
The configured control variable must not appear here.}

\apiheading{Notes}
No work is invalidated when the merged parameter values compare equal to the stored values.
In affine mode, this method raises \code{ImmutableConfigurationError}, including when called without arguments.
Construct a new model to change parameters incorporated into \code{H_d} or \code{H_c}.

\end{methodbox}

\begin{methodbox}{\code{set_control()}}
\begin{minted}{python}
ControlModel.set_control(
    control_name: str | None=None,
    pulse_initial: float | None=None,
    pulse_final: float | None=None,
    initial_state: int | None=None,
    final_state: int | None=None,
    alpha: float | None=None,
    beta: float | None=None,
    dia_alpha: float | None=None,
    dia_beta: float | None=None,
    norm: float | None=None,
    num_steps: int | None=None,
    allow_endpoint_passages: bool | None=None
) -> None
\end{minted}

Validate a candidate configuration before committing any changes.
Omitted or \code{None}-valued arguments retain the current settings, subject to the initial defaults described below.

\apiheading{Parameters}
\apiitem{control_name}{\code{str | None}}{Control keyword for a callable Hamiltonian, or a label for the scalar coefficient in an affine Hamiltonian.
A non-empty name must be configured before computing the eigensystem or pulse in callable mode.
It is optional in affine mode, where plots then label the axis \code{"control"}.
Default: \code{None}.}

\apiitem{pulse_initial}{\code{float | None}}{Finite initial control value.
Default: \code{None}.}

\apiitem{pulse_final}{\code{float | None}}{Finite final control value, distinct from \code{pulse_initial}.
Default: \code{None}.}

\apiitem{initial_state}{\code{int | None}}{Non-negative initial energy-level index.
Its upper bound is checked during a full solve.
Default: \code{None}.}

\apiitem{final_state}{\code{int | None}}{Non-negative final energy-level index.
If still unset when \code{initial_state} becomes available, it is initialized to \code{initial_state}; an existing value is retained.
A value different from the index in which the state followed adiabatically from \code{initial_state} ends (\code{initial_state} unless the state passes a true level crossing) selects a diabatic transfer and must be reachable through anticrossings towards it, which is checked during a full solve.
Default: \code{None}.}

\apiitem{alpha}{\code{float | None}}{Finite real exponent $\alpha$ of the energy gap $|E_n-E_k|$.
Must be supplied before a full solve.
Default: \code{None}.}

\apiitem{beta}{\code{float | None}}{Finite real exponent $\beta$ of the matrix element $|\langle k|\partial_x H|n\rangle|$.
Must be supplied before a full solve.
Default: \code{None}.}

\apiitem{dia_alpha}{\code{float | None}}{Finite real diabatic exponent $\tilde\alpha$ of the energy gap $|E_n-E_k|$.
Required for a diabatic transfer, which is checked during a full solve.
Default: \code{None}.}

\apiitem{dia_beta}{\code{float | None}}{Finite real diabatic exponent $\tilde\beta$ of the matrix element $|\langle k|\partial_x H|n\rangle|$.
Required for a diabatic transfer, which is checked during a full solve.
Default: \code{None}.}

\apiitem{norm}{\code{float | None}}{Order $p\geq1$, or \code{np.inf}, of the norm that combines the transitions out of the occupied state, Eq.~\eqref{eqn: transition norm}, in the adiabatic metric and in the di--ad tensor.
With $(\alpha,\beta)=(4,2)$, $p=1$ gives the multilevel FAQUAD condition of Eq.~\eqref{eqn: faquad}, $p=2$ its $\ell^2$ form, and $p=\infty$ the generalized FAQUAD of Ref.~\cite{garcia-ripollQuantumControlFrequencyTunable2020}; all orders agree for a single contributing transition.
Default: \code{None}, which retains the current value, initially \code{2.0}.}

\apiitem{num_steps}{\code{int | None}}{Number $N$ of control-grid points, at least 3.
Default: \code{None}, which retains an existing value, or selects \code{1025} on first configuration.}

\apiitem{allow_endpoint_passages}{\code{bool | None}}{Whether a gap minimum at \code{pulse_initial} or \code{pulse_final} also counts as an anticrossing to pass diabatically, for a deliberately truncated passage.
A passage at the start relabels the initial state before any evolution, so its transitions never enter the metric.
With \code{False}, a diabatic transfer needs the sweep to bracket each anticrossing on the way to \code{final_state}.
Default: \code{None}, which retains the current value, initially \code{False}.}

\apiheading{Notes}
Finite negative or zero metric exponents are accepted; the resulting metric must still satisfy the numerical checks at solve time.
Calling \code{set_control()} without arguments only establishes \code{num_steps} if it was unset.

\end{methodbox}

\begin{methodbox}{\code{solve_problem()}}
\begin{minted}{python}
ControlModel.solve_problem(
    pulse_accuracy: int | None=None,
    solver: Callable[..., Any] | None=None,
    solver_kwargs: dict[str, Any] | None=None,
    metric_integrator: Callable[..., Any] | None=None,
    metric_integrator_kwargs: dict[str, Any] | None=None
) -> None
\end{minted}

Compute the eigensystem and derivative matrix elements, construct the metric, and integrate the normalized pulse trajectory.
Reuse cached stages whose inputs have not changed.

\apiheading{Parameters}
\apiitem{pulse_accuracy}{\code{int | None}}{Number $M$ of output pulse samples, an integer of at least 3.
This is a sample count, not an ODE error tolerance.
Default: \code{None}, which reuses the value of the previous solve, or \code{1000} on the first solve; this also applies to the automatic re-solves triggered by \code{control_sol} and \code{synthesize_pulse()} after a configuration change.}

\apiitem{solver}{\code{Callable[..., Any] | None}}{ODE integrator.
Default: \code{None}, which retains the configured solver, initially \code{scipy.integrate.solve_ivp}.}

\apiitem{solver_kwargs}{\code{dict[str, Any] | None}}{Solver options; a supplied dictionary replaces the previous one.
Initially empty, in which case $\tau(x)$ is obtained analytically from the antiderivative of the PCHIP interpolant, without step-size control (see the box ``\code{solve_problem()}: metric and solver conventions'').
A non-empty dictionary makes \code{solve_ivp} integrate the inverse equation numerically, with defaults \code{method="RK45"}, \code{atol=1e-12}, and \code{rtol=1e-10} for keys that are not supplied.
Its independent variable is then the control $x$, not $\tau$: step options such as \code{max_step}, \code{first_step}, and \code{min_step} are in units of the control, and \code{atol} and \code{rtol} apply to $\tau(x)$.
The keys \code{dense_output}, \code{events}, and \code{t_eval} are reserved.
A key that the chosen method does not accept raises \code{ValidationError}.
Default: \code{None}, which retains the previous dictionary.}

\apiitem{metric_integrator}{\code{Callable[..., Any] | None}}{Integrator called as \code{integrator(sqrt_metric, dx=dx, **kwargs)}.
It must return one finite positive real scalar.
Default: \code{None}, which retains the configured integrator, initially \code{scipy.integrate.romb}.}

\apiitem{metric_integrator_kwargs}{\code{dict[str, Any] | None}}{Additional metric-integrator options; a supplied dictionary replaces the previous one.
Do not supply \code{dx}, which is managed internally.
Default: \code{None}, which retains the previous dictionary.}

\apiheading{Notes}
This method does not create a \code{PulseControl} object or assign a physical duration.
The solver, the metric integrator, their options, and \code{pulse_accuracy} are stored on the model, so a choice passed once is also used by every later call and by the automatic re-solves of \code{control_sol}, \code{tau}, and \code{synthesize_pulse()}; the defaults are restored with \code{solve_problem(solver=solve_ivp, solver_kwargs={}, metric_integrator=romb, metric_integrator_kwargs={})}.
Missing exponents, endpoints, or diabatic exponents raise \code{MissingControlParameterError}.
A transition that $\partial_x H$ does not couple (matrix element at most $10^{-8}$ times the largest one at that grid point) contributes $|E_n-E_k|^{-\alpha}$ for a zero matrix-element exponent ($0^0=1$) and nothing for a negative one, as discussed below Eq.~\eqref{eqn: generalized qgt}.
A degenerate gap (below $10^{-12}$ times the spectral bandwidth) in a transition with a positive gap exponent raises \code{MetricComputationError}, as do non-finite, negative, or identically zero metrics.
The exception is a true level crossing on a grid point: the two states are degenerate at that point only and are not coupled by $\partial_x H$, and for a nonzero matrix-element exponent their contribution is set to zero, its value on either side.
Locally zero or numerically singular metric values emit \code{NumericalStabilityWarning}; an unusable speed may subsequently raise \code{MetricComputationError}.

\end{methodbox}

\begin{methodbox}{\code{solve_problem()}: metric and solver conventions}
For equal initial and final indices, $k_\mathrm{i}=k_\mathrm{f}$, the implementation uses
\[
 G(x)=\sum_{n\ne k_\mathrm{i}}
 \frac{|\langle k_\mathrm{i}(x)|\partial_x H(x)|n(x)\rangle|^{\beta}}
 {|E_n(x)-E_{k_\mathrm{i}}(x)|^{\alpha}},
\]
so that only transitions out of the occupied state contribute.
For unequal indices, the occupied level $k(x)$ is followed along the sweep,
starting from $k_\mathrm{i}$, and again only transitions out of it contribute,
\[
 G(x)=\sum_{n\ne k(x)}
 \frac{|\langle k(x)|\partial_x H(x)|n(x)\rangle|^{\beta_{kn}(x)}}
 {|E_n(x)-E_k(x)|^{\alpha_{kn}(x)}},
\]
\[
 \alpha_{kn}=\tilde\alpha+\xi_{kn}(\alpha-\tilde\alpha),\qquad
 \beta_{kn}=\tilde\beta+\xi_{kn}(\beta-\tilde\beta),
\]
with $\tilde\alpha=$ \code{dia_alpha} and $\tilde\beta=$ \code{dia_beta}.
This is the di--ad tensor of Eq.~\eqref{eqn: di-ad tensor}, and $\xi_{kn}$ is its di--ad transition matrix.
Both sums are written for the default \code{norm=2}; for another order $p$, each sum $\sum_n r_n$ becomes $(\sum_n r_n^{p/2})^{2/p}$ over the same terms $r_n$, and $\max_n r_n$ for $p=\infty$, Eq.~\eqref{eqn: transition norm}.
The terms are scaled by the largest one at each grid point before they are raised to the power $p/2$, so that large orders and exponents do not overflow.
Every interior local minimum of the gap between $k$ and its neighbor towards
\code{final_state} is a
diabatic anticrossing: $k$ advances by one level at the first such minimum, until
\code{final_state} is reached.
A minimum counts only if the gap rises above it on both sides, before it drops
below it again, by more than $10^{-12}$ times the largest spectral bandwidth;
gap changes below this tolerance are ignored, so a flat gap also bounds a
diabatic anticrossing where it begins.
A minimum at an end of the control grid counts only with
\code{allow_endpoint_passages=True}.
All other anticrossings, including those after
\code{final_state} is reached, are adiabatic ($\xi_{kn}=1$).
Around a diabatic
anticrossing, $\xi_{kn}=0$ from the minimum to halfway (along the control grid)
to the nearest gap maximum on each side, and then rises to $1$ at that maximum
with a $C^1$ smoothstep.
The ends of the control grid are not maxima, so no
blending happens towards them.
Coincident anticrossings, such as a fan of levels,
are passed at the same grid point.
The transition list is recomputed whenever
the eigensystem or the state indices change.
If \code{final_state} cannot be
reached, for instance because the sweep does not bracket an anticrossing,
\code{InvalidControlParameterError} is raised.

The normalization is the path length of Eq.~\eqref{eqn: path length}, and the continuous pulse equation is
\[
 \mathcal{L}^{(\alpha,\beta)}=\int_{\min(x_\mathrm{i},x_\mathrm{f})}^{\max(x_\mathrm{i},x_\mathrm{f})}\sqrt{G(x)}\,dx,
 \qquad
 \frac{dx}{d\tau}=\operatorname{sgn}(x_\mathrm{f}-x_\mathrm{i})
 \frac{\mathcal{L}^{(\alpha,\beta)}}{\sqrt{G(x)}}.
\]
The integral is evaluated by the metric integrator on the sampled metric; the code stores the result as \code{_a_tilde}.
Numerically, the inverse rate $\sqrt{G}/\mathcal{L}^{(\alpha,\beta)}$ is interpolated with a shape-preserving cubic (PCHIP) interpolator.

\apiheading{Built-in solver}
With the default \code{solve_ivp}, the forward equation is not integrated
directly.
Since it is autonomous, its inverse
\[
 \frac{d\tau}{dx}=\operatorname{sgn}(x_\mathrm{f}-x_\mathrm{i})\,\frac{\sqrt{G(x)}}{\mathcal{L}^{(\alpha,\beta)}}
\]
is integrated from $x_\mathrm{i}$ to $x_\mathrm{f}$ and $\tau(x)$ is inverted afterward.
The speed
$\mathcal{L}^{(\alpha,\beta)}/\sqrt{G}$ can span many orders of magnitude, e.g., for large
$\alpha$ and $\beta$, so the forward equation may require steps below the
floating-point spacing, whereas the inverse is well conditioned.
The right-hand side is the PCHIP interpolant $h(x)$ of $\sqrt{G}/\mathcal{L}^{(\alpha,\beta)}$ on the control grid.
Without \code{solver_kwargs}, $\tau(x)$ is obtained exactly from the piecewise-polynomial antiderivative of this interpolant, which is deterministic and free of step-size control.
With \code{solver_kwargs}, \code{solve_ivp} integrates the inverse equation with dense output (default \code{method="RK45"}, \code{atol=1e-12}, \code{rtol=1e-10}, since errors in $\tau$ are amplified by $dx/d\tau$ when inverting).
The step options therefore refer to $x$; for instance, \code{max_step=1e-2} on a sweep from $x=-500$ to $500$ takes at least $10^5$ steps.
In both cases, $\tau(x)$ is evaluated on $\max(N, 8M)$ uniformly spaced control values and divided by $\tau(x_\mathrm{f})$, so that the sweep ends exactly at $\tau=1$.
Nodes in flat stretches of $\tau(x)$, where the metric is numerically negligible and $\tau$ does not increase, are removed before $\tau(x)$ is inverted, and $x(\tau)$ is obtained by PCHIP interpolation on the uniform grid of $M$ samples in $\tau\in[0,1]$.
Both endpoints are set exactly.

\apiheading{Custom solver contract}
A different callable is invoked as
\begin{minted}{python}
solver(fun, [0.0, 1.0], [pulse_initial], t_eval=tau, **solver_kwargs)
\end{minted}
It must return either \code{(t, y)} or an object exposing \code{t} and
\code{y}.
If present, \code{success} must be true.
The time vector must
have $M$ finite, strictly increasing entries spanning $[0,1]$; the state array
must have shape \code{(M,)}, \code{(1, M)}, or \code{(M, 1)} and finite
entries.
Custom solvers integrate the forward equation
\[
\frac{dx}{d\tau}=\operatorname{sgn}(x_\mathrm{f}-x_\mathrm{i})\,\frac{I_h}{h(x)},
\]
where $h$ is the PCHIP interpolant of $\sqrt{G}/\mathcal{L}^{(\alpha,\beta)}$ above and $I_h=\left|\int_{x_\mathrm{i}}^{x_\mathrm{f}}h\,dx\right|\approx1$ is its exact integral, so that the exact solution ends at $x(1)=x_\mathrm{f}$.
For a fixed sampled metric, the quadrature value $\mathcal{L}^{(\alpha,\beta)}$ is an intermediate scale in both solver paths: it cancels algebraically in the normalized inverse time and in the forward ratio $I_h/h(x)$.
Custom solvers therefore use the interpolant's own integral for normalization, avoiding an endpoint mismatch from a different quadrature of the raw metric samples.
The speed is evaluated with the state clipped to the control range and vanishes beyond $x_\mathrm{f}$, so an overshooting step cannot leave the grid.
The result is accepted only if, at every sample, it deviates from the exact $\tau(x)$ of the interpolant by at most $10^{-2}$ in $\tau$; otherwise a \code{SolverError} is raised.
A solver that returns other time points than the requested uniform \code{t_eval} grid is resampled onto it by PCHIP interpolation, so that \code{tau} is always uniform, as \code{synthesize_pulse()}, \code{Dynamics}, and \code{PulseControl} assume.
Both endpoints are then set exactly.
Since the forward equation is badly conditioned for large exponents, the default solver is preferable there.
\end{methodbox}

\begin{methodbox}{\code{synthesize_pulse()}}
\begin{minted}{python}
ControlModel.synthesize_pulse(duration: float) -> PulseControl
\end{minted}

Create a \code{PulseControl} from the optimized sample array and a physical duration.
Automatically solve the problem if its ODE result is unavailable or invalid.

\apiheading{Parameters}
\apiitem{duration}{\code{float}}{Finite positive total physical duration.}

\apiheading{Returns}
\apiitem{pulse}{\code{PulseControl}}{New pulse object containing copies of the control samples on a uniform grid from zero to \code{duration}.}

\apiheading{Notes}
Changing \code{duration} rescales the time grid without recomputing the metric-based waveform.

\end{methodbox}

\begin{methodbox}{\code{plot_eigenvalues()}}
\begin{minted}{python}
ControlModel.plot_eigenvalues(
    fig: Figure | SubFigure | None=None,
    ax: Axes | None=None,
    legend: bool=True,
    legend_kwargs: dict[str, Any] | None=None,
    xlabel: str | None=None, ylabel: str | None=None,
    title: str | None=None,
    show: bool=True,
    **plot_kwargs: Any
) -> tuple[Figure | SubFigure, Axes]
\end{minted}

Plot each energy branch against the control grid.
Only \code{pulse_initial}, \code{pulse_final}, \code{num_steps}, and, in callable mode, \code{control_name} are needed; the eigensystem, including Hamiltonian derivatives, is computed on demand.

\apiheading{Parameters}
\apiitem{fig}{\code{Figure | SubFigure | None}}{Existing figure or subfigure.
If \code{ax} is absent and \code{fig} is supplied, its single axis is used, or a subplot is added if it has none; a figure with several axes requires \code{ax}.
Default: \code{None}.}

\apiitem{ax}{\code{Axes | None}}{Existing axes.
If supplied, the plot is drawn on it, \code{fig} is ignored, and its parent figure or subfigure is returned.
An unsupported \code{ax}, or an unsupported \code{fig} when \code{ax} is absent, raises \code{ValidationError}.
Default: \code{None}.}

\apiitem{legend}{\code{bool}}{Draw a legend.
Default: \code{True}.}

\apiitem{legend_kwargs}{\code{dict[str, Any] | None}}{Options forwarded to \code{ax.legend()}.
Default: \code{None}.}

\apiitem{xlabel}{\code{str | None}}{Horizontal-axis label.
Default: \code{None}, which uses \code{control_name}.}

\apiitem{ylabel}{\code{str | None}}{Vertical-axis label.
Default: \code{None}, which uses \code{"Energy"}.}

\apiitem{title}{\code{str | None}}{Plot title.
Default: \code{None}, which uses \code{"Hamiltonian eigenenergies"}.}

\apiitem{show}{\code{bool}}{Whether to call \code{pyplot.show()}.
Pass \code{False} to keep drawing on the figure.
Default: \code{True}.}

\apiitem{plot_kwargs}{\code{Any}}{Additional keyword arguments forwarded to \code{ax.plot()}.
The \code{label} keyword is supplied internally, so do not pass \code{label} here.}

\apiheading{Returns}
\apiitem{fig, ax}{\code{tuple[Figure | SubFigure, Axes]}}{Figure or subfigure and axes containing all energy branches.}

\end{methodbox}

\begin{methodbox}{\code{plot_metric_tensor()}}
\begin{minted}{python}
ControlModel.plot_metric_tensor(
    fig: Figure | SubFigure | None=None,
    ax: Axes | None=None, legend: bool=True,
    legend_kwargs: dict[str, Any] | None=None,
    xlabel: str | None=None, ylabel: str | None=None,
    title: str | None=None, show: bool=True,
    **plot_kwargs: Any
) -> tuple[Figure | SubFigure, Axes]
\end{minted}

Compute the eigensystem and metric if needed, then plot $G$ against the control grid.
Full control configuration is required, including metric exponents and state indices.

\apiheading{Parameters}
\apiitem{fig}{\code{Figure | SubFigure | None}}{Existing figure or subfigure.
If \code{ax} is absent and \code{fig} is supplied, its single axis is used, or a subplot is added if it has none; a figure with several axes requires \code{ax}.
Default: \code{None}.}

\apiitem{ax}{\code{Axes | None}}{Existing axes.
If supplied, the plot is drawn on it, \code{fig} is ignored, and its parent figure or subfigure is returned.
An unsupported \code{ax}, or an unsupported \code{fig} when \code{ax} is absent, raises \code{ValidationError}.
Default: \code{None}.}

\apiitem{legend}{\code{bool}}{Draw a legend.
Default: \code{True}.}

\apiitem{legend_kwargs}{\code{dict[str, Any] | None}}{Options forwarded to \code{ax.legend()}.
Default: \code{None}.}

\apiitem{xlabel}{\code{str | None}}{Horizontal-axis label.
Default: \code{None}, which uses \code{control_name}.}

\apiitem{ylabel}{\code{str | None}}{Vertical-axis label.
Default: \code{None}, which uses \code{"G tensor"}.}

\apiitem{title}{\code{str | None}}{Plot title.
Default: \code{None}, which uses \code{"G tensor"}.}

\apiitem{show}{\code{bool}}{Whether to call \code{pyplot.show()}.
Pass \code{False} to keep drawing on the figure.
Default: \code{True}.}

\apiitem{plot_kwargs}{\code{Any}}{Additional keyword arguments forwarded to \code{ax.plot()}.
The \code{label} keyword is supplied internally, so do not pass \code{label} here.}

\apiheading{Returns}
\apiitem{fig, ax}{\code{tuple[Figure | SubFigure, Axes]}}{Figure or subfigure and axes containing the metric curve.}

\apiheading{Notes}
This method does not solve the pulse ODE.

\end{methodbox}

\begin{methodbox}{\code{print_summary()}}
\begin{minted}{python}
ControlModel.print_summary() -> None
\end{minted}

Print the Hamiltonian and derivative availability, fixed parameters, control settings, metric exponents, and computation-status flags.
The status marks are emojis; if the encoding of \code{sys.stdout} cannot represent them (e.g., cp1252), they are printed as \code{[OK]}, \code{[--]}, and \code{->}.

\end{methodbox}

\subsection{\code{PulseControl}}
\begin{apisignature}
class PulseControl(pulse: ArrayLike, duration: float)
\end{apisignature}

Stores a real-valued pulse sampled uniformly in physical time.
Construction is independent of \code{ControlModel}: any suitable one-dimensional sample array can be supplied.
In this section, $M$ is the number of samples, which equals \code{pulse_accuracy} for pulses from \code{synthesize_pulse()}.
Processing methods return arrays and leave the stored pulse unchanged.

\apiheading{Parameters}
\apiitem{pulse}{\code{ArrayLike}}{One-dimensional finite samples, with at least two entries.
Array-like inputs are converted with \code{np.asarray()}.
Complex input is accepted only when its imaginary part is negligible relative to the real part, $\max|\mathrm{Im}| \le 10^{-10} \max|\mathrm{Re}|$, so the check does not depend on units; the real part is then stored as a \code{float} copy.}

\apiitem{duration}{\code{float}}{Finite positive physical duration.}

\apiheading{Attributes}
\apiitem{pulse}{\code{NDArray[np.float64]}, shape \code{(M,)}}{Read-only property returning a copy of the real samples.}

\apiitem{times}{\code{NDArray[np.float64]}, shape \code{(M,)}}{Read-only property returning a copy of \code{np.linspace(0, duration, M)}.}

\apiitem{duration}{\code{float}}{Read-only total physical duration.}

Invalid sample shape, size, values, or duration raise \code{ValidationError}.

\apiheading{Methods}
\apiitem{discretized_pulse(...)}{\code{tuple[NDArray[np.float64], NDArray[np.float64]]}}{Resample by linear interpolation at a smaller or larger set of knots.}

\apiitem{fourier_spectrum(...)}{\code{tuple[NDArray[np.float64], NDArray[np.float64], NDArray[np.float64]]}}{Compute a short-time Fourier magnitude spectrum.}

\apiitem{filtered_pulse(...)}{\code{tuple[NDArray[np.float64], NDArray[np.float64]]}}{Apply a low-pass Butterworth filter, zero-phase by default or causal.}

\apiitem{plot_pulse(...)}{\code{tuple[Figure | SubFigure, Axes]}}{Plot the physical-time waveform.}

\apiitem{export_pulse(...)}{\code{Path}}{Write time and amplitude samples to disk.}

\Needspace{12\baselineskip}
\apiheading{Detailed methods}
\begin{methodbox}{\code{discretized_pulse()}}
\begin{minted}{python}
PulseControl.discretized_pulse(
    linear_steps: int=4
) -> tuple[NDArray[np.float64], NDArray[np.float64]]
\end{minted}

Linearly interpolate the stored pulse and evaluate it at uniformly spaced output knots, including both endpoints.

\apiheading{Parameters}
\apiitem{linear_steps}{\code{int}}{Number of returned sample points, not number of line segments.
Must be an integer of at least 2.
Default: \code{4}.}

\apiheading{Returns}
\apiitem{new_time}{\code{NDArray[np.float64]}, shape \code{(linear_steps,)}}{Physical-time knot locations from zero to \code{duration}.}

\apiitem{approx_sol}{\code{NDArray[np.float64]}, shape \code{(linear_steps,)}}{Interpolated pulse values at the knots.}

\apiheading{Notes}
The output defines \code{linear_steps - 1} segments.
Non-integer values (including booleans) and values below 2 raise \code{ValidationError}.

\end{methodbox}

\begin{methodbox}{\code{fourier_spectrum()}}
\begin{minted}{python}
PulseControl.fourier_spectrum(
    window_len: int=256,
    hop: int=32,
    detrend: bool=False,
    padding: str="hold",
    mfft: int | None=None
) -> tuple[NDArray[np.float64], NDArray[np.float64], NDArray[np.float64]]
\end{minted}

Compute a short-time Fourier transform (STFT) using a periodic Hann window, and return its amplitude spectrum.
The magnitude scaling and the doubled positive frequencies of the one-sided spectrum make it the physical amplitude: a sinusoid of amplitude $A$ within a window gives $A$ at its frequency, and a constant $c$ gives $c$ at zero frequency.

\apiheading{Parameters}
\apiitem{window_len}{\code{int}}{Window length in samples.
Must satisfy \code{2 <= window_len < M}; choose a smaller value for short pulses.
Default: \code{256}.}

\apiitem{hop}{\code{int}}{Advance between windows in samples.
Must satisfy \code{1 <= hop <= window_len}.
Default: \code{32}.}

\apiitem{detrend}{\code{bool}}{Subtract the mean of each window segment before the transform, which removes the DC part of the spectrum.
Default: \code{False}.}

\apiitem{padding}{\code{str}}{How the pulse is extended beyond its ends for the windows that overhang them.
\code{"hold"} holds the first and last values, as an arbitrary waveform generator holds its output before and after a pulse.
\code{"zeros"} pads with zeros, which adds the jump from the end values to zero to the spectrum; it was the default up to version 0.3.0.
\code{"even"} and \code{"odd"} mirror the pulse.
Other values raise \code{ValidationError}.
Default: \code{"hold"}.}

\apiitem{mfft}{\code{int | None}}{Length of the FFT, at least \code{window_len}.
Each window is zero-padded to this length, which gives \code{mfft // 2 + 1} frequency bins; this interpolates the spectrum on a finer frequency grid, while the frequency resolution is still set by the window duration.
Default: \code{None}, i.e., \code{window_len}.}

\apiheading{Returns}
\apiitem{frequencies}{\code{NDArray[np.float64]}, shape \code{(F,)}}{Non-negative frequency bins in cycles per unit physical time, not angular frequencies.}

\apiitem{times}{\code{NDArray[np.float64]}, shape \code{(K,)}}{STFT window-center times on the physical-time scale.}

\apiitem{magnitude}{\code{NDArray[np.float64]}, shape \code{(F, K)}}{Amplitude spectrum in the units of the control, not its square or a power spectral density.}

\apiheading{Notes}
The sample spacing is \code{duration/(M - 1)}.
Boundary windows are extended as set by \code{padding}, so window-center times extend up to \code{window_len/2} samples outside the original interval.

\end{methodbox}

\begin{methodbox}{\code{filtered_pulse()}}
\begin{minted}{python}
PulseControl.filtered_pulse(
    cutoff_freq: float,
    filter_order: int=3,
    padding: str | None="hold",
    causal: bool=False,
) -> tuple[NDArray[np.float64], NDArray[np.float64]]
\end{minted}

Design a low-pass Butterworth filter in second-order-section form and apply it forward and backward for zero-phase filtering, or forward only for causal filtering.

\apiheading{Parameters}
\apiitem{cutoff_freq}{\code{float}}{Positive cutoff in cycles per unit time, strictly below \code{(M - 1)/(2 * duration)}, the Nyquist frequency.}

\apiitem{filter_order}{\code{int}}{Positive integer specifying the Butterworth design order.
Default: \code{3}.}

\apiitem{padding}{\code{str | None}}{Extension of the pulse beyond its ends for zero-phase filtering.
\code{"hold"} holds the first and last values until the impulse response of the filter has decayed, as an arbitrary waveform generator holds its output before and after a pulse.
\code{"odd"}, \code{"even"}, \code{"constant"}, and \code{None} are passed to \code{scipy.signal.sosfiltfilt} as \code{padtype}.
Default: \code{"hold"}.}

\apiitem{causal}{\code{bool}}{Apply the filter forward only, as a transmission line would, starting in the steady state of the first value.
\code{padding} is then not used.
Default: \code{False}.}

\apiheading{Returns}
\apiitem{pulse_times}{\code{NDArray[np.float64]}, shape \code{(M,)}}{Copy of the original physical-time sample grid.}

\apiitem{filtered_pulse}{\code{NDArray[np.float64]}, shape \code{(M,)}}{Filtered sample values.}

\apiheading{Notes}
Forward-backward filtering applies the designed filter twice: it has no phase shift but is acausal, its magnitude response is the square of the designed one (gain $1/2$, i.e., $-6\,$dB, at \code{cutoff_freq}), and its effective order is \code{2 * filter_order}.
Causal filtering has the designed gain $1/\sqrt{2}$ ($-3\,$dB) at \code{cutoff_freq} and a phase lag.
With a \code{padtype} of \code{sosfiltfilt}, both ends of the pulse are padded with $3(2n_\mathrm{s}+1)$ samples, where $n_\mathrm{s}$ is the number of second-order sections ($3$ fewer for an odd order, e.g., 12 for order 3 and 15 for order 4), so the pulse needs more samples than that; a shorter pulse raises \code{ValidationError}.
Low-pass filtering moves the endpoints of a pulse that is steep there, whereas \code{Dynamics} still measures fidelities against the eigenstates at \code{pulse_initial} and \code{pulse_final}; a \code{GeodesiQWarning} is issued when an endpoint moves by more than $1\%$ of the pulse range.
This method does not create a new \code{PulseControl} object; construct one from the returned samples if desired.

\end{methodbox}

\begin{methodbox}{\code{plot_pulse()}}
\begin{minted}{python}
PulseControl.plot_pulse(
    fig: Figure | SubFigure | None=None,
    ax: Axes | None=None,
    xlabel: str | None=None,
    ylabel: str | None=None,
    title: str | None=None,
    show: bool=True,
    **plot_kwargs: Any
) -> tuple[Figure | SubFigure, Axes]
\end{minted}

Plot the stored control samples against physical time.

\apiheading{Parameters}
\apiitem{fig}{\code{Figure | SubFigure | None}}{Existing figure or subfigure.
If \code{ax} is absent and \code{fig} is supplied, its single axis is used, or a subplot is added if it has none; a figure with several axes requires \code{ax}.
Default: \code{None}.}

\apiitem{ax}{\code{Axes | None}}{Existing axes.
If supplied, the plot is drawn on it, \code{fig} is ignored, and its parent figure or subfigure is returned.
Default: \code{None}.}

\apiitem{xlabel}{\code{str | None}}{Horizontal-axis label.
Default: \code{None}, which uses \code{"Time $t$"}.}

\apiitem{ylabel}{\code{str | None}}{Vertical-axis label.
Default: \code{None}, which uses \code{"Control Pulse"}.}

\apiitem{title}{\code{str | None}}{Plot title.
Default: \code{None}, which sets no title.}

\apiitem{show}{\code{bool}}{Whether to call \code{pyplot.show()}.
Pass \code{False} to keep drawing on the figure.
Default: \code{True}.}

\apiitem{plot_kwargs}{\code{Any}}{Keyword options forwarded to \code{ax.plot()}, such as \code{color}, \code{linewidth}, or \code{linestyle}.}

\apiheading{Returns}
\apiitem{fig, ax}{\code{tuple[Figure | SubFigure, Axes]}}{Figure or subfigure and axes containing the pulse.}

\end{methodbox}

\begin{methodbox}{\code{export_pulse()}}
\begin{minted}{python}
PulseControl.export_pulse(filename: str | Path, file_extension: str="npz", overwrite: bool=False) -> Path
\end{minted}

Export stored time and amplitude samples.
The suffix \code{.<file_extension>} is appended unless \code{filename} already ends with it, so dots in the name are kept (\code{"pulse_tf1.5"} is written to \code{pulse_tf1.5.npz}).
Without \code{overwrite}, the file is created only if it does not exist, in one step with the write.

\apiheading{Parameters}
\apiitem{filename}{\code{str | Path}}{Output path.
Parent directories must already exist.}

\apiitem{file_extension}{\code{str}}{One of \code{"npz"}, \code{"txt"}, or \code{"csv"}.
A single leading dot is removed and matching is case-insensitive.
Other values, including non-strings, raise \code{ValidationError}.
Default: \code{"npz"}.}

\apiitem{overwrite}{\code{bool}}{Permit replacing an existing target file; otherwise an existing file raises \code{IOErrorGeodesiQ}.
Default: \code{False}.}

\apiheading{Returns}
\apiitem{output_path}{\code{Path}}{Path of the written file.}

\apiheading{Notes}
All formats use the column names \code{times} and \code{pulse}: NPZ files contain arrays with these names, TXT files are whitespace-delimited with the header \code{times pulse}, and CSV files are comma-delimited with the header \code{times,pulse}.
All formats keep full double precision (TXT and CSV use the format \texttt{\%.17g}).

\end{methodbox}

\subsection{\code{Dynamics}}
\begin{apisignature}
class Dynamics(model: ControlModel, duration: float | None=None, *,
               times: ArrayLike | None=None, pulse: ArrayLike | None=None,
               hbar: float=1.0, order: int | None=None,
               exact_hamiltonian: bool=False,
               options: dict[str, Any] | None=None)
\end{apisignature}

Simulates the evolution under the model's \texttt{geodesiq} pulse, rescaled to \code{duration}, or under a custom pulse given by the samples \code{pulse} at the strictly increasing times \code{times}.
Construction prepares a \texttt{QuTiP} \code{QobjEvo} representing $H(t)/\hbar$.
For affine models, it uses the supplied drift and control matrices directly, with the sampled control waveform interpolated as set by \code{order}.
For callable models, it samples the Hamiltonian, subtracts the sample mean, and uses a singular-value decomposition to obtain constant Hermitian operators and sampled coefficients.
Components with singular values exceeding $10^{-10}$ times the largest singular value are retained, and the coefficients are interpolated in the same way.
With \code{exact_hamiltonian=True}, it instead evaluates the Hamiltonian at the interpolated pulse at every solver step.

\apiheading{Parameters}
\apiitem{model}{\code{ControlModel}}{Configured model supplying the Hamiltonian, control grid, pulse solution, and state indices.
Access to its lazy properties may trigger eigensystem and pulse calculations during construction.}

\apiitem{duration}{\code{float | None}}{Finite positive physical duration of the optimal pulse.
Required unless \code{times} and \code{pulse} are given; mutually exclusive with them.
Default: \code{None}.}

\apiitem{times}{\code{ArrayLike | None}}{One-dimensional, real, strictly increasing physical times of a custom pulse.
Must be given together with \code{pulse}.
Keyword-only.
Default: \code{None}.}

\apiitem{pulse}{\code{ArrayLike | None}}{Real control values of a custom pulse sampled at \code{times}, e.g., the output of \code{filtered_pulse()} or \code{discretized_pulse()}.
Complex input is accepted under the same relative check as in \code{PulseControl}.
The endpoint eigenstates are taken at \code{model.pulse_initial} and \code{model.pulse_final} if these are set, and at the first and last pulse values otherwise.
A pulse that starts or ends further than $1\%$ of the control range from \code{model.pulse_initial} or \code{model.pulse_final} emits a \code{GeodesiQWarning}, since the fidelity then also measures this endpoint mismatch.
Keyword-only.
Default: \code{None}.}

\apiitem{hbar}{\code{float}}{Finite positive reduced Planck constant in the chosen energy--time units.
Keyword-only.
Default: \code{1.0}.}

\apiitem{order}{\code{int | None}}{Interpolation of the pulse between its samples, used for the propagation and for every Hamiltonian evaluated between samples: \code{0} holds each sample until the next one, \code{1} is linear, and \code{3} is a cubic spline.
For a callable model without \code{exact_hamiltonian}, it interpolates the decomposed coefficients instead.
Other values raise \code{ValidationError}.
Keyword-only.
Default: \code{None}, which uses \code{3} for the optimal pulse (\code{duration}) and \code{1} for a custom pulse.}

\apiitem{exact_hamiltonian}{\code{bool}}{For a callable model, evaluate \code{H_func} at the interpolated pulse at every solver step instead of interpolating the decomposition entry-wise in $H$.
Exact for any pulse, but slower to propagate; affine models are always exact.
Keyword-only.
Default: \code{False}.}

\apiitem{options}{\code{dict[str, Any] | None}}{Solver settings forwarded to all \texttt{QuTiP} propagation calls, including \code{atol}, \code{rtol}, \code{nsteps}, \code{max_step}, and \code{method}.
The dictionary is deep-copied at construction and for each solver call; each method overrides output-storage settings as needed for its documented return values.
Supported options depend on the integration method and are checked when evolution is requested.
\code{state_fidelity()} and the final-only propagator output only the end of the pulse; since \code{nsteps} bounds the steps between two outputs, it is then multiplied by the number of sample intervals.
Keyword-only.
Default: \code{None}, which retains \texttt{QuTiP}'s numerical defaults.}

Time-integration failures raise \code{SolverError} with the original exception chained; rejected solver configurations raise \code{ValidationError}.
For accuracy checks, tighten \code{atol} and \code{rtol} and choose \code{max_step} below half the width of the narrowest pulse feature: output sample times alone do not bound internal solver steps.
The default state normalization can hide norm drift; set \code{normalize_output=False} to inspect probability conservation.
Propagated operators are not normalized, so their unitarity should be checked separately.
These options control time integration, independently of eigensystem and metric sampling, pulse interpolation, and SVD truncation.

\apiheading{Attributes}
\apiitem{times}{\code{NDArray[np.float64]}, shape \code{(M,)}}{Read-only property returning a copy of the physical time grid of the pulse used in the evolution, where $M$ is its number of samples (\code{pulse_accuracy} for the optimal pulse, the length of \code{times} for a custom pulse).}

\apiitem{pulse}{\code{NDArray[np.float64]}, shape \code{(M,)}}{Read-only property returning a copy of the control samples of that pulse.}

\apiitem{duration}{\code{float}}{Read-only total duration of that pulse.}

\apiitem{interpolation_error}{\code{float}}{Read-only estimate of the propagator error caused by interpolating the decomposed Hamiltonian of a callable model between the pulse samples, $\sum_k \Delta t_k \lVert H_\mathrm{interp} - H(x(t))\rVert_\mathrm{F}/\hbar$ at the interval midpoints, which estimates the bound $\lVert U - U_\mathrm{exact}\rVert \le \int \lVert H_\mathrm{interp} - H(x(t))\rVert\,\mathrm{d}t/\hbar$.
A value above $10^{-3}$ emits a \code{NumericalStabilityWarning} at construction.
Zero for affine models, \code{order=0}, and \code{exact_hamiltonian=True}, which propagate $H(x(t))$ itself.}

Invalid \code{duration}, \code{times}, \code{pulse}, or \code{hbar} inputs, or combining \code{duration} with a custom pulse, raise \code{ValidationError}; giving neither, or only one of \code{times} and \code{pulse}, raises \code{MissingArgsError}.
Supplying a \code{model} that is not a \code{ControlModel} also raises \code{ValidationError}.
Missing pulse or Hamiltonian data raise \code{ConfigurationError}.
With \code{duration}, the model is solved when needed: \code{MissingControlParameterError} reports missing control parameters, and errors of \code{solve_problem()} propagate unchanged.
Pulse samples, configured state indices, boundary control values, and the Hamiltonian (copies of \code{H_d} and \code{H_c}, or of the fixed parameters and \code{control_name} of a callable model) are captured during construction.
The propagation, the endpoint eigenstates, and the adiabatic basis of \code{populations()} therefore all use the same Hamiltonian, and later changes to the model do not affect an existing \code{Dynamics} object.
Recreate \code{Dynamics} to use a changed model; state kept inside \code{H_func} itself (e.g., in a closure) is not copied.

\apiheading{Methods}
\apiitem{time_evolution_operator()}{\code{list[qt.Qobj]}}{Compute propagators along the pulse.}

\apiitem{state_fidelity(...)}{\code{float}}{Compute the state-transfer fidelity $|\langle\psi_\mathrm{target}|\Psi(t_\mathrm{f})\rangle|^2$ (squared \texttt{QuTiP} fidelity), optionally with dissipation.}

\apiitem{average_gate_fidelity(...)}{\code{list[float]}}{Compare one or more operations with a target gate.}

\apiitem{populations(...)}{\code{NDArray[np.float64]}}{Compute adiabatic or diabatic state populations along the pulse.}

\apiitem{plot_populations(...)}{\code{tuple[Figure | SubFigure, Axes]}}{Plot the state populations against physical time.}

\Needspace{12\baselineskip}
\apiheading{Detailed methods}
\begin{methodbox}{\code{time_evolution_operator()}}
\begin{minted}{python}
Dynamics.time_evolution_operator(final_only: bool=False) -> list[qt.Qobj]
\end{minted}

Call the \texttt{QuTiP} propagator using the \code{QobjEvo} prepared during construction, which represents $H(t)/\hbar$.
Its sampled time-dependent coefficients are interpolated as set by \code{order}; with \code{exact_hamiltonian=True}, the Hamiltonian is evaluated at the interpolated pulse instead.

\apiheading{Parameters}
\apiitem{final_only}{\code{bool}}{If \code{True}, return only the propagator at the end of the pulse, which avoids storing one dense matrix per time sample.
Nothing is output in between, so the result agrees with the last full propagator within the solver tolerance.
Default: \code{False}.}

\apiheading{Returns}
\apiitem{propagators}{\code{list[qt.Qobj]}}{Evolution operators \code{U(t_j, 0)} at the pulse time points, or \code{[U(t_f, 0)]} if \code{final_only=True}.
The result is always wrapped as a list, including when \texttt{QuTiP} returns a single object.}

\apiheading{Notes}
This method describes closed-system evolution and accepts no collapse operators.
It uses the solver \code{options} supplied to the \code{Dynamics} constructor, as do \code{state_fidelity()}, \code{populations()}, and \code{average_gate_fidelity()} when it computes propagators.

\end{methodbox}

\begin{methodbox}{\code{state_fidelity()}}
\begin{minted}{python}
Dynamics.state_fidelity(
    initial_state: np.ndarray | int | qt.Qobj | None=None,
    final_state: np.ndarray | int | qt.Qobj | None=None,
    c_ops: Sequence[qt.Qobj | qt.QobjEvo | np.ndarray] | None=None
) -> float
\end{minted}

Evolve the initial state using the prepared \code{QobjEvo} and compare the final state with a target.
Each state argument may be omitted, an integer index, or an explicit state, independently of the other.
If an omitted state requires an unconfigured model state index, or the Hamiltonian dimension is unavailable, \code{ConfigurationError} is raised.
An explicit state is an array or \code{Qobj} holding a ket, a bra (converted to a ket), or a density matrix of the model dimension $d$; kets must be normalized and density matrices Hermitian with unit trace (within $10^{-6}$), and other inputs raise \code{ValidationError}.

\apiheading{Parameters}
\apiitem{initial_state}{\code{np.ndarray | int | qt.Qobj | None}}{Initial state.
If omitted, the model's \code{initial_state} index is used; an integer selects that energy level at the initial control value; an explicit state (array or \code{Qobj}) is used directly.
Default: \code{None}.}

\apiitem{final_state}{\code{np.ndarray | int | qt.Qobj | None}}{Target state, resolved in the same way: the model's \code{final_state} index if omitted, an energy level at the final control value for an integer, or an explicit state.
Integer indices refer to the eigensystem at the final control value, not computational-basis labels.
Default: \code{None}.}

\apiitem{c_ops}{\code{Sequence[qt.Qobj | qt.QobjEvo | np.ndarray] | None}}{Collapse operators.
Supply a list or tuple of $d\times d$ operators (\code{Qobj}, \code{np.ndarray}, or \code{QobjEvo}, which may be mixed); arrays are converted to \code{Qobj}, and other types or shapes raise \code{ValidationError}.
Any rates must already be included in the operators, for example through a square-root rate factor.
Default: \code{None}, which gives closed evolution.}

\apiheading{Returns}
\apiitem{fidelity}{\code{float}}{\code{qt.fidelity(target_state, evolved_final_state)**2}, i.e., the squared Uhlmann fidelity $\mathcal{F}=\big(\operatorname{Tr}\sqrt{\sqrt{\rho}\,\rho_\mathrm{target}\sqrt{\rho}}\big)^2$ between the evolved state $\rho$ and the target $\rho_\mathrm{target}$.
For pure states this is the transition probability \code{|<target|psi_final>|**2}.}

\apiheading{Notes}
The evolution calls \code{qt.mesolve}, which solves the Schr\"odinger equation (it defers to \code{qt.sesolve}) for a ket without collapse operators, and the Lindblad master equation otherwise.
An omitted or integer state whose gap to a neighboring level is at most $10^{-8}$ times the spectral bandwidth emits a \code{NumericalStabilityWarning}.
Its eigenvector is then an arbitrary vector of the degenerate subspace, and so is the fidelity; such a state should be passed explicitly.
The same check applies to the initial state of \code{populations()}.

\end{methodbox}

\begin{methodbox}{\code{average_gate_fidelity()}}
\begin{minted}{python}
Dynamics.average_gate_fidelity(
    gate: qt.Qobj | list[qt.Qobj] | None=None,
    *,
    target_gate: qt.Qobj | np.ndarray,
    final_only: bool=False
) -> list[float]
\end{minted}

Calculate the \texttt{QuTiP} average gate fidelity of each supplied operation, or for the pulse propagators if \code{gate} is omitted.

\apiheading{Parameters}
\apiitem{gate}{\code{qt.Qobj | list[qt.Qobj] | None}}{Single operation or list of operations.
A raw \code{np.ndarray} is not accepted for \code{gate}.
Each operation must be $d\times d$, otherwise \code{ValidationError} is raised.
Default: \code{None}, which computes \code{time_evolution_operator()}.}

\apiitem{target_gate}{\code{qt.Qobj | np.ndarray}}{Target unitary operation of shape $d\times d$ (otherwise \code{ValidationError}).
Arrays are converted to \code{Qobj}.
Required: there is no default target, so that the fidelity is never computed against the identity by accident.
Keyword-only.}

\apiitem{final_only}{\code{bool}}{Used only when \code{gate} is omitted: if \code{True}, evaluate only the final propagator (see \code{time_evolution_operator()}).
Keyword-only.
Default: \code{False}.}

\apiheading{Returns}
\apiitem{gate_fid}{\code{list[float]}}{One average gate fidelity per operation; a single gate produces a one-element list.}

\apiheading{Notes}
Operations are compared as $d\times d$ matrices, so their \texttt{QuTiP} tensor-product dimensions need not match.
A supplied list may represent arbitrary operations; the wrapper does not verify a time ordering.
This method has no \code{c_ops} argument; the automatically generated propagators are closed-system operations.

\end{methodbox}

\begin{methodbox}{\code{populations()}}
\begin{minted}{python}
Dynamics.populations(
    initial_state: np.ndarray | int | qt.Qobj | None=None,
    basis: Literal["adiabatic", "diabatic"]="adiabatic",
    c_ops: Sequence[qt.Qobj | qt.QobjEvo | np.ndarray] | None=None
) -> NDArray[np.float64]
\end{minted}

Evolve an initial state and return the population of every state at each pulse time sample, using the constructor's solver \code{options}.

\apiheading{Parameters}
\apiitem{initial_state}{\code{np.ndarray | int | qt.Qobj | None}}{Initial state, with the normalization and dimension requirements of \code{state_fidelity()}.
An integer selects an eigenstate at the initial control value; an explicit ket, bra, or density matrix is used directly, with a bra converted to a ket.
Default: \code{None}, which uses the model's configured initial energy-level index.}

\apiitem{basis}{\code{Literal["adiabatic", "diabatic"]}}{Basis used to measure populations.
The value \code{"adiabatic"} uses the instantaneous Hamiltonian eigenstates at each time, sorted by increasing energy, and \code{"diabatic"} uses the fixed basis in which the Hamiltonian matrix is written.
Adiabatic populations assigned to individual exactly degenerate levels are not unique.
Default: \code{"adiabatic"}.}

\apiitem{c_ops}{\code{Sequence[qt.Qobj | qt.QobjEvo | np.ndarray] | None}}{Collapse operators, following \code{state_fidelity()}; a list or tuple of compatible operators enables Lindblad evolution.
Invalid containers, elements, or dimensions raise \code{ValidationError}.
Default: \code{None}, which gives closed evolution.}

\apiheading{Returns}
\apiitem{populations}{\code{NDArray[np.float64]}, shape \code{(len(times), d)}}{The entry \code{populations[i, n]} is the population of state \code{n} at physical time \code{times[i]} in the selected basis.}

\apiheading{Errors}
An invalid basis or initial state raises \code{ValidationError}.
If the model's initial state index is needed but not configured, or its Hamiltonian dimension is unavailable, \code{ConfigurationError} is raised.
Time-integration failures raise \code{SolverError}; rejected solver options or a missing time sample in the evolution result raise \code{ValidationError}.

\end{methodbox}

\begin{methodbox}{\code{plot_populations()}}
\begin{minted}{python}
Dynamics.plot_populations(
    fig: Figure | SubFigure | None=None,
    ax: Axes | None=None,
    initial_state: np.ndarray | int | qt.Qobj | None=None,
    basis: Literal["adiabatic", "diabatic"]="adiabatic",
    c_ops: Sequence[qt.Qobj | qt.QobjEvo | np.ndarray] | None=None,
    states: Sequence[int] | None=None,
    legend: bool=True, legend_kwargs: dict[str, Any] | None=None,
    xlabel: str | None=None, ylabel: str | None=None,
    title: str | None=None, show: bool=True,
    **plot_kwargs: Any
) -> tuple[Figure | SubFigure, Axes]
\end{minted}

Call \code{populations()} and plot the selected columns against physical time.

\apiheading{Parameters}
\apiitem{fig}{\code{Figure | SubFigure | None}}{Existing figure or subfigure.
If \code{ax} is absent and \code{fig} is supplied, its single axis is used, or a subplot is added if it has none; a figure with several axes requires \code{ax}.
Default: \code{None}.}

\apiitem{ax}{\code{Axes | None}}{Existing axes.
If supplied, the plot is drawn on it, \code{fig} is ignored, and its parent figure or subfigure is returned.
Default: \code{None}.}

\apiitem{initial_state}{\code{np.ndarray | int | qt.Qobj | None}}{Initial state, as in \code{populations()}.
Default: \code{None}.}

\apiitem{basis}{\code{Literal["adiabatic", "diabatic"]}}{Population basis, as in \code{populations()}.
Default: \code{"adiabatic"}.}

\apiitem{c_ops}{\code{Sequence[qt.Qobj | qt.QobjEvo | np.ndarray] | None}}{Collapse operators, as in \code{populations()}.
Default: \code{None}.}

\apiitem{states}{\code{Sequence[int] | None}}{State indices to plot, in the supplied order.
Each index must be an integer in $[0, d)$; booleans and out-of-range indices raise \code{ValidationError}.
Default: \code{None}, which plots all states.}

\apiitem{legend}{\code{bool}}{Draw a legend.
Default: \code{True}.}

\apiitem{legend_kwargs}{\code{dict[str, Any] | None}}{Options forwarded to \code{ax.legend()}.
Default: \code{None}.}

\apiitem{xlabel}{\code{str | None}}{Horizontal-axis label.
Default: \code{None}, which uses \code{"Time $t$"}.}

\apiitem{ylabel}{\code{str | None}}{Vertical-axis label.
Default: \code{None}, which uses \code{"Population"}.}

\apiitem{title}{\code{str | None}}{Plot title.
Default: \code{None}, which uses \code{"Adiabatic populations"} or \code{"Diabatic populations"}.}

\apiitem{show}{\code{bool}}{Whether to call \code{pyplot.show()}.
Pass \code{False} to keep drawing on the figure.
Default: \code{True}.}

\apiitem{plot_kwargs}{\code{Any}}{Additional keyword arguments forwarded to \code{ax.plot()} for each selected state.
The \code{label} keyword is supplied internally, so do not pass \code{label} here.}

\apiheading{Returns}
\apiitem{fig, ax}{\code{tuple[Figure | SubFigure, Axes]}}{Figure or subfigure and axes containing the population curves.}

\apiheading{Notes}
The validation, degeneracy caveat, and solver settings of \code{populations()} apply.

\end{methodbox}

\end{appendix}

\bibliography{references}

\end{document}

%% file: Extras/summary_fig_v4.tex
\begin{tikzpicture}[font=\fontsize{9}{11}\selectfont, >=Stealth]
\definecolor{controlblue}{HTML}{7570B3}
\definecolor{pulsepurple}{HTML}{D95F02}
\definecolor{dynamicsgreen}{HTML}{1B9E77}
\definecolor{navyblue}{RGB}{28, 54, 94}
\definecolor{lightbg}{RGB}{248, 249, 250}
\definecolor{bordergrey}{RGB}{185, 190, 196}
\definecolor{darktext}{RGB}{44, 62, 80}

\tikzset{
    card/.style={
        draw=bordergrey, fill=white, line width=0.6pt, rounded corners=4pt,
        inner xsep=6pt, inner ysep=5pt, text=darktext, align=left, anchor=north west
    },
    optionalcard/.style={card, draw=pulsepurple!60, dash pattern=on 3pt off 2pt},
    tag/.style={
        font=\fontsize{10}{12}\selectfont\bfseries, text=white,
        rounded corners=2.5pt, inner xsep=5pt, inner ysep=2pt, anchor=north west
    },
    verb/.style={font=\fontsize{9}{11}\selectfont\bfseries, text=darktext, inner sep=0pt, anchor=north west},
    note/.style={font=\fontsize{8}{10}\selectfont\itshape, text=darktext!75, inner sep=0pt, anchor=west},
    flowlabel/.style={font=\fontsize{8}{10}\selectfont, text=darktext, inner sep=1.5pt},
    iobox/.style={
        draw=bordergrey, fill=lightbg, line width=0.6pt, rounded corners=3pt, inner xsep=5pt, inner ysep=2.5pt,
        font=\fontsize{8}{10}\selectfont, text=darktext, align=center, text width=3.9cm, minimum height=0.95cm,
        anchor=south
    },
    arrow/.style={-{Stealth[length=2mm,width=1.7mm]}, line width=0.8pt, draw=navyblue, line cap=round},
    bypass/.style={arrow, draw=navyblue!75, dash pattern=on 3pt off 2pt, rounded corners=4pt},
    feedback/.style={arrow, draw=dynamicsgreen!80!black, dash pattern=on 3pt off 2pt, rounded corners=4pt},
    dataarrow/.style={-{Stealth[length=1.8mm,width=1.5mm]}, line width=0.6pt, dash pattern=on 2pt off 1.5pt,
                      draw=darktext!70},
}

\newcommand{\cardbullet}[1]{\par\hangindent=0.85em\hangafter=1\noindent\makebox[0.85em][l]{\textbullet}#1\par}
\newcommand{\cardbody}[3]{%
    \begin{minipage}[t][3.75cm][t]{3.9cm}
        \raggedright
        \vspace{0.98cm}
        \cardbullet{#1}\vspace{0.15em}
        \cardbullet{#2}\vspace{0.15em}
        \cardbullet{#3}
    \end{minipage}%
}
\def\gap{0.95cm}

\node[card] (model) at (0, 0) {\cardbody
    {Hamiltonian $\hat{H}(x;\boldsymbol{\gamma})$, initial and target eigenstates}
    {Metric exponents $(\alpha,\beta;\tilde\alpha,\tilde\beta)$}
    {\texttt{solve\_problem()} $\rightarrow$ optimal pulse $x_\mathrm{gq}(t)$}};
\node[tag, fill=controlblue] (t1) at ([shift={(5pt,-4pt)}]model.north west) {ControlModel};
\node[verb] at ([yshift=-3pt]t1.south west) {Define \& solve};

\node[optionalcard] (pulse) at ([xshift=\gap]model.north east) {\cardbody
    {Pulse profile and Fourier spectrum}
    {Low-pass filtering, ramp-like discretization}
    {Export of the adjusted pulse $x^\mathrm{adj}_\mathrm{gq}(t)$}};
\node[tag, fill=pulsepurple] (t2) at ([shift={(5pt,-4pt)}]pulse.north west) {PulseControl};
\node[verb] (v2) at ([yshift=-3pt]t2.south west) {Shape \& export};
\node[note] at ([xshift=4pt]v2.east) {(optional)};

\node[card] (dynamics) at ([xshift=\gap]pulse.north east) {\cardbody
    {Closed (Schr\"odinger) or open (Lindblad) evolution under $x(t)$}
    {State fidelity $\mathcal{F}=|\langle\psi_\mathrm{target}|\Psi(t_\mathrm{f})\rangle|^2$}
    {Average gate fidelity, populations}};
\node[tag, fill=dynamicsgreen] (t3) at ([shift={(5pt,-4pt)}]dynamics.north west) {Dynamics};
\node[verb] at ([yshift=-3pt]t3.south west) {Simulate \& validate};

\node[iobox] (input) at ([yshift=0.55cm]model.north)
    {\textbf{Input:} $\hat{H}(x;\boldsymbol{\gamma})$, $|\psi_{k_\mathrm{i}}\rangle$, $|\psi_\mathrm{target}\rangle$, $(\alpha,\beta;\tilde\alpha,\tilde\beta)$};
\draw[dataarrow] (input) -- (model);
\node[iobox] (output) at ([yshift=0.55cm]dynamics.north) {\textbf{Output:} $x(t)$, $\mathcal{F}[x]$, $\hat{U}(t_\mathrm{f})$};
\draw[dataarrow] (dynamics) -- (output);

\coordinate (branch) at ($(model.east)!0.4!(pulse.west)$);
\coordinate (merge) at ($(pulse.east)!0.6!(dynamics.west)$);
\draw[arrow] (model.east) -- (pulse.west);
\fill[navyblue] (branch) circle (1.4pt);
\node[flowlabel, anchor=north] at ([yshift=-2pt]$(model.east)!0.5!(pulse.west)$) {$x_\mathrm{gq}$};
\draw[arrow] (pulse.east) -- (dynamics.west);
\fill[navyblue] (merge) circle (1.4pt);
\node[flowlabel, anchor=north] at ([yshift=-2pt]$(pulse.east)!0.5!(dynamics.west)$) {$x^\mathrm{adj}_\mathrm{gq}$};

\draw[bypass] (branch) -- ([yshift=0.4cm]branch |- pulse.north) coordinate (skipL)
    -- node[flowlabel, anchor=south] {skip PulseControl: $x_\mathrm{gq}(t)$ directly to Dynamics}
       (skipL -| merge) -- ([yshift=4pt]merge);

\draw[feedback] (dynamics.south) -- ([yshift=-0.4cm]dynamics.south) coordinate (fbR)
    -- node[flowlabel, anchor=north] {fidelity feedback: new exponents or settings}
       (fbR -| model.south) -- (model.south);
\end{tikzpicture}

%% file: Extras/lz_output.tex
\newcommand{\notebookname}{landau\_zener.ipynb}%
\newcommand{\outputimage}{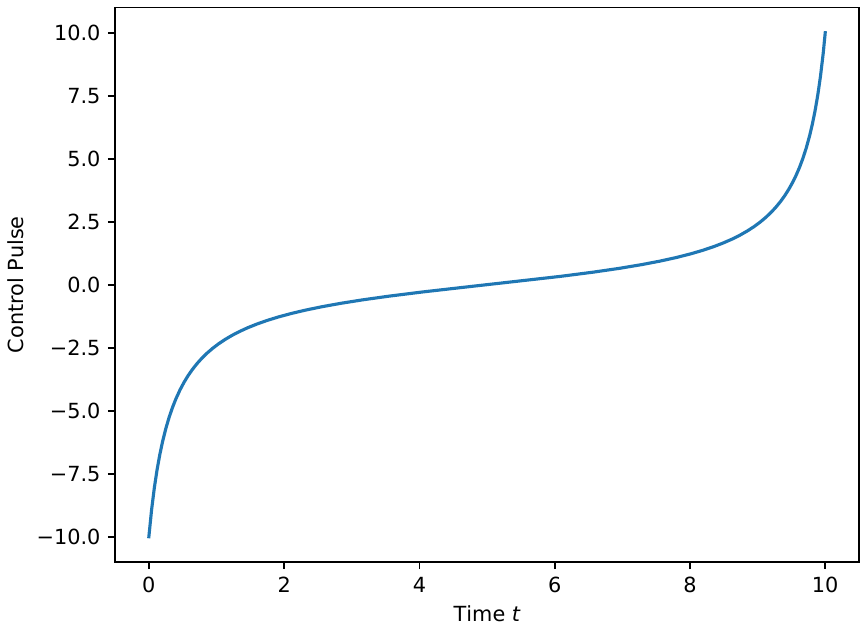}%
\ifdefined\outputwidth\else\newlength{\outputwidth}\fi%
\setlength{\outputwidth}{299.5pt}%
\newcommand{\notebookcode}{%
    pulse = model.synthesize\_pulse(duration=\nbnum{10})\\
    fig, ax = pulse.plot\_pulse()%
}%
\input{Extras/notebook_window}%

%% file: Extras/dqd_energies.tex
\newcommand{\notebookname}{dqd\_model.ipynb}%
\newcommand{\outputimage}{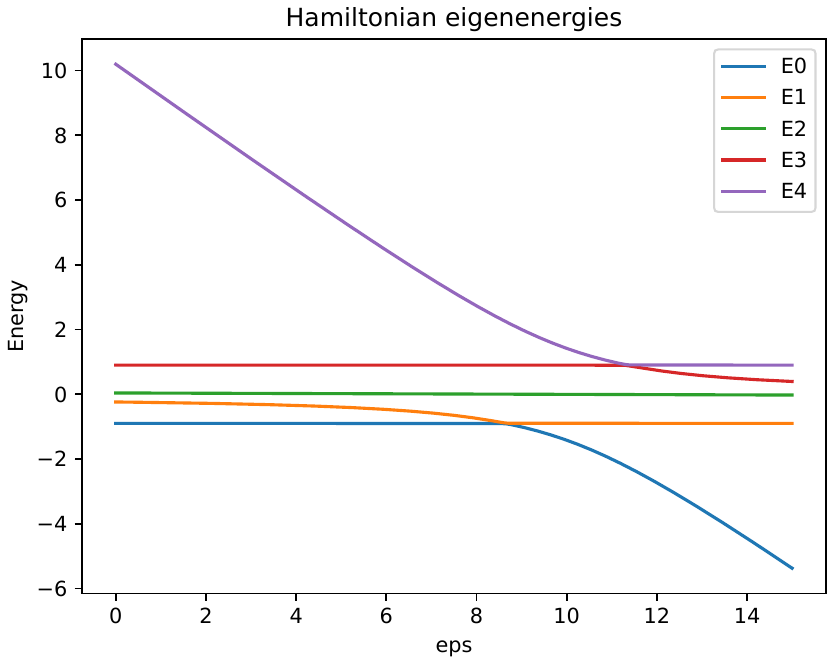}%
\ifdefined\outputwidth\else\newlength{\outputwidth}\fi%
\setlength{\outputwidth}{288.0pt}%
\newcommand{\notebookcode}{%
    dqd\_model = gq.ControlModel(dqd\_hamiltonian)\\
    dqd\_model.set\_parameters(\\
    \hspace*{2em}U=\nbnum{10}, tc=\nbnum{1}, Ez=\nbnum{0.9}, dEz=\nbnum{0.1}, dEx=\nbnum{0.01}\\
    )\\
    dqd\_model.set\_control(\\
    \hspace*{2em}control\_name=\nbstr{\textquotedbl eps\textquotedbl},\\
    \hspace*{2em}pulse\_initial=\nbnum{15}, pulse\_final=\nbnum{0},\\
    \hspace*{2em}initial\_state=\nbnum{0}, alpha=\nbnum{2}, beta=\nbnum{2},\\
    )\\
    dqd\_model.plot\_eigenvalues()%
}%
\input{Extras/notebook_window}%

%% file: main.bbl
\begin{thebibliography}{100}
\providecommand{\url}[1]{\texttt{#1}}
\providecommand{\urlprefix}{URL }
\expandafter\ifx\csname urlstyle\endcsname\relax
  \providecommand{\doi}[1]{doi:\discretionary{}{}{}#1}\else
  \providecommand{\doi}{doi:\discretionary{}{}{}\begingroup
  \urlstyle{rm}\Url}\fi
\providecommand{\eprint}[2][]{\url{#2}}

\bibitem{glaserTrainingSchrodingersCat2015}
S.~J. Glaser, U.~Boscain, T.~Calarco, C.~P. Koch, W.~K{\"o}ckenberger,
  R.~Kosloff, I.~Kuprov, B.~Luy, S.~Schirmer, T.~{Schulte-Herbr{\"u}ggen},
  D.~Sugny and F.~K. Wilhelm,
\newblock \emph{Training {{Schr\"odinger}}'s cat: Quantum optimal control},
\newblock Eur. Phys. J. D \textbf{69}(12), 279 (2015),
\newblock \doi{10.1140/epjd/e2015-60464-1}.

\bibitem{izmalkovObservationMacroscopicLandauZener2004}
A.~Izmalkov, M.~Grajcar, E.~Il'ichev, N.~Oukhanski, T.~Wagner, H.-G. Meyer,
  W.~Krech, M.~H.~S. Amin, A.~M. van~den Brink and A.~M. Zagoskin,
\newblock \emph{Observation of macroscopic {{Landau--Zener}} transitions in a
  superconducting device},
\newblock EPL \textbf{65}(6), 844 (2004),
\newblock \doi{10.1209/epl/i2003-10200-6}.

\bibitem{berkeTransmonPlatformQuantum2022}
C.~Berke, E.~Varvelis, S.~Trebst, A.~Altland and D.~P. DiVincenzo,
\newblock \emph{Transmon platform for quantum computing challenged by chaotic
  fluctuations},
\newblock Nat. Commun. \textbf{13}(1), 2495 (2022),
\newblock \doi{10.1038/s41467-022-29940-y}.

\bibitem{fernandez-fernandezFlyingSpinQubits2024}
D.~{Fern{\'a}ndez-Fern{\'a}ndez}, Y.~Ban and G.~Platero,
\newblock \emph{Flying {{Spin Qubits}} in {{Quantum Dot Arrays Driven}} by
  {{Spin-Orbit Interaction}}},
\newblock Quantum \textbf{8}, 1533 (2024),
\newblock \doi{10.22331/q-2024-11-21-1533}.

\bibitem{fernandez-fernandezLongrangeTransportAsymmetric2025}
D.~{Fern{\'a}ndez-Fern{\'a}ndez}, J.~C. Bayer, R.~J. Haug and G.~Platero,
\newblock \emph{Long-range transport in asymmetric quadruple quantum dot
  configurations},
\newblock Commun. Phys. \textbf{8}(1), 410 (2025),
\newblock \doi{10.1038/s42005-025-02319-3}.

\bibitem{fernandez-fernandezSpinorbitenabledRealizationArbitrary2025}
D.~{Fern{\'a}ndez-Fern{\'a}ndez}, Y.~Matsumoto, L.~M.~K. Vandersypen,
  G.~Platero and S.~Bosco,
\newblock \emph{Spin-orbit-enabled realization of arbitrary two-qubit gates on
  moving spins} (2025), \eprint{2508.08394}.

\bibitem{farinaSiteresolvedMagnonTriplon2025}
P.~{Cova Fari{\~n}a}, D.~Jirovec, X.~Zhang, E.~Morozova, S.~D. Oosterhout,
  S.~Reale, T.-K. Hsiao, G.~Scappucci, M.~Veldhorst and L.~M.~K. Vandersypen,
\newblock \emph{Site-resolved magnon and triplon dynamics on a programmable
  quantum dot spin ladder} (2025), \eprint{2506.08663}.

\bibitem{jirovecManybodyInterferometrySemiconductor2026}
D.~Jirovec, S.~Reale, P.~{Cova Fari{\~n}a}, C.~{Ventura-Meinersen}, M.~P.
  Nguyen, X.~Zhang, S.~D. Oosterhout, G.~Scappucci, M.~Veldhorst,
  M.~{Rimbach-Russ}, S.~Bosco and L.~M.~K. Vandersypen,
\newblock \emph{Many-body interferometry with semiconductor spins},
\newblock Science \textbf{392}(6794), 183 (2026),
\newblock \doi{10.1126/science.aed4177}.

\bibitem{karchDynamicalPreparationU12026}
S.~Karch, M.~Will, I.~P. Rodriguez, N.~Liebster, S.~Huh, M.~Knap, F.~Pollmann,
  C.~Kuhlenkamp, I.~Bloch and M.~Aidelsburger,
\newblock \emph{Dynamical preparation of {{U}}(1) quantum spin liquids in an
  analogue quantum simulator} (2026), \eprint{2604.24744}.

\bibitem{bernienProbingManybodyDynamics2017}
H.~Bernien, S.~Schwartz, A.~Keesling, H.~Levine, A.~Omran, H.~Pichler, S.~Choi,
  A.~S. Zibrov, M.~Endres, M.~Greiner, V.~Vuleti{\'c} and M.~D. Lukin,
\newblock \emph{Probing many-body dynamics on a 51-atom quantum simulator},
\newblock Nature \textbf{551}(7682), 579 (2017),
\newblock \doi{10.1038/nature24622}.

\bibitem{khanejaTimeOptimalControl2001}
N.~Khaneja, R.~Brockett and S.~J. Glaser,
\newblock \emph{Time optimal control in spin systems},
\newblock Phys. Rev. A \textbf{63}(3), 032308 (2001),
\newblock \doi{10.1103/PhysRevA.63.032308}.

\bibitem{carliniTimeOptimalQuantumEvolution2006}
A.~Carlini, A.~Hosoya, T.~Koike and Y.~Okudaira,
\newblock \emph{Time-{{Optimal Quantum Evolution}}},
\newblock Phys. Rev. Lett. \textbf{96}(6), 060503 (2006),
\newblock \doi{10.1103/PhysRevLett.96.060503}.

\bibitem{pontryaginMathematicalTheoryOptimal2017}
L.~S. Pontryagin, V.~G. Boltyanskii, R.~V. Gamkrelidze and E.~F. Mishchenko,
\newblock \emph{Mathematical {{Theory}} of {{Optimal Processes}}},
\newblock Interscience, New York (1962).

\bibitem{carliniTimeoptimalUnitaryOperations2007}
A.~Carlini, A.~Hosoya, T.~Koike and Y.~Okudaira,
\newblock \emph{Time-optimal unitary operations},
\newblock Phys. Rev. A \textbf{75}(4), 042308 (2007),
\newblock \doi{10.1103/PhysRevA.75.042308}.

\bibitem{boscainIntroductionPontryaginMaximum2021}
U.~Boscain, M.~Sigalotti and D.~Sugny,
\newblock \emph{Introduction to the {{Pontryagin Maximum Principle}} for
  {{Quantum Optimal Control}}},
\newblock PRX Quantum \textbf{2}(3), 030203 (2021),
\newblock \doi{10.1103/PRXQuantum.2.030203}.

\bibitem{koikeQuantumBrachistochrone2022}
T.~Koike,
\newblock \emph{Quantum brachistochrone},
\newblock Phil. Trans. R. Soc. A \textbf{380}(2239), 20210273 (2022),
\newblock \doi{10.1098/rsta.2021.0273}.

\bibitem{rezakhaniQuantumAdiabaticBrachistochrone2009}
A.~T. Rezakhani, W.-J. Kuo, A.~Hamma, D.~A. Lidar and P.~Zanardi,
\newblock \emph{Quantum {{Adiabatic Brachistochrone}}},
\newblock Phys. Rev. Lett. \textbf{103}(8), 080502 (2009),
\newblock \doi{10.1103/PhysRevLett.103.080502}.

\bibitem{stepanenkoTimeOptimalTransferQuantum2025}
A.~A. Stepanenko, K.~S. Chernova and M.~A. Gorlach,
\newblock \emph{Time-{{Optimal Transfer}} of the {{Quantum State}} in {{Long
  Qubit Arrays}}},
\newblock Phys. Rev. Lett. \textbf{135}(1), 010803 (2025),
\newblock \doi{10.1103/mxrq-7rl4}.

\bibitem{khanejaOptimalControlCoupled2005}
N.~Khaneja, T.~Reiss, C.~Kehlet, T.~{Schulte-Herbr{\"u}ggen} and S.~J. Glaser,
\newblock \emph{Optimal control of coupled spin dynamics: Design of {{NMR}}
  pulse sequences by gradient ascent algorithms},
\newblock J. Magn. Reson. \textbf{172}(2), 296 (2005),
\newblock \doi{10.1016/j.jmr.2004.11.004}.

\bibitem{goodwinAcceleratedNewtonRaphsonGRAPE2023}
D.~L. Goodwin and M.~S. Vinding,
\newblock \emph{Accelerated {{Newton--Raphson GRAPE}} methods for optimal
  control},
\newblock Phys. Rev. Res. \textbf{5}(1), L012042 (2023),
\newblock \doi{10.1103/PhysRevResearch.5.L012042}.

\bibitem{doriaOptimalControlTechnique2011}
P.~Doria, T.~Calarco and S.~Montangero,
\newblock \emph{Optimal control technique for many-body quantum dynamics},
\newblock Phys. Rev. Lett. \textbf{106}(19), 190501 (2011),
\newblock \doi{10.1103/PhysRevLett.106.190501}.

\bibitem{canevaChoppedRandombasisQuantum2011}
T.~Caneva, T.~Calarco and S.~Montangero,
\newblock \emph{Chopped random-basis quantum optimization},
\newblock Phys. Rev. A \textbf{84}(2), 022326 (2011),
\newblock \doi{10.1103/PhysRevA.84.022326}.

\bibitem{fauquenotOpenClosedLoop2025}
S.~Fauquenot, A.~Sarkar and S.~Feld,
\newblock \emph{Open and {{Closed Loop Approaches}} for {{Energy Efficient
  Quantum Optimal Control}}},
\newblock Adv. Quantum Technol. \textbf{8}(10), 2400690 (2025),
\newblock \doi{10.1002/qute.202400690}.

\bibitem{katiraee-farUnifiedEvolutionaryOptimization2025}
S.~R. {Katiraee-Far}, Y.~Matsumoto, B.~Undseth, M.~{De Smet}, V.~Gualtieri,
  C.~{Ventura-Meinersen}, I.~{Fern{\'a}ndez de Fuentes}, K.~Capannelli,
  M.~{Rimbach-Russ}, G.~Scappucci, L.~M.~K. Vandersypen and E.~Greplova,
\newblock \emph{Unified evolutionary optimization for high-fidelity spin qubit
  operations} (2025), \eprint{2503.12256}.

\bibitem{martinisFastAdiabaticQubit2014}
J.~M. Martinis and M.~R. Geller,
\newblock \emph{Fast adiabatic qubit gates using only {$\sigma_z$} control},
\newblock Phys. Rev. A \textbf{90}(2), 022307 (2014),
\newblock \doi{10.1103/PhysRevA.90.022307}.

\bibitem{rimbach-russSimpleFrameworkSystematic2023}
M.~{Rimbach-Russ}, S.~G.~J. Philips, X.~Xue and L.~M.~K. Vandersypen,
\newblock \emph{Simple framework for systematic high-fidelity gate operations},
\newblock Quantum Sci. Technol. \textbf{8}(4), 045025 (2023),
\newblock \doi{10.1088/2058-9565/acf786}.

\bibitem{wuSimultaneousHighFidelitySingleQubit2025}
Y.-H. Wu, L.~C. Camenzind, P.~B{\"u}tler, I.~K. Jin, A.~Noiri, K.~Takeda,
  T.~Nakajima, T.~Kobayashi, G.~Scappucci, H.-S. Goan and S.~Tarucha,
\newblock \emph{Simultaneous {{High-Fidelity Single-Qubit Gates}} in a {{Spin
  Qubit Array}}},
\newblock Nat. Phys.  (2026),
\newblock \doi{10.1038/s41567-026-03466-2}.

\bibitem{polatPulseShapingUltrafast2026}
{\.I}.~Polat, R.~W.~J. Overwater, M.~{Rimbach-Russ} and F.~Sebastiano,
\newblock \emph{Pulse shaping for ultra-fast adiabatic quantum gates},
\newblock npj Quantum Inf. \textbf{12}(1), 130 (2026),
\newblock \doi{10.1038/s41534-026-01245-8}.

\bibitem{bergmannCoherentPopulationTransfer1998}
K.~Bergmann, H.~Theuer and B.~W. Shore,
\newblock \emph{Coherent population transfer among quantum states of atoms and
  molecules},
\newblock Rev. Mod. Phys. \textbf{70}(3), 1003 (1998),
\newblock \doi{10.1103/RevModPhys.70.1003}.

\bibitem{motzoiSimplePulsesElimination2009}
F.~Motzoi, J.~M. Gambetta, P.~Rebentrost and F.~K. Wilhelm,
\newblock \emph{Simple {{Pulses}} for {{Elimination}} of {{Leakage}} in
  {{Weakly Nonlinear Qubits}}},
\newblock Phys. Rev. Lett. \textbf{103}(11), 110501 (2009),
\newblock \doi{10.1103/PhysRevLett.103.110501}.

\bibitem{berryTransitionlessQuantumDriving2009}
M.~V. Berry,
\newblock \emph{Transitionless quantum driving},
\newblock J. Phys. A: Math. Theor. \textbf{42}(36), 365303 (2009),
\newblock \doi{10.1088/1751-8113/42/36/365303}.

\bibitem{degrandiAdiabaticPerturbationTheory2010}
C.~De~Grandi and A.~Polkovnikov,
\newblock \emph{Adiabatic perturbation theory: From {{Landau--Zener}} problem
  to quenching through a quantum critical point},
\newblock In \emph{Quantum {{Quenching}}, {{Annealing}} and {{Computation}}},
  vol. 802 of \emph{Lecture {{Notes}} in {{Physics}}}, pp. 75--114. Springer,
  Berlin, Heidelberg,
\newblock \doi{10.1007/978-3-642-11470-0_4} (2010).

\bibitem{ivakhnenkoNonadiabaticLandauZener2023}
O.~V. Ivakhnenko, S.~N. Shevchenko and F.~Nori,
\newblock \emph{Nonadiabatic
  {{Landau}}--{{Zener}}--{{St\"uckelberg}}--{{Majorana}} transitions, dynamics,
  and interference},
\newblock Phys. Rep. \textbf{995}, 1 (2023),
\newblock \doi{10.1016/j.physrep.2022.10.002}.

\bibitem{vuthaSimpleApproachLandauZener2010}
A.~C. Vutha,
\newblock \emph{A simple approach to the {{Landau--Zener}} formula},
\newblock Eur. J. Phys. \textbf{31}(2), 389 (2010),
\newblock \doi{10.1088/0143-0807/31/2/016},
\newblock \eprint{1001.3322}.

\bibitem{motzoiOptimalControlMethods2011}
F.~Motzoi, J.~M. Gambetta, S.~T. Merkel and F.~K. Wilhelm,
\newblock \emph{Optimal control methods for rapidly time-varying
  {{Hamiltonians}}},
\newblock Phys. Rev. A \textbf{84}(2), 022307 (2011),
\newblock \doi{10.1103/PhysRevA.84.022307}.

\bibitem{chenLewisRiesenfeldInvariantsTransitionless2011}
X.~Chen, E.~Torrontegui and J.~G. Muga,
\newblock \emph{{{Lewis--Riesenfeld}} invariants and transitionless quantum
  driving},
\newblock Phys. Rev. A \textbf{83}(6), 062116 (2011),
\newblock \doi{10.1103/PhysRevA.83.062116}.

\bibitem{ribeiroSystematicMagnusBasedApproach2017}
H.~Ribeiro, A.~Baksic and A.~A. Clerk,
\newblock \emph{Systematic {{Magnus-Based Approach}} for {{Suppressing
  Leakage}} and {{Nonadiabatic Errors}} in {{Quantum Dynamics}}},
\newblock Phys. Rev. X \textbf{7}(1), 011021 (2017),
\newblock \doi{10.1103/PhysRevX.7.011021}.

\bibitem{selsMinimizingIrreversibleLosses2017}
D.~Sels and A.~Polkovnikov,
\newblock \emph{Minimizing irreversible losses in quantum systems by local
  counterdiabatic driving},
\newblock Proc. Natl. Acad. Sci. U.S.A. \textbf{114}(20), E3909 (2017),
\newblock \doi{10.1073/pnas.1619826114}.

\bibitem{theisCounteractingSystemsDiabaticities2018}
L.~S. Theis, F.~Motzoi, S.~Machnes and F.~K. Wilhelm,
\newblock \emph{Counteracting systems of diabaticities using {{DRAG}} controls:
  {{The}} status after 10 years},
\newblock EPL \textbf{123}(6), 60001 (2018),
\newblock \doi{10.1209/0295-5075/123/60001}.

\bibitem{banFastLongrangeCharge2018}
Y.~Ban, X.~Chen and G.~Platero,
\newblock \emph{Fast long-range charge transfer in quantum dot arrays},
\newblock Nanotechnology \textbf{29}(50), 505201 (2018),
\newblock \doi{10.1088/1361-6528/aae0ce}.

\bibitem{takahashiHamiltonianEngineeringAdiabatic2019}
K.~Takahashi,
\newblock \emph{Hamiltonian {{Engineering}} for {{Adiabatic Quantum
  Computation}}: {{Lessons}} from {{Shortcuts}} to {{Adiabaticity}}},
\newblock J. Phys. Soc. Jpn. \textbf{88}(6), 061002 (2019),
\newblock \doi{10.7566/JPSJ.88.061002}.

\bibitem{banSpinEntangledState2019}
Y.~Ban, X.~Chen, S.~Kohler and G.~Platero,
\newblock \emph{Spin {{Entangled State Transfer}} in {{Quantum Dot Arrays}}:
  {{Coherent Adiabatic}} and {{Speed-Up Protocols}}},
\newblock Adv. Quantum Technol. \textbf{2}(10), 1900048 (2019),
\newblock \doi{10.1002/qute.201900048}.

\bibitem{guery-odelinShortcutsAdiabaticityConcepts2019}
D.~{Gu{\'e}ry-Odelin}, A.~Ruschhaupt, A.~Kiely, E.~Torrontegui,
  S.~{Mart{\'i}nez-Garaot} and J.~G. Muga,
\newblock \emph{Shortcuts to adiabaticity: {{Concepts}}, methods, and
  applications},
\newblock Rev. Mod. Phys. \textbf{91}(4), 045001 (2019),
\newblock \doi{10.1103/RevModPhys.91.045001}.

\bibitem{setiawanAnalyticDesignAccelerated2021}
F.~Setiawan, P.~Groszkowski, H.~Ribeiro and A.~A. Clerk,
\newblock \emph{Analytic {{Design}} of {{Accelerated Adiabatic Gates}} in
  {{Realistic Qubits}}: {{General Theory}} and {{Applications}} to
  {{Superconducting Circuits}}},
\newblock PRX Quantum \textbf{2}(3), 030306 (2021),
\newblock \doi{10.1103/PRXQuantum.2.030306}.

\bibitem{takahashiDynamicalInvariantFormalism2022}
K.~Takahashi,
\newblock \emph{Dynamical invariant formalism of shortcuts to adiabaticity},
\newblock Phil. Trans. R. Soc. A \textbf{380}(2239), 20220301 (2022),
\newblock \doi{10.1098/rsta.2022.0301}.

\bibitem{glasbrennerLandauZenerFormula2023}
E.~P. Glasbrenner and W.~P. Schleich,
\newblock \emph{The {{Landau}}--{{Zener}} formula made simple},
\newblock J. Phys. B: At. Mol. Opt. Phys. \textbf{56}(10), 104001 (2023),
\newblock \doi{10.1088/1361-6455/acc774}.

\bibitem{dengisAcceleratedCreationNOON2025}
S.~Dengis, S.~Wimberger and P.~Schlagheck,
\newblock \emph{Accelerated creation of {{NOON}} states with ultracold atoms
  via counterdiabatic driving},
\newblock Phys. Rev. A \textbf{111}(3), L031301 (2025),
\newblock \doi{10.1103/PhysRevA.111.L031301}.

\bibitem{romeroOptimizingEdgestateTransfer2024}
S.~V. Romero, X.~Chen, G.~Platero and Y.~Ban,
\newblock \emph{Optimizing edge-state transfer in a {{Su--Schrieffer--Heeger}}
  chain via hybrid analog-digital strategies},
\newblock Phys. Rev. Appl. \textbf{21}(3), 034033 (2024),
\newblock \doi{10.1103/PhysRevApplied.21.034033}.

\bibitem{liuAcceleratedAdiabaticPassage2024}
X.-F. Liu, Y.~Matsumoto, T.~Fujita, A.~Ludwig, A.~D. Wieck and A.~Oiwa,
\newblock \emph{Accelerated adiabatic passage of a single electron spin qubit
  in quantum dots},
\newblock Phys. Rev. Lett. \textbf{132}(2), 027002 (2024),
\newblock \doi{10.1103/PhysRevLett.132.027002}.

\bibitem{xuImprovingCoherentPopulation2019}
J.~Xu, Y.-X. Du and W.~Huang,
\newblock \emph{Improving coherent population transfer via a stricter adiabatic
  condition},
\newblock Phys. Rev. A \textbf{100}(2), 023848 (2019),
\newblock \doi{10.1103/PhysRevA.100.023848}.

\bibitem{fehseGeneralizedFastQuasiadiabatic2023}
F.~Fehse, M.~David, M.~{Pioro-Ladri{\`e}re} and W.~A. Coish,
\newblock \emph{Generalized fast quasiadiabatic population transfer for
  improved qubit readout, shuttling, and noise mitigation},
\newblock Phys. Rev. B \textbf{107}(24), 245303 (2023),
\newblock \doi{10.1103/PhysRevB.107.245303}.

\bibitem{limaPartialLandauZenerTransitions2025}
J.~R.~F. Lima and G.~Burkard,
\newblock \emph{Partial {{Landau--Zener}} transitions and applications to qubit
  shuttling},
\newblock Phys. Rev. B \textbf{111}(23), 235439 (2025),
\newblock \doi{10.1103/y4xf-zjjx}.

\bibitem{richermeExperimentalPerformanceQuantum2013}
P.~Richerme, C.~Senko, J.~Smith, A.~Lee, S.~Korenblit and C.~Monroe,
\newblock \emph{Experimental performance of a quantum simulator: {{Optimizing}}
  adiabatic evolution and identifying many-body ground states},
\newblock Phys. Rev. A \textbf{88}(1), 012334 (2013),
\newblock \doi{10.1103/PhysRevA.88.012334}.

\bibitem{rolandQuantumSearchLocal2002}
J.~Roland and N.~J. Cerf,
\newblock \emph{Quantum search by local adiabatic evolution},
\newblock Phys. Rev. A \textbf{65}(4), 042308 (2002),
\newblock \doi{10.1103/PhysRevA.65.042308}.

\bibitem{martinez-garaotFastQuasiadiabaticDynamics2015}
S.~{Mart{\'i}nez-Garaot}, A.~Ruschhaupt, J.~Gillet, {\relax Th}.~Busch and
  J.~G. Muga,
\newblock \emph{Fast quasiadiabatic dynamics},
\newblock Phys. Rev. A \textbf{92}(4), 043406 (2015),
\newblock \doi{10.1103/PhysRevA.92.043406}.

\bibitem{baksicSpeedingAdiabaticQuantum2016}
A.~Baksic, H.~Ribeiro and A.~A. Clerk,
\newblock \emph{Speeding up {{Adiabatic Quantum State Transfer}} by {{Using
  Dressed States}}},
\newblock Phys. Rev. Lett. \textbf{116}(23), 230503 (2016),
\newblock \doi{10.1103/PhysRevLett.116.230503}.

\bibitem{stepanenkoQuantumAdvantageTransfer2026}
A.~Stepanenko, K.~Chernova and M.~Gorlach,
\newblock \emph{Quantum advantage in transfer of quantum states} (2026),
  \eprint{2604.05915}.

\bibitem{tangExploringGroundStates2024}
J.~Tang, R.~Xu, Y.~Ding, X.~Xu, Y.~Ban, M.-H. Yung, A.~{P{\'e}rez-Obiol},
  G.~Platero and X.~Chen,
\newblock \emph{Exploring ground states of {{Fermi--Hubbard}} model on
  honeycomb lattices with counterdiabaticity},
\newblock npj Quantum Mater. \textbf{9}(1), 87 (2024),
\newblock \doi{10.1038/s41535-024-00697-5}.

\bibitem{tangWeightedNestedCommutators2026}
J.~Tang, X.~Chen and Z.-Y. Wei,
\newblock \emph{Weighted {{Nested Commutators}} for {{Scalable Counterdiabatic
  State Preparation}}} (2026), \eprint{2603.25625}.

\bibitem{goerzKrotovPythonImplementation2019}
M.~H. Goerz, D.~Basilewitsch, F.~{Gago-Encinas}, M.~G. Krauss, K.~P. Horn,
  D.~M. Reich and C.~P. Koch,
\newblock \emph{Krotov: {{A Python}} implementation of {{Krotov}}'s method for
  quantum optimal control},
\newblock SciPost Phys. \textbf{7}(6), 080 (2019),
\newblock \doi{10.21468/SciPostPhys.7.6.080}.

\bibitem{rossignoloQuOCSQuantumOptimal2023}
M.~Rossignolo, T.~Reisser, A.~Marshall, P.~Rembold, A.~Pagano, P.~J. Vetter,
  R.~S. Said, M.~M. M{\"u}ller, F.~Motzoi, T.~Calarco, F.~Jelezko and
  S.~Montangero,
\newblock \emph{{{QuOCS}}: {{The}} quantum optimal control suite},
\newblock Comput. Phys. Commun. \textbf{291}, 108782 (2023),
\newblock \doi{10.1016/j.cpc.2023.108782}.

\bibitem{wittlerIntegratedToolSet2021}
N.~Wittler, F.~Roy, K.~Pack, M.~Werninghaus, A.~S. Roy, D.~J. Egger, S.~Filipp,
  F.~K. Wilhelm and S.~Machnes,
\newblock \emph{Integrated tool set for control, calibration, and
  characterization of quantum devices applied to superconducting qubits},
\newblock Phys. Rev. Appl. \textbf{15}(3), 034080 (2021),
\newblock \doi{10.1103/PhysRevApplied.15.034080}.

\bibitem{puzzuoliQiskitDynamicsPython2023}
D.~Puzzuoli, C.~J. Wood, D.~J. Egger, B.~Rosand and K.~Ueda,
\newblock \emph{Qiskit {{Dynamics}}: {{A Python}} package for simulating the
  time dynamics of quantum systems},
\newblock J. Open Source Softw. \textbf{8}(90), 5853 (2023),
\newblock \doi{10.21105/joss.05853}.

\bibitem{lambertQuTiP5Quantum2026}
N.~Lambert, E.~Gigu{\`e}re, P.~Menczel, B.~Li, P.~Hopf, G.~Su{\'a}rez, M.~Gali,
  J.~Lishman, R.~Gadhvi, R.~Agarwal, A.~Galicia, N.~Shammah \emph{et~al.},
\newblock \emph{{{QuTiP}} 5: {{The Quantum Toolbox}} in {{Python}}},
\newblock Phys. Rep. \textbf{1153}, 1 (2026),
\newblock \doi{10.1016/j.physrep.2025.10.001}.

\bibitem{chenSpeedingQuantumAdiabatic2022}
J.-F. Chen,
\newblock \emph{Speeding up quantum adiabatic processes with a dynamical
  quantum geometric tensor},
\newblock Phys. Rev. Res. \textbf{4}(2), 023252 (2022),
\newblock \doi{10.1103/PhysRevResearch.4.023252}.

\bibitem{ventura-meinersenQuantumGeometricProtocols2025}
C.~{Ventura-Meinersen}, S.~Bosco and M.~{Rimbach-Russ},
\newblock \emph{Quantum geometric protocols for fast high-fidelity adiabatic
  state transfer},
\newblock EPJ Quantum Technol. \textbf{12}(1), 125 (2025),
\newblock \doi{10.1140/epjqt/s40507-025-00426-2}.

\bibitem{meinersenUnifyingAdiabaticStatetransfer2025}
C.~{Ventura-Meinersen}, D.~{Fern{\'a}ndez-Fern{\'a}ndez}, G.~Platero and
  M.~{Rimbach-Russ},
\newblock \emph{Unifying adiabatic state-transfer protocols with ({$\alpha$},
  {$\beta$})-hypergeometries} (2025), \eprint{2504.08031}.

\bibitem{ventura-meinersenMultilevelSpectralNavigation2026a}
C.~{Ventura-Meinersen}, E.~Valvo, S.~Bosco and M.~{Rimbach-Russ},
\newblock \emph{Multilevel spectral navigation with geometric
  diabatic-adiabatic control},
\newblock Phys. Rev. Res. \textbf{8}(3), L032034 (2026),
\newblock \doi{10.1103/hfv7-3pxt}.

\bibitem{dengisTimeindependentCounterdiabaticDriving2026}
S.~Dengis and P.~Schlagheck,
\newblock \emph{Time-independent counterdiabatic driving for emergent two-level
  subspaces in many-body systems},
\newblock Phys. Rev. A  (2026),
\newblock \doi{10.1103/vscj-25yr}.

\bibitem{kolodrubetzClassifyingMeasuringGeometry2013}
M.~Kolodrubetz, V.~Gritsev and A.~Polkovnikov,
\newblock \emph{Classifying and measuring geometry of a quantum ground state
  manifold},
\newblock Phys. Rev. B \textbf{88}(6), 064304 (2013),
\newblock \doi{10.1103/PhysRevB.88.064304}.

\bibitem{houLocalGeometryQuantum2024}
X.-Y. Hou, Z.~Zhou, X.~Wang, H.~Guo and C.-C. Chien,
\newblock \emph{Local geometry and quantum geometric tensor of mixed states},
\newblock Phys. Rev. B \textbf{110}(3), 035144 (2024),
\newblock \doi{10.1103/PhysRevB.110.035144}.

\bibitem{liskaHiddenSymmetriesBianchi2021}
D.~Liska and V.~Gritsev,
\newblock \emph{Hidden symmetries, the {{Bianchi}} classification and geodesics
  of the quantum geometric ground-state manifolds},
\newblock SciPost Phys. \textbf{10}(1), 020 (2021),
\newblock \doi{10.21468/SciPostPhys.10.1.020}.

\bibitem{sidhuGeometricPerspectiveQuantum2020}
J.~S. Sidhu and P.~Kok,
\newblock \emph{Geometric perspective on quantum parameter estimation},
\newblock AVS Quantum Sci. \textbf{2}(1), 014701 (2020),
\newblock \doi{10.1116/1.5119961}.

\bibitem{valvoProbingElectrostaticDisorder2026}
E.~Valvo, C.~{Ventura-Meinersen}, M.~Jakob, S.~Bosco, T.~Vakhtel and
  M.~{Rimbach-Russ},
\newblock \emph{Probing {{Electrostatic Disorder}} via {$g$}-{{Tensor
  Geometry}}} (2026), \eprint{2604.09475}.

\bibitem{lambertClassicalQuantumInformation2023}
J.~Lambert and E.~S. S{\o}rensen,
\newblock \emph{From classical to quantum information geometry: A guide for
  physicists},
\newblock New J. Phys. \textbf{25}(8), 081201 (2023),
\newblock \doi{10.1088/1367-2630/aceb14}.

\bibitem{camposvenutiQuantumCriticalScaling2007}
L.~Campos~Venuti and P.~Zanardi,
\newblock \emph{Quantum {{Critical Scaling}} of the {{Geometric Tensors}}},
\newblock Phys. Rev. Lett. \textbf{99}(9), 095701 (2007),
\newblock \doi{10.1103/PhysRevLett.99.095701}.

\bibitem{ventura-meinersenGeometricSignaturesOnset2026}
C.~{Ventura-Meinersen}, E.~Valvo, S.~Bosco, F.~Machado and M.~{Rimbach-Russ},
\newblock \emph{Geometric signatures of the onset of many-body ergodicity}
  (2026), \eprint{2608.14818}.

\bibitem{dashEfficiencyNeuralQuantum2025}
S.~Dash, L.~Gravina, F.~Vicentini, M.~Ferrero and A.~Georges,
\newblock \emph{Efficiency of neural quantum states in light of the quantum
  geometric tensor},
\newblock Commun. Phys. \textbf{8}(1), 92 (2025),
\newblock \doi{10.1038/s42005-025-02005-4}.

\bibitem{dawidMachineLearningQuantum2025}
A.~Dawid, J.~Arnold, B.~Requena, A.~Gresch, M.~P{\l}odzie{\'n}, K.~Donatella,
  K.~A. Nicoli, P.~Stornati, R.~Koch, M.~B{\"u}ttner, R.~Oku{\l}a,
  G.~{Mu{\~n}oz-Gil} \emph{et~al.},
\newblock \emph{Machine {{Learning}} in {{Quantum Sciences}}},
\newblock Cambridge University Press, Cambridge,
\newblock ISBN 978-1-009-50493-5,
\newblock \doi{10.1017/9781009504942} (2025).

\bibitem{albashAdiabaticQuantumComputation2018}
T.~Albash and D.~A. Lidar,
\newblock \emph{Adiabatic quantum computation},
\newblock Rev. Mod. Phys. \textbf{90}(1), 015002 (2018),
\newblock \doi{10.1103/RevModPhys.90.015002}.

\bibitem{demirplakAdiabaticPopulationTransfer2003}
M.~Demirplak and S.~A. Rice,
\newblock \emph{Adiabatic population transfer with control fields},
\newblock J. Phys. Chem. A \textbf{107}(46), 9937 (2003),
\newblock \doi{10.1021/jp030708a}.

\bibitem{fernandez-fernandezQuantumControlHole2022}
D.~{Fern{\'a}ndez-Fern{\'a}ndez}, Y.~Ban and G.~Platero,
\newblock \emph{Quantum {{Control}} of {{Hole Spin Qubits}} in {{Double Quantum
  Dots}}},
\newblock Phys. Rev. Appl. \textbf{18}(5), 054090 (2022),
\newblock \doi{10.1103/PhysRevApplied.18.054090}.

\bibitem{palmeroGenerationCatStates2019}
M.~Palmero, M.~{\'A}. Sim{\'o}n and D.~Poletti,
\newblock \emph{Towards generation of cat states in trapped ions set-ups via
  {{FAQUAD}} protocols and dynamical decoupling},
\newblock Entropy \textbf{21}(12), 1207 (2019),
\newblock \doi{10.3390/e21121207}.

\bibitem{garcia-ripollQuantumControlFrequencyTunable2020}
J.~J. {Garc{\'i}a-Ripoll}, A.~{Ruiz-Chamorro} and E.~Torrontegui,
\newblock \emph{Quantum control of frequency-tunable transmon superconducting
  qubits},
\newblock Phys. Rev. Appl. \textbf{14}(4), 044035 (2020),
\newblock \doi{10.1103/PhysRevApplied.14.044035}.

\bibitem{provostRiemannianStructureManifolds1980}
J.~P. Provost and G.~Vallee,
\newblock \emph{Riemannian structure on manifolds of quantum states},
\newblock Commun. Math. Phys. \textbf{76}(3), 289 (1980),
\newblock \doi{10.1007/BF02193559}.

\bibitem{kolodrubetzGeometryNonadiabaticResponse2017}
M.~Kolodrubetz, D.~Sels, P.~Mehta and A.~Polkovnikov,
\newblock \emph{Geometry and non-adiabatic response in quantum and classical
  systems},
\newblock Phys. Rep. \textbf{697}, 1 (2017),
\newblock \doi{10.1016/j.physrep.2017.07.001}.

\bibitem{bukovGeometricSpeedLimit2019}
M.~Bukov, D.~Sels and A.~Polkovnikov,
\newblock \emph{Geometric {{Speed Limit}} of {{Accessible Many-Body State
  Preparation}}},
\newblock Phys. Rev. X \textbf{9}(1), 011034 (2019),
\newblock \doi{10.1103/PhysRevX.9.011034}.

\bibitem{tomkaAccuracyAdiabaticimpulseApproximation2018}
M.~Tomka, L.~Campos~Venuti and P.~Zanardi,
\newblock \emph{Accuracy of the adiabatic-impulse approximation for closed and
  open quantum systems},
\newblock Phys. Rev. A \textbf{97}(3), 032121 (2018),
\newblock \doi{10.1103/PhysRevA.97.032121}.

\bibitem{suzukiGeneralizedAdiabaticImpulse2022}
T.~Suzuki and H.~Nakazato,
\newblock \emph{Generalized adiabatic impulse approximation},
\newblock Phys. Rev. A \textbf{105}(2), 022211 (2022),
\newblock \doi{10.1103/PhysRevA.105.022211}.

\bibitem{johanssonQuTiPOpensourcePython2012}
J.~R. Johansson, P.~D. Nation and F.~Nori,
\newblock \emph{{{QuTiP}}: {{An}} open-source {{Python}} framework for the
  dynamics of open quantum systems},
\newblock Comput. Phys. Commun. \textbf{183}(8), 1760 (2012),
\newblock \doi{10.1016/j.cpc.2012.02.021}.

\bibitem{johanssonQuTiP2Python2013}
J.~R. Johansson, P.~D. Nation and F.~Nori,
\newblock \emph{{{QuTiP}} 2: {{A Python}} framework for the dynamics of open
  quantum systems},
\newblock Comput. Phys. Commun. \textbf{184}(4), 1234 (2013),
\newblock \doi{10.1016/j.cpc.2012.11.019}.

\bibitem{burkardSemiconductorSpinQubits2023}
G.~Burkard, T.~D. Ladd, A.~Pan, J.~M. Nichol and J.~R. Petta,
\newblock \emph{Semiconductor spin qubits},
\newblock Rev. Mod. Phys. \textbf{95}(2), 025003 (2023),
\newblock \doi{10.1103/RevModPhys.95.025003}.

\bibitem{kolokProtocolsMeasureNonAbelian2024}
B.~Kolok and A.~P{\'a}lyi,
\newblock \emph{Protocols to measure the non-{{Abelian Berry}} phase by pumping
  a spin qubit through a quantum-dot loop},
\newblock Phys. Rev. B \textbf{109}(4), 045438 (2024),
\newblock \doi{10.1103/PhysRevB.109.045438}.

\bibitem{nguyenSuppressingSpinQubit2026}
D.~Q.~L. Nguyen, M.~{Rimbach-Russ} and S.~Bosco,
\newblock \emph{Suppressing spin qubit decoherence during shuttling via
  confinement modulation} (2026), \eprint{2605.00611}.

\bibitem{meinekeSpinQubitLeapfrogging2026}
N.~Meineke and G.~Burkard,
\newblock \emph{Spin-qubit leapfrogging: {{Dynamics}} of shuttling electrons on
  top of another},
\newblock Phys. Rev. Res. \textbf{8}(4), 043043 (2026),
\newblock \doi{10.1103/l1vq-9zhr}.

\bibitem{jakobFastReadoutQuantum2025}
M.~Jakob, K.~Laubscher, P.~{Del Vecchio}, A.~Chatterjee, V.~Fatemi and
  S.~Bosco,
\newblock \emph{Fast readout of quantum dot spin qubits via {{Andreev}} spins},
\newblock Phys. Rev. Appl. \textbf{26}(3), 034009 (2026),
\newblock \doi{10.1103/qcch-2hhy}.

\bibitem{seedhousePauliBlockadeSilicon2021}
A.~E. Seedhouse, T.~Tanttu, R.~C. Leon, R.~Zhao, K.~Y. Tan, B.~Hensen, F.~E.
  Hudson, K.~M. Itoh, J.~Yoneda, C.~H. Yang, A.~Morello, A.~Laucht
  \emph{et~al.},
\newblock \emph{Pauli {{Blockade}} in {{Silicon Quantum Dots}} with
  {{Spin-Orbit Control}}},
\newblock PRX Quantum \textbf{2}(1), 010303 (2021),
\newblock \doi{10.1103/PRXQuantum.2.010303}.

\bibitem{niegemannParitySingletTripletHighFidelity2022}
D.~J. Niegemann, V.~{El-Homsy}, B.~Jadot, M.~Nurizzo, B.~{Cardoso-Paz},
  E.~Chanrion, M.~Dartiailh, B.~Klemt, V.~Thiney, C.~B{\"a}uerle, P.-A.
  Mortemousque, B.~Bertrand \emph{et~al.},
\newblock \emph{Parity and {{Singlet-Triplet High-Fidelity Readout}} in a
  {{Silicon Double Quantum Dot}} at 0.5 {{K}}},
\newblock PRX Quantum \textbf{3}(4), 040335 (2022),
\newblock \doi{10.1103/PRXQuantum.3.040335}.

\bibitem{jirovecSinglettripletHoleSpin2021}
D.~Jirovec, A.~Hofmann, A.~Ballabio, P.~M. Mutter, G.~Tavani, M.~Botifoll,
  A.~Crippa, J.~Kukucka, O.~Sagi, F.~Martins, J.~{Saez-Mollejo}, I.~Prieto
  \emph{et~al.},
\newblock \emph{A singlet-triplet hole spin qubit in planar {{Ge}}},
\newblock Nat. Mater. \textbf{20}(8), 1106 (2021),
\newblock \doi{10.1038/s41563-021-01022-2}.

\bibitem{krzywdaDecoherenceElectronSpin2025}
J.~A. Krzywda and {\L}.~Cywi{\'n}ski,
\newblock \emph{Decoherence of electron spin qubit during transfer between two
  semiconductor quantum dots at low magnetic fields},
\newblock Phys. Rev. B \textbf{111}(11), 115305 (2025),
\newblock \doi{10.1103/PhysRevB.111.115305}.

\bibitem{jirovecDynamicsHoleSingletTriplet2022}
D.~Jirovec, P.~M. Mutter, A.~Hofmann, A.~Crippa, M.~Rychetsky, D.~L. Craig,
  J.~Kukucka, F.~Martins, A.~Ballabio, N.~Ares, D.~Chrastina, G.~Isella
  \emph{et~al.},
\newblock \emph{Dynamics of {{Hole Singlet-Triplet Qubits}} with {{Large}}
  {$g$}-{{Factor Differences}}},
\newblock Phys. Rev. Lett. \textbf{128}(12), 126803 (2022),
\newblock \doi{10.1103/PhysRevLett.128.126803}.

\bibitem{ungererStrongCouplingMicrowave2024}
J.~H. Ungerer, A.~Pally, A.~Kononov, S.~Lehmann, J.~Ridderbos, P.~P. Potts,
  C.~Thelander, K.~A. Dick, V.~F. Maisi, P.~Scarlino, A.~Baumgartner and
  C.~Sch{\"o}nenberger,
\newblock \emph{Strong coupling between a microwave photon and a
  singlet-triplet qubit},
\newblock Nat. Commun. \textbf{15}(1), 1068 (2024),
\newblock \doi{10.1038/s41467-024-45235-w}.

\bibitem{geyerAnisotropicExchangeInteraction2024}
S.~Geyer, B.~Het{\'e}nyi, S.~Bosco, L.~C. Camenzind, R.~S. Eggli, A.~Fuhrer,
  D.~Loss, R.~J. Warburton, D.~M. Zumb{\"u}hl and A.~V. Kuhlmann,
\newblock \emph{Anisotropic exchange interaction of two hole-spin qubits},
\newblock Nat. Phys. \textbf{20}(7), 1152 (2024),
\newblock \doi{10.1038/s41567-024-02481-5}.

\bibitem{kellyIdentifyingMitigatingErrors2025}
E.~G. Kelly, L.~Massai, B.~Het{\'e}nyi, M.~{Pita-Vidal}, A.~Orekhov,
  C.~Carlsson, I.~Seidler, K.~Tsoukalas, L.~Sommer, M.~Aldeghi, S.~W. Bedell,
  S.~Paredes \emph{et~al.},
\newblock \emph{Identification and mitigation of errors in hole spin qubit
  readout},
\newblock Nat. Electron. \textbf{9}(10), 1089 (2026),
\newblock \doi{10.1038/s41928-026-01705-1}.

\bibitem{vanriggelen-doelmanCoherentSpinQubit2024}
F.~{van Riggelen-Doelman}, C.-A. Wang, S.~L. {de Snoo}, W.~I.~L. Lawrie, N.~W.
  Hendrickx, M.~{Rimbach-Russ}, A.~Sammak, G.~Scappucci, C.~D{\'e}prez and
  M.~Veldhorst,
\newblock \emph{Coherent spin qubit shuttling through germanium quantum dots},
\newblock Nat. Commun. \textbf{15}(1), 5716 (2024),
\newblock \doi{10.1038/s41467-024-49358-y}.

\bibitem{vanriggelenPhaseFlipCode2022}
F.~{van Riggelen}, W.~I.~L. Lawrie, M.~Russ, N.~W. Hendrickx, A.~Sammak,
  M.~Rispler, B.~M. Terhal, G.~Scappucci and M.~Veldhorst,
\newblock \emph{Phase flip code with semiconductor spin qubits},
\newblock npj Quantum Inf. \textbf{8}(1), 124 (2022),
\newblock \doi{10.1038/s41534-022-00639-8}.

\bibitem{boscoFullyTunableLongitudinal2022}
S.~Bosco, P.~Scarlino, J.~Klinovaja and D.~Loss,
\newblock \emph{Fully tunable longitudinal spin-photon interactions in {{Si}}
  and {{Ge}} quantum dots},
\newblock Phys. Rev. Lett. \textbf{129}(6), 066801 (2022),
\newblock \doi{10.1103/PhysRevLett.129.066801}.

\bibitem{saez-mollejoExchangeAnisotropiesMicrowavedriven2025}
J.~{Saez-Mollejo}, D.~Jirovec, Y.~Schell, J.~Kukucka, S.~Calcaterra,
  D.~Chrastina, G.~Isella, M.~{Rimbach-Russ}, S.~Bosco and G.~Katsaros,
\newblock \emph{Exchange anisotropies in microwave-driven singlet-triplet
  qubits},
\newblock Nat. Commun. \textbf{16}(1), 3862 (2025),
\newblock \doi{10.1038/s41467-025-58969-y}.

\bibitem{nguyenSingleStepHighFidelityThreeQubit2025}
M.~T. Nguyen, M.~{Rimbach-Russ}, L.~M. Vandersypen and S.~Bosco,
\newblock \emph{Single-{{Step High-Fidelity Three-Qubit Gates}} by
  {{Anisotropic Chiral Interactions}}},
\newblock PRX Quantum \textbf{6}(3), 030326 (2025),
\newblock \doi{10.1103/kp8s-py9m}.

\bibitem{valvoElectricallyTuneableVariability2025}
E.~Valvo, M.~Jakob, P.~{Del Vecchio}, M.~{Rimbach-Russ} and S.~Bosco,
\newblock \emph{Electrically {{Tuneable Variability}} in {{Germanium Hole Spin
  Qubits}}} (2025), \eprint{2512.12702}.

\bibitem{takedaQuantumErrorCorrection2022}
K.~Takeda, A.~Noiri, T.~Nakajima, T.~Kobayashi and S.~Tarucha,
\newblock \emph{Quantum error correction with silicon spin qubits},
\newblock Nature \textbf{608}(7924), 682 (2022),
\newblock \doi{10.1038/s41586-022-04986-6}.

\bibitem{xueQuantumLogicSpin2022}
X.~Xue, M.~Russ, N.~Samkharadze, B.~Undseth, A.~Sammak, G.~Scappucci and
  L.~M.~K. Vandersypen,
\newblock \emph{Quantum logic with spin qubits crossing the surface code
  threshold},
\newblock Nature \textbf{601}(7893), 343 (2022),
\newblock \doi{10.1038/s41586-021-04273-w}.

\bibitem{desmetHighfidelitySinglespinShuttling2025}
M.~De~Smet, Y.~Matsumoto, A.-M.~J. Zwerver, L.~Tryputen, S.~L. {de Snoo}, S.~V.
  Amitonov, S.~R. {Katiraee-Far}, A.~Sammak, N.~Samkharadze, {\"O}.~G{\"u}l,
  R.~N.~M. Wasserman, E.~Greplov{\'a} \emph{et~al.},
\newblock \emph{High-fidelity single-spin shuttling in silicon},
\newblock Nat. Nanotechnol. \textbf{20}(7), 866 (2025),
\newblock \doi{10.1038/s41565-025-01920-5}.

\bibitem{weinbergQuSpinPythonPackage2017}
P.~Weinberg and M.~Bukov,
\newblock \emph{{{QuSpin}}: A {{Python}} package for dynamics and exact
  diagonalisation of quantum many body systems part {{I}}: Spin chains},
\newblock SciPost Phys. \textbf{2}(1), 003 (2017),
\newblock \doi{10.21468/SciPostPhys.2.1.003}.

\bibitem{weinbergQuSpinPythonPackage2019}
P.~Weinberg and M.~Bukov,
\newblock \emph{{{QuSpin}}: A {{Python}} package for dynamics and exact
  diagonalisation of quantum many body systems. {{Part II}}: Bosons, fermions
  and higher spins},
\newblock SciPost Phys. \textbf{7}(2), 020 (2019),
\newblock \doi{10.21468/SciPostPhys.7.2.020}.

\bibitem{virtanenSciPy10Fundamental2020}
P.~Virtanen, R.~Gommers, T.~E. Oliphant, M.~Haberland, T.~Reddy, D.~Cournapeau,
  E.~Burovski, P.~Peterson, W.~Weckesser, J.~Bright, S.~J. {van der Walt},
  M.~Brett \emph{et~al.},
\newblock \emph{{{SciPy}} 1.0: Fundamental algorithms for scientific computing
  in {{Python}}},
\newblock Nat. Methods \textbf{17}(3), 261 (2020),
\newblock \doi{10.1038/s41592-019-0686-2}.

\bibitem{diotalleviSymPTComprehensiveTool2024}
G.~F. Diotallevi, L.~Reascos and M.~Benito,
\newblock \emph{{{SymPT}}: A comprehensive tool for automating effective
  {{Hamiltonian}} derivations} (2024), \eprint{2412.10240}.

\bibitem{dayPymablockAlgorithmPackage2025}
I.~A. Day, S.~Miles, H.~K. Kerstens, D.~Varjas and A.~R. Akhmerov,
\newblock \emph{Pymablock: An algorithm and a package for quasi-degenerate
  perturbation theory},
\newblock SciPost Phys. Codebases \textbf{50} (2025),
\newblock \doi{10.21468/SciPostPhysCodeb.50},
\newblock \eprint{2404.03728}.

\bibitem{finzgarCounterdiabaticDrivingPerformance2025}
J.~R. Fin{\v z}gar, S.~Notarnicola, M.~Cain, M.~D. Lukin and D.~Sels,
\newblock \emph{Counterdiabatic {{Driving}} with {{Performance Guarantees}}},
\newblock Phys. Rev. Lett. \textbf{135}(18), 180602 (2025),
\newblock \doi{10.1103/pqhl-nbtk}.

\bibitem{morawetzUniversalCounterdiabaticDriving2025}
S.~Morawetz and A.~Polkovnikov,
\newblock \emph{Universal {{Counterdiabatic Driving}} in {{Krylov Space}}},
\newblock PRX Quantum \textbf{6}(4), 040320 (2025),
\newblock \doi{10.1103/wbbs-s8fs}.

\bibitem{harrisArrayProgrammingNumPy2020}
C.~R. Harris, K.~J. Millman, S.~J. {van der Walt}, R.~Gommers, P.~Virtanen,
  D.~Cournapeau, E.~Wieser, J.~Taylor, S.~Berg, N.~J. Smith, R.~Kern, M.~Picus
  \emph{et~al.},
\newblock \emph{Array programming with {{NumPy}}},
\newblock Nature \textbf{585}(7825), 357 (2020),
\newblock \doi{10.1038/s41586-020-2649-2}.

\bibitem{hunterMatplotlib2DGraphics2007}
J.~D. Hunter,
\newblock \emph{Matplotlib: {{A 2D}} graphics environment},
\newblock Comput. Sci. Eng. \textbf{9}(3), 90 (2007),
\newblock \doi{10.1109/MCSE.2007.55}.

\bibitem{rachDressingChoppedrandombasisOptimization2015}
N.~Rach, M.~M. M{\"u}ller, T.~Calarco and S.~Montangero,
\newblock \emph{Dressing the chopped-random-basis optimization: {{A}}
  bandwidth-limited access to the trap-free landscape},
\newblock Phys. Rev. A \textbf{92}(6), 062343 (2015),
\newblock \doi{10.1103/PhysRevA.92.062343}.

\end{thebibliography}
